\pdfoutput=1

\documentclass[11pt,twoside,a4paper,cmspaper,final,collab]{cms-tdr}
\def\svnVersion{229122}\def\cmsCernNo{2013-149}\def\cmsCernDate{\today}\def\cmsMessage{Published in the Journal of High Energy Physics as \href{http://dx.doi.org/10.1009/JHEP02(2014)013}{\doi{10.1009/JHEP02(2014)013.}}}

\begin{document}\cmsNoteHeader{SMP-12-002}

\hyphenation{had-ron-i-za-tion}
\hyphenation{cal-or-i-me-ter}
\hyphenation{de-vices}

\RCS$Revision: 221370 $
\RCS$HeadURL: svn+ssh://svn.cern.ch/reps/tdr2/papers/SMP-12-002/trunk/SMP-12-002.tex $
\RCS$Id: SMP-12-002.tex 221370 2013-12-18 13:18:16Z isabel $
\newlength\cmsFigWidth
\ifthenelse{\boolean{cms@external}}{\setlength\cmsFigWidth{0.85\columnwidth}}{\setlength\cmsFigWidth{0.4\textwidth}}
\ifthenelse{\boolean{cms@external}}{\providecommand{\cmsLeft}{top}}{\providecommand{\cmsLeft}{left}}
\ifthenelse{\boolean{cms@external}}{\providecommand{\cmsRight}{bottom}}{\providecommand{\cmsRight}{right}}
\newcommand{\AccEff}{\ensuremath{\mathcal{C}}\xspace}%
\newcommand{\AccEffnorm}{\ensuremath{\mathcal{C}^\text{norm}}\xspace}%
\newcommand{\pp}{\ensuremath{{\Pp\Pp}}}%
\newcommand{\rts}{\ensuremath{\sqrt{s}}}%
\newcommand{\ra}{\ensuremath{\rightarrow}}%
\newcommand{\MN}{\ensuremath{\mu\nu}}%
\renewcommand{\EE}{\ensuremath{\Pep\Pem}}%
\renewcommand{\MM}{\ensuremath{\Pgmp\Pgmm}}%
\newcommand{\EN}{\Pe\cPgn}%
\newcommand{\LN}{\ensuremath{\ell\nu}}%
\newcommand{\MW}{\ensuremath{{m}_\PW}}%
\newcommand{\MZ}{\ensuremath{{m}_\cPZ}}%
\providecommand{\MT}{\ensuremath{M_\cmsSymbolFace{T}}\xspace}%
\newcommand{\MLL}{\ensuremath{{M}_{\ell\ell}}}%

\newcommand{\ubar}{\cPaqu}%
\newcommand{\dbar}{\cPaqd}%
\newcommand{\sbar}{\cPaqs}%
\newcommand{\cbar}{\cPaqc}%
\newcommand{\tbar}{\cPaqt}%
\newcommand{\ppWc}{\ensuremath{\Pp\Pp \to \PW + \cPqc + \mathrm{X}}}%
\newcommand{\ppWpc}{\ensuremath{\Pp\Pp \to \PWp + \cbar}}%
\newcommand{\ppWmc}{\ensuremath{\Pp\Pp \to \PWm + \cPqc}}%
\newcommand{\ppWj}{\ensuremath{\Pp\Pp \to \PW + \text{jets} + \mathrm{X}}}%

\newcommand{\PWmn}{\ensuremath{\PW \ra \MN}}%
\newcommand{\PWpmn}{\ensuremath{\PWp \ra \mu^+\nu}}%
\newcommand{\PWmmn}{\ensuremath{\PWm \ra \mu^-\cPagn}}%
\newcommand{\Zmm}{\ensuremath{\cPZ \ra \MM}}%
\newcommand{\Wtn}{\ensuremath{\PW \ra \tau\nu}}%
\newcommand{\Ztt}{\ensuremath{\cPZ \ra \tau^+\tau^-}}%
\newcommand{\ppZmm}{\pp \ra \cPZ(\gamma^*) + X \ra \MM + X}%
\newcommand{\ppWmn}{\pp \ra \PW + X \ra \MN + X}%
\newcommand{\ppWpmn}{\pp \ra \PWp + X \ra \mu^+\nu + X}%
\newcommand{\ppWmmn}{\pp \ra \PWm + X \ra \mu^-\cPagn + X}%

\newcommand{\Wen}{\ensuremath{\PW \ra \EN}}%
\newcommand{\PWpen}{\ensuremath{\PWp \ra e^+\nu}}%
\newcommand{\PWmen}{\ensuremath{\PWm \ra e^-\cPagn}}%
\newcommand{\Zee}{\ensuremath{\cPZ \ra \EE}}%
\newcommand{\ppZee}{\pp \ra \cPZ(\gamma^*) + X \ra \EE + X}%
\newcommand{\ppWen}{\pp \ra \PW + X \ra \EN + X}%
\newcommand{\ppWpen}{\pp \ra \PWp + X \ra e^+\nu  + X}%
\newcommand{\ppWmen}{\pp \ra \PWm + X \ra e^-\cPagn + X}%

\newcommand{\Wln}{\ensuremath{\PW \ra \LN}}%
\newcommand{\PWpln}{\ensuremath{\PWp \ra \ell^+\nu}}%
\newcommand{\PWmln}{\ensuremath{\PWm \ra \ell^-\cPagn}}%
\newcommand{\Zll}{\ensuremath{\cPZ \ra \ell^+ \ell^-}}%
\newcommand{\ppZll}{\pp \ra \cPZ(\gamma^*) + X \ra \ell^+ \ell^- + X}%
\newcommand{\ppWln}{\pp \ra \PW + X \ra \LN + X}%
\newcommand{\ppWpln}{\pp \ra \PWp + X \ra \ell^+\nu  + X}%
\newcommand{\ppWmln}{\pp \ra \PWm + X \ra \ell^-\cPagn + X}%

\newcommand{\PWptn}{\ensuremath{\PWp \ra \tau^+\nu_\tau}}%
\newcommand{\PWmtn}{\ensuremath{\PWm \ra \tau^-\cPagn}}%

\newcommand{\Wev}{\Wen}
\newcommand{\PWmv}{\PWmn}
\newcommand{\Ztautau}{\Ztt}
\renewcommand{\MET}{\met}
\newcommand{\gammaZ}  {\ensuremath{\gamma^{*}\cPZ}}
\newcommand{\gammaZmm}{\ensuremath{\gamma^{*}\cPZ\rightarrow \MM}}
\newcommand{\gammaZee}{\ensuremath{\gamma^{*}\cPZ\rightarrow \Pep \Pem}}
\newcommand{\gammaZtt}{\ensuremath{\gamma^{*}\cPZ\rightarrow \tau^{+}  \tau^{-}}}
\newcommand{\gammaZll}{\ensuremath{\gamma^{*}\cPZ\rightarrow \ell^{+}  \ell^{-}}}
\newcommand{\mz}{\ensuremath{m_{\cPZ}}}
\newcommand{\hta}{\ensuremath{\eta$}}
\newcommand{\fh}{\ensuremath{\phi$}}
\newcommand{\etot}{\ensuremath{\epsilon_\text{tot}}}

\newcommand{\SXY}{\ensuremath{\sigma_{XY}}}
\newcommand{\pth}{\hat{p}_{\perp}}
\newcommand{\pTrel}{p_{T,\text{rel}}\xspace}
\newcommand{\ETmissperp}{\ensuremath{E_{\mathrm{T},\perp}^{\text{miss}}}\xspace}
\newcommand{\ETmisspara}{\ensuremath{E_{\mathrm{T},\parallel}^{\text{miss}}}\xspace}
\newcommand{\Lint}{\ensuremath{\mathcal{L}_{\text{int}}}}
\newcommand{\IECAL}    {I_{\text{ECAL}}}%
\newcommand{\IHCAL}    {I_{\text{HCAL}}}%
\newcommand{\ITRK}     {I_{\text{trk}}}%
\newcommand{\IRelComb} {I^{\text{rel}}_{\text{comb}}}%
\newcommand{\IRel}     {I^{\text{rel}}}%
\newcommand{\IRelmu}   {I^\mu_{\text{rel}}}%
\newcommand{\IRele}    {I^{\Pe}_{\text{rel}}}%
\newcommand{\IComb}    {I_{\text{comb}}}%
\newcommand{\Nev}    {N_{\text{ev}}}
\newcommand{\Nsig}   {N_{\text{sig}}}
\newcommand{\Nsel}   {N_{\text{sel}}}
\newcommand{\Nbg}    {N_{\text{bg}}}
\newcommand{\rhoeff} {\rho_{\text{eff}}}
\newcommand{\effmc}  {\epsilon_{\text{sim}}}
\newcommand{\effdt}  {\epsilon_{\text{data}}}

\newcommand{\OSSS} {\ensuremath{\mathrm{OS}-\mathrm{SS}}}%
\newcommand{\Wc}   {\ensuremath{\PW + \cPqc}}
\newcommand{\Wuds} {\ensuremath{\PW + \mathrm{light}}}
\newcommand{\Wj}   {\ensuremath{\PW + \text{jets}}}
\newcommand{\Wcc}  {\ensuremath{\PW + \ccbar}}
\newcommand{\Wbb}  {\ensuremath{\PW + \bbbar}}
\newcommand{\WbX}  {\ensuremath{\PW + \cPqb+\mathrm{X}}}
\newcommand{\PWpj}  {\ensuremath{\PWp + \text{jets}}}
\newcommand{\PWmj}  {\ensuremath{\PWm + \text{jets}}}
\newcommand{\PWpc}  {\ensuremath{\PWp + \cPaqc}}
\newcommand{\PWmc}  {\ensuremath{\PWm + \cPqc}}
\newcommand{\SWpc}  {\sigma(\PWpc)}
\newcommand{\SWmc}  {\sigma(\PWmc)}
\newcommand{\SWc}  {\ensuremath{\sigma(\Wc)}}
\newcommand{\SWcdifft} {\ensuremath{\frac{\rd\sigma(\Wc)}{\rd\abs{\eta}} }}
\newcommand{\SWcdiff}  {\ensuremath{\frac{1}{\sigma(\Wc)} \frac{\rd{}\sigma(\Wc)}{\rd{}\abs{\eta}} }}
\newcommand{\SWcdifftline} {\ensuremath{d\sigma(\Wc)/\rd\abs{\eta} }}
\newcommand{\SWcdiffline}  {\ensuremath{(1/\sigma(\Wc))\,\rd\sigma(\Wc)/\rd\abs{\eta} }}
\newcommand{\SWj}  {\ensuremath{\sigma(\Wj)}}
\newcommand{\Rcpm}  {\ensuremath{R_\cPqc^{\pm}}}
\newcommand{\Rc}  {\ensuremath{R_c}}
\newcommand{\rhocpm}  {\ensuremath{\rho_c^{\pm}}}
\newcommand{\rhoc}  {\ensuremath{\rho_c}}

\newcommand{\Reff}   {\ensuremath{R_{\epsilon}}}
\newcommand{\Lmu}    {\ensuremath{L_{\mu}}}

\newcommand{\Dpm}     {\ensuremath{\mathrm{D^{\pm}}}}
\newcommand{\Dz}      {\ensuremath{\mathrm{D^0}}}
\newcommand{\Ds}      {\ensuremath{\mathrm{D_{s}}}}
\newcommand{\Lambdac} {\ensuremath{Lambda_\cPqc}}
\newcommand{\Dstar}   {\ensuremath{{\mathrm{D}^{\ast\pm}(2010)}}}

\newcommand{\Dp}     {\ensuremath{\mathrm{D^+}}}
\newcommand{\Dstarp} {\ensuremath{{\mathrm{D}^{\ast+}(2010)}}}
\newcommand{\Dm}     {\ensuremath{\mathrm{D^-}}}
\newcommand{\Dstarm} {\ensuremath{{\mathrm{D}^{\ast-}(2010)}}}
\newcommand{\Dzbar}  {\ensuremath{\bar{\mathrm{D}}^0}}

\newcommand{\rec}  {\text{rec}}

\newcommand{\FEWZ} {\textsc{fewz}\xspace}
\cmsNoteHeader{SMP-12-002} 
\title{Measurement of associated W + charm production in pp collisions at $\sqrt{s} = 7\TeV$}

\date{\today}

\abstract{
       Measurements are presented of the associated production of a W boson and a
charm-quark jet (W + c) in pp collisions at a center-of-mass energy of 7\TeV.
The analysis is conducted with a data sample corresponding to a total
integrated luminosity of 5\fbinv, collected by the CMS detector at the LHC.
W boson candidates are identified by their
decay into a charged lepton (muon or electron) and a neutrino.
The W + c measurements are performed for charm-quark jets in the kinematic region
$\pt^\text{jet}>25\GeV$, $\abs{\eta^\text{jet}}<2.5$, for
two different thresholds for the transverse momentum of the lepton from the W-boson decay, and
in the pseudorapidity range $\abs{\eta^\ell}<2.1$.
Hadronic and inclusive semileptonic decays of charm hadrons are used to
measure the following total cross sections:
$\sigma(\Pp\Pp \rightarrow \mathrm{\PW+c + X}) \times \mathcal{B}(\PW \rightarrow \ell\nu ) = 107.7 \pm 3.3\stat \pm 6.9\syst\unit{pb}~(\pt^\ell>25\GeV)$ and
$\sigma(\Pp\Pp \rightarrow \mathrm{W+c + X}) \times \mathcal{B}(\PW \rightarrow \ell\nu ) = 84.1 \pm 2.0\stat \pm 4.9\syst\unit{pb}~(\pt^\ell>35\GeV)$,
and the cross section ratios
$\sigma(\Pp\Pp \rightarrow \mathrm{\PWp + \cPaqc +X})/\sigma(\Pp\Pp \rightarrow \mathrm{\PWm + c + X}) = 0.954 \pm 0.025\stat \pm 0.004\syst~(\pt^\ell>25\GeV)$ and
$\sigma(\Pp\Pp \rightarrow \mathrm{\PWp + \cPaqc +X})/\sigma(\Pp\Pp \rightarrow \mathrm{\PWm + c + X}) = 0.938 \pm 0.019\stat \pm 0.006\syst~(\pt^\ell>35\GeV)$.
Cross sections and cross section ratios are also measured differentially with respect to
the absolute value of the pseudorapidity of the lepton from the W-boson decay.
These are the first measurements from the LHC directly sensitive to the strange quark
and antiquark content of the proton.
Results are compared with theoretical predictions
and are consistent with the predictions based on global fits of parton distribution functions.
}

\hypersetup{%
pdfauthor={CMS Collaboration},%
pdftitle={Measurement of associated W + charm production in pp collisions at sqrt(s) = 7 TeV},%
pdfsubject={CMS},%
pdfkeywords={CMS, physics, EWK, Standard Model Physics}}

\maketitle

\section{Introduction}

\begin{figure}[htbp]
\begin{center}
    \includegraphics[width=6.cm]{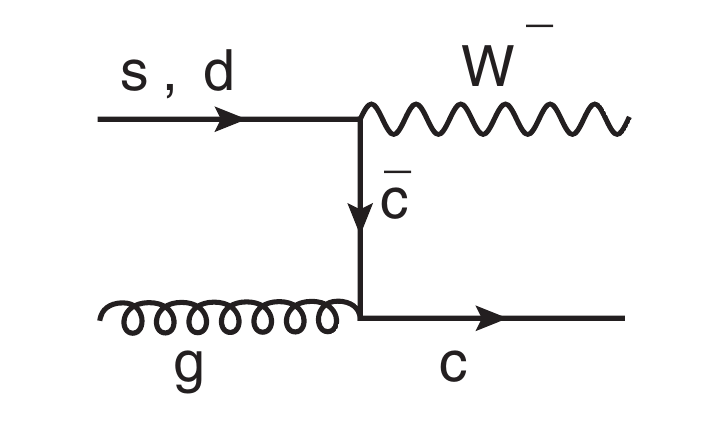} \hspace{2.cm}
    \includegraphics[width=6.cm]{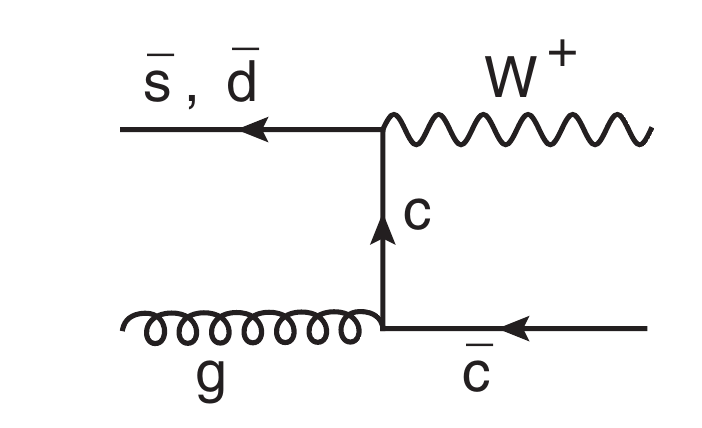}
\caption{Main diagrams at the hard-scattering level for associated W + c
      production at the LHC.}
\label{fig:wcprod}
\end{center}
\end{figure}

The study of associated production of a $\PW$ boson and a charm (c) quark at hadron colliders
(hereafter referred to as $\Wc$ production) provides direct access to the strange-quark content
of the proton at an energy scale of the order of the $\PW$-boson mass
($Q^2{\sim} (100\GeV)^2)$~\cite{Baur, Charm_Kusina,Charm_Eleni}.
This sensitivity is due to the dominance of $\cPqs\Pg \to \PWmc$ and
$\ensuremath{\mathrm{\cPaqs g}}\to \PWpc$ contributions at the hard-scattering level
(Fig.~\ref{fig:wcprod}). Recent work~\cite{Ball:2012wy} indicates that precise
measurements of this process at the Large Hadron Collider (LHC) may
significantly reduce the uncertainties in the strange quark and antiquark parton distribution functions
(PDFs) and help resolve existing ambiguities and limitations of
low-energy neutrino deep-inelastic scattering (DIS) data~\cite{NNPDF:NuTeV}. More precise knowledge of the PDFs is essential
for many present and future precision analyses, such as the measurement of the
W-boson mass~\cite{Rojo_Wmass}.
An asymmetry between the strange quark and antiquark PDFs has also been proposed as an explanation of the NuTeV
anomaly~\cite{NNPDF:NuTeV}, making it crucial to measure observables related to this asymmetry with high precision.

$\Wc$ production receives contributions at a few percent level from the
processes $\cPqd\Pg\to \PWmc$ and $\cPaqd\Pg\to \PWpc$,
which are Cabibbo suppressed~\cite{Cabibbo}.
Overall, the $\PWmc$ yield is expected to be slightly larger than the
$\PWpc$ yield at the LHC because of the participation of down valence quarks in the initial
state.  A key property of the $\Pq\Pg\to \PW+\cPqc$ reaction
is the presence of a charm quark and a W boson with opposite-sign charges.

The $\ppWc$ process is a sizable background for
signals involving bottom or top quarks and missing transverse energy in the final state.
Particularly relevant cases are top-quark studies and third-generation
squark searches.
Measurements of the $\Pp\Pap\to\Wc+\mathrm{X}$ cross section
and of the cross section ratio $\sigma({\Pp\Pap\to\PW+\cPqc\text{-jet}+\mathrm{X}})/\sigma({\Pp\Pap\to\PW+\text{jets}+\mathrm{X}})$
have been performed with a relative precision of about 20--30\%
at the Tevatron~\cite{CDF-first,CDF,D0} hadron collider using semileptonic charm hadron decays.

We present a detailed study of the $\ppWc$ process with the
Compact Muon Solenoid (CMS) detector, using a data sample corresponding to a total integrated luminosity of $5 \fbinv$ collected in 2011
at a center-of-mass energy of $7\TeV$.
We measure the total cross section and the cross section ratio $\Rcpm=\SWpc/\SWmc$
using the muon and electron decay channels of the W boson.
Charm-quark jets are identified within the fiducial region of transverse momentum $\pt^\text{jet}>25\GeV$ and
pseudorapidity $\abs{\eta^\text{jet}}<2.5$ using exclusive hadronic,  inclusive hadronic, and
semileptonic decays of charm hadrons.
Furthermore, the cross section and the $\Rcpm$ ratio are measured
as a function of the pseudorapidity of the lepton from the W decay, thus probing a wide range in
the Bjorken $x$ variable, which at leading order can be interpreted as the momentum fraction of the proton carried by the interacting parton.

This paper is organized as follows:
the CMS detector is briefly described in
Section~\ref{sec:detector} and the general analysis strategy is outlined
in Section~\ref{sec:AnalysisStrategy}. The samples used to carry out the measurement
and the event selection criteria are presented in Sections~\ref{sec:samples}
and~\ref{sec:selection}. Section~\ref{sec:total_xsec} details the measurement of the
total cross section and Sections~\ref{sec:differential_xec} and \ref{sec:ratio} are
devoted to studies of the differential cross section and the charge ratio.
Results and comparisons with theoretical predictions
are discussed in Section~\ref{sec:results}.
Finally, we summarize the results of this paper in Section~\ref{sec:summary}.

\section{CMS detector\label{sec:detector}}

The central feature of the CMS apparatus is a superconducting solenoid of 6\unit{m} internal diameter,
providing a magnetic field of 3.8\unit{T}. Within the field volume are a silicon pixel and
strip tracker, an electromagnetic calorimeter (ECAL), and a brass/scintillator hadron calorimeter
(HCAL). Muons are detected in gas-ionization detectors embedded in the steel flux return yoke of the magnet.

The CMS experiment uses a right-handed coordinate system with the origin at the
nominal interaction point, the $x$ axis pointing to the center of
the LHC ring, the $y$ axis pointing up (perpendicular to the LHC plane),
and the $z$ axis along the anticlockwise-beam direction. The polar
angle $\theta$ is measured from the positive $z$ axis and the
azimuthal angle $\phi$ is measured in the $x$-$y$ plane.
The pseudorapidity is given by $\eta = -\ln(\tan(\theta/2))$.

The tracker measures charged-particle trajectories in the pseudorapidity range $\abs{\eta} \le2.5$.
It consists of 1440 silicon pixel and 15\,148 silicon strip detector modules. It provides an impact
parameter resolution of $15\mum$ and a transverse momentum ($\pt$) resolution of about 1\% for
charged particles with \pt around $40\GeV$.
The ECAL consists of nearly 76\,000 lead tungstate crystals, which provide
coverage in pseudorapidity $\abs{\eta} \le1.479$ in a cylindrical barrel region and
$1.479\le \abs{\eta}\le3.0$
in two endcap regions (EE). A preshower detector, consisting of two planes of silicon sensors
interleaved with a total of three radiation lengths of lead, is located in front of the EE. The ECAL has an ultimate
energy resolution of better than 0.5\% for unconverted photons with transverse energies (\ET) above
$100\GeV$. The energy resolution is 3\% or better for the range of electron energies relevant for
this analysis. The HCAL is a sampling device with brass as passive material
and scintillator as active material. The combined calorimeter cells are grouped in projective
towers of granularity $\Delta \eta \times \Delta \phi = 0.087 \times 0.087$ at central rapidities and
$ 0.175\times 0.175$ at forward
rapidities.
Muons are detected in the pseudorapidity range $\abs{\eta} \le2.4$, with detection planes based on
three technologies: drift tubes, cathode strip chambers, and resistive-plate chambers. A high-$\pt$
muon originating from the interaction point produces track segments in typically three or four
muon stations. Matching these segments to tracks measured in the inner tracker results in a $\pt$
resolution between 1\% and 2\% for $\pt$ values up to $100\GeV$.
The first level of the CMS trigger system, composed of custom hardware processors, is
designed to select the most interesting events in less than 1 $\mus$ using information from the
calorimeters and muon detectors. The high-level trigger processor farm further decreases
the event rate to a few hundred hertz before data storage.
A more detailed description of CMS can be found elsewhere~\cite{JINST}.

\section{Analysis strategy~\label{sec:AnalysisStrategy}}

We study $\Wc$ associated production in final states containing
a $\Wln$ decay (where $\ell = \mu$ or $\Pe$) and a leading jet
with charm-quark content. Jets originating from a $\cPqc$ ($\cPaqc$) parton
are identified using one of the three following signatures:
a displaced secondary vertex with three tracks and an invariant mass consistent with a
$\Dp\to \PKm\pi^+\pi^+$ ($\Dm\to \PKp\pi^-\pi^-$) decay; a displaced secondary vertex with two tracks
consistent with a $\Dz\to \PKm\pi^+$ ($\Dzbar\to \PKp\pi^-$) decay and associated with a
previous $\Dstarp\to\Dz\pi^+$ ($\Dstarm\to\Dzbar\pi^-$) decay at the primary vertex;
or a semileptonic decay leading to a well-identified muon. In total, since
 both electron and muon channels are considered in the $\PW$-boson decay,
six different final states are explored.

The $\Dpm$, $\Dstar$, and $\cPqc \to\ell\nu+{\rm X}$ decays provide a direct measurement of
the charm-quark jet charge, which is a powerful tool to disentangle the $\Wc$
signal component from most of the background processes.
We define two types of distributions: opposite-sign distributions,
denoted by OS, are built on samples containing a $\PW$ boson and a charm-quark jet
with an opposite-charge sign; same-sign distributions, denoted by SS,
are built from samples where the $\PW$ boson and the charm-quark jet have the same charge sign.
The final distributions used in the analysis are obtained by subtracting the SS distribution from the OS distribution
(referred to as $\OSSS$) for any
given variable. This subtraction has no effect on the signal at leading order.
In contrast, $\Wcc$ and $\Wbb$ events provide the same
OS and SS contributions and are suppressed in
$\OSSS$ distributions.
Moreover, any $\OSSS$ asymmetry present in
$\ttbar$, single-top-quark, or $\Wuds$-quark jet backgrounds is found to be negligible
according to simulations. As a consequence, $\OSSS$
distributions are largely dominated by the $\Wc$ component, allowing for many detailed
studies of the $\ppWc$ process.

Using displaced secondary vertices
is a simple way to suppress backgrounds, such as Drell--Yan events, $\Wuds$-quark jet, and
multijet final states with no heavy-flavour content. It also reduces backgrounds
containing b-hadron decays, which often lead to secondary vertices with a higher track multiplicity than a
typical D-meson decay.

The sample containing semileptonic charm decays is complementary; it is a larger data sample
but is more affected by backgrounds, in particular Drell--Yan events.
Exclusive identification of $\Dpm$ and $\Dstar$ final states allows for a precise accounting
of systematic uncertainties in charm branching fractions and acceptances for cross section measurements.
However, only charge identification is strictly required for studies that are independent of
the overall $\Wc$ normalization, such as relative differential measurements or
measurements of the $\SWpc/\SWmc$ ratio.

In order to improve the statistical precision, we also employ
inclusive selections of charm hadron decays, \ie without requiring the identification of the full final state,
thus allowing for decays with one or more neutral particles.
Inclusive samples of events with three-track and two-track secondary vertices are selected by loosening the invariant mass constraints.
Even with these relaxed criteria, simulations predict that the
background contributions to the $\OSSS$ subtracted distributions in these inclusive samples are small compared with
the signal yield.

\section{Data and Monte Carlo samples and signal definition\label{sec:samples}}

The analysis reported in this paper was performed with a data sample of proton-proton collisions at $\sqrt{s} = 7\TeV$
collected with the CMS detector in 2011.
A detailed data certification process~\cite{DQM} guarantees
that the data set available for analysis, corresponding to an integrated luminosity $\lumi= 5.0 \pm 0.1\fbinv$,
fulfills the quality requirements for all detectors used in this analysis.
Candidate events for the muon decay channel of the $\PW$ boson are selected online by a single-muon trigger
that requires a reconstructed muon with $\pt>24\GeV$.
Candidate events for the electron channel are selected by a variety of electron triggers.
Trigger conditions were tightened throughout the 2011 data run to cope with the
increasing instantaneous luminosity of the LHC collider. Most of the data used in this analysis are selected by requiring
an electron candidate with transverse energy $\et >32\GeV$.

Muon and electron candidates are reconstructed following standard CMS algorithms~\cite{CMS-PAPER-MUO-10-004, EGAMMA-PAS}.
Jets, missing transverse energy, and related quantities are computed using particle-flow techniques~\cite{CMS-PAS-PFT-2010-003}
in which a full reconstruction of the event is developed from the individual particle signals in the different subdetectors.
Jets are reconstructed from the particle-flow candidates
using an anti-\kt clustering algorithm~\cite{antikt}
with a distance parameter of 0.5.
Charged particles with tracks not originating at the primary vertex are not considered for the
jet clustering,
and the extra energy clustered in jets from the presence of additional pp interactions (pileup events)
is subtracted from the jet energy~\cite{PUsubtraction, PUsubtraction2}.
Finally, energy corrections derived from data and simulated samples are applied
to correct for $\eta$ and $\pt$ dependent detector effects~\cite{CMS-PAPER-JME-10-011}.

Large samples of events simulated with Monte Carlo (MC) techniques are used to evaluate signal
and background efficiencies.
 The $\PW$-boson signal ($\PWmn$ and $\Wen$) as well as other electroweak processes (such as $\cPZ \rightarrow \mu\mu$, $\cPZ \rightarrow \Pe\Pe$,
$\PW \rightarrow \tau\nu$, and $\cPZ \rightarrow \tau\tau$ production) are generated
with the $\MADGRAPH$~\cite{MADGRAPH5} (v5.1.1) event generator,
interfaced to the \PYTHIA~\cite{Pythia} (v6.4.24) program for parton shower simulation. The $\MADGRAPH$ generator
produces parton-level events with a vector boson and up to four partons in the final state on the basis
of matrix element calculations. It has been shown to reproduce successfully the observed jet multiplicity and kinematic properties of
$\Wj$ final states at the LHC energy regime~\cite{VJETS-paper}.
The matching matrix element/parton shower scale $m^2$ is equal to $(10\GeV)^2$ and the factorization and renormalization scales are set to $Q^2 = M^2_{\PW/\cPZ} + p^2_{\mathrm{T,}\PW/\cPZ}$.
Constraints on the phase space at the generator level are not imposed, except for
the condition $M_{\ell\ell}>10\GeV$ in the case of $\cPZ(\gamma^*)$ production.

Potential backgrounds in this analysis come from $\ttbar$ and single-top-quark production.
A sample of $\ttbar$ events is generated with the \MADGRAPH generator interfaced to \PYTHIA.
Single-top-quark events are generated in the $t$-channel, $s$-channel, and $\cPqt\PW$ associated modes
with the next-to-leading-order (NLO) generator \POWHEG~\cite{POWHEG} (v1.0),
interfaced with \PYTHIA.
The PDF set used in these \POWHEG productions is CT10~\cite{CT10}.
We also consider the small contributions from diboson ($\PW\PW$, $\PW\cPZ$, $\cPZ\cPZ$) events and quantum chromodynamics (QCD)
multijet events using \PYTHIA.
All leading-order (LO) generations use the CTEQ6L1 PDF set~\cite{CTEQ6L1} with
parameters set for the underlying event according to the Z2 tune~\cite{Z2Tune}.

Cross sections for single W and Z production processes are normalized
to the predictions from $\FEWZ$~\cite{fewz} evaluated at
next-to-next-to-leading order (NNLO) using  the MSTW08NNLO~\cite{MSTW08} PDF set.
The $\ttbar$ cross section is taken at NNLO from Ref.~\cite{Czakon:2013goa}.
For the rest of the processes, cross sections are normalized to the NLO
cross section predictions from \MCFM~\cite{MCFM} using the MSTW08NLO PDF set.
The QCD multijet cross section is evaluated at LO.

Several minimum-bias interactions, as expected from the projected running conditions of the accelerator,
are superimposed on the hard scattering to simulate
the real experimental conditions
of multiple pp collisions occurring simultaneously.
To reach an optimal agreement with the experimental data, the simulated distributions
are reweighted according to the actual number of interactions (an average of nine) occurring
given the instantaneous luminosity for each bunch crossing.
Generated events are processed through the full
$\GEANTfour$~\cite{GEANT4} detector simulation, trigger emulation, and event
reconstruction chain of the CMS experiment.
Predictions derived from the MC-simulated samples are normalized to the integrated luminosity of the data sample.

At the hard-scattering level we identify $\Wc$ signal events as those containing an
odd number of charm partons in the final state. This choice provides a simple
operational definition of the process and ensures
that pure QCD splittings of the $\Pg \ra\ccbar$ type
are associated with the background.
Events containing b quarks in the final state are always classified as $\WbX$ in order
to correctly identify $\cPqb\to \cPqc$ decays.
The $\Wc$ signal reference is defined at the hard-scattering
level of $\MADGRAPH$, which provides an implicit parton-jet matching
for a jet separation parameter of $R = \sqrt{{(\Delta\eta)}^2+{(\Delta\phi)}^2} = 1$ that is suitable for
comparisons with the NLO theoretical predictions of \MCFM at the ${\lesssim} 1\%$
level. The phase space definition at the generator level is chosen in order to approximately match
the experimental selections used in the analysis. For charm partons
we require $\pt^{\cPqc}>25\GeV$, $\abs{\eta^\cPqc}<2.5$.
Differential measurements are performed as a function of the absolute value of the lepton pseudorapidity $\abs{\eta^\ell}$,
whereas total cross sections and average ratios require $\abs{\eta^\ell} < 2.1$. Potential dependencies on the center-of-mass energy
of the hard scattering process are explored by considering two different transverse momentum thresholds for the charged leptons from the $\PW$-boson decay:
$\pt^\ell > 25\GeV$ and $\pt^\ell > 35\GeV$. The $\pt^\ell > 25\GeV$ case is analyzed in the $\PWmn$ channel only.

\section{Event selection\label{sec:selection}}

The selection of $\PW$-boson candidates closely follows the criteria used in the analysis of inclusive $\PWmn$ and $\Wen$
production~\cite{WZ-paper}.
The leptonic decay of a $\PW$ boson into a muon or an electron, and a neutrino is characterized by the presence of a high-transverse momentum, isolated lepton.
The neutrino escapes detection causing an apparent imbalance in the transverse energy of the event.
Experimentally, the magnitude of the vector momentum imbalance in the plane perpendicular to the beam direction defines the missing transverse energy of an event, $\ETmiss$.
In $\PW$-boson events, this variable is an estimator of the transverse energy of the undetected neutrino.

Muon tracks are required to have a transverse momentum $\pt^\mu > 25\GeV$
and to be measured in the pseudorapidity range $\abs{\eta^\mu} < 2.1$.
A muon isolation variable, $\IRelmu$, is defined as the sum of the transverse energies of neutral particles and momenta of
charged particles (except for the muon itself) in a $\Delta R = \sqrt{{(\Delta\eta)}^2+{(\Delta\phi)}^2}=0.4$ cone
around the direction of the muon, and normalized to the muon transverse momentum.
The muon is required to be isolated from any other detector activity according to the criterion
$\IRelmu<0.12$.

Electron candidates with $\pt^\Pe > 35\GeV$
are accepted in the pseudorapidity range $\abs{\eta^\Pe} < 2.1$ with the exception of the region
$1.44 < \abs{\eta^\Pe} < 1.57$ where service infrastructure for the detector is located, thus degrading the performance.
The electron isolation variable, $\IRele$, is defined as the sum of the transverse components of ECAL and HCAL energy
deposits (excluding the footprint of the electron candidate) and transverse momenta of
tracks reconstructed in the inner tracker in a $\Delta R=0.3$ cone around the electron direction,
and normalized to the electron $\pt$.
An isolated electron must satisfy $\IRele<0.05$.

The background arising from Drell--Yan processes is reduced by removing
events containing additional muons (electrons) with $\pt^\ell>25~(20)\GeV$ in the pseudorapidity region
$\abs{\eta^\Pgm} < 2.4$ ($\abs{\eta^\Pe} < 2.5$).
Finally, the reconstructed transverse mass, $\MT$,  which is built from the transverse momentum of the isolated lepton, $\pt^\ell$, and the missing
transverse energy in the event, $\MT \equiv \sqrt{\smash[b]{2~\pt^\ell~\ETmiss~[1-\cos(\phi_\ell-\phi_{\ETmiss})]}}$,
where $\phi_\ell$ and $\phi_{\ETmiss}$ are the azimuthal angles of the lepton and the $\ETmiss$ vector,
must be large. In the muon channel, $\MT$ must be greater than $40\GeV$. A higher threshold is set in the electron channel, $\MT > 55\GeV$, since a
condition on this variable ($\MT > 50\GeV$) is already included in the online trigger selection.
This requirement reduces the QCD multijet background to a negligible level in the muon channel.
Residual QCD background in the electron channel is estimated from the experimental $\ETmiss$ distribution.
It is found to be negligible after subtraction of the SS component.

A $\Wj$ sample is selected by demanding the presence
of at least one jet with $\pt^\text{jet}>25\GeV$ in the
pseudorapidity range $\abs{\eta^\text{jet}}<2.5$, thus ensuring that the jet
passes through the tracker volume, and hence achieving the best possible jet $\pt$ resolution.
A $\Wc$ candidate sample is further selected by searching for a distinct signature of a charmed particle decay among the constituents of the leading
jet associated with the $\PW$ boson, as introduced in Section~\ref{sec:AnalysisStrategy}.
For that purpose, events with a secondary vertex consistent with the decay of a relatively long-lived quark are kept.
Secondary vertices are reconstructed using an adaptive vertex finder~\cite{Vertex}
algorithm with well understood performance~\cite{CMS-PAPER-BTV-12-001}. This algorithm is
stable with respect to alignment uncertainties and is an essential component of
the vertex-based b-tagging algorithms used in the CMS experiment.
In its default implementation, used in this analysis, tracks within a
$\Delta R=0.3$ cone around the jet axis, that have a transverse momentum larger than $1\GeV$
and a probability of originating from the primary vertex below $50\%$ are
considered to come from a secondary vertex. Finally, only secondary vertices
with a transverse decay length significance with respect to
the primary vertex position larger than 3 are kept.

A search for $\Dpm$ and $\Dz$ charm meson decays is carried out in those events having
reconstructed secondary vertices with three or two tracks, respectively.
In addition, a $\Wc$ candidate sample with the charm quark decaying semileptonically is selected from the
events with an identified muon among the
particles constituting the jet. These samples are described
in more detail in the following subsections.

\subsection{Selection of exclusive \texorpdfstring{$\Dpm$}{D +/-} decays~\label{sec:Dpm}}

We identify $\Dpm\to \PK^\mp \pi^\pm\pi^\pm$ decays in the selected $\Wj$ sample using
secondary vertices with three tracks and
a reconstructed invariant mass within $50\MeV$ of the $\Dpm$ mass,
$1869.5 \pm 0.4\MeV$~\cite{PDG}. The kaon mass is assigned to
the track that has opposite sign to the total charge of the three-prong vertex and
the remaining tracks are assumed to have the mass of a charged pion.
This assignment is correct in more than 99\% of the cases, since
the fraction of double Cabibbo-suppressed decays is very small:
${\mathcal{B}}(\PDp\to \PKp \pi^+\pi^-)/{\mathcal{B}}(\Dp\to \PKm \pi^+\pi^+) = 0.00577 \pm 0.00022$~\cite{PDG}.

\begin{figure}[htb]
\begin{center}
    \includegraphics[width=0.49\textwidth]{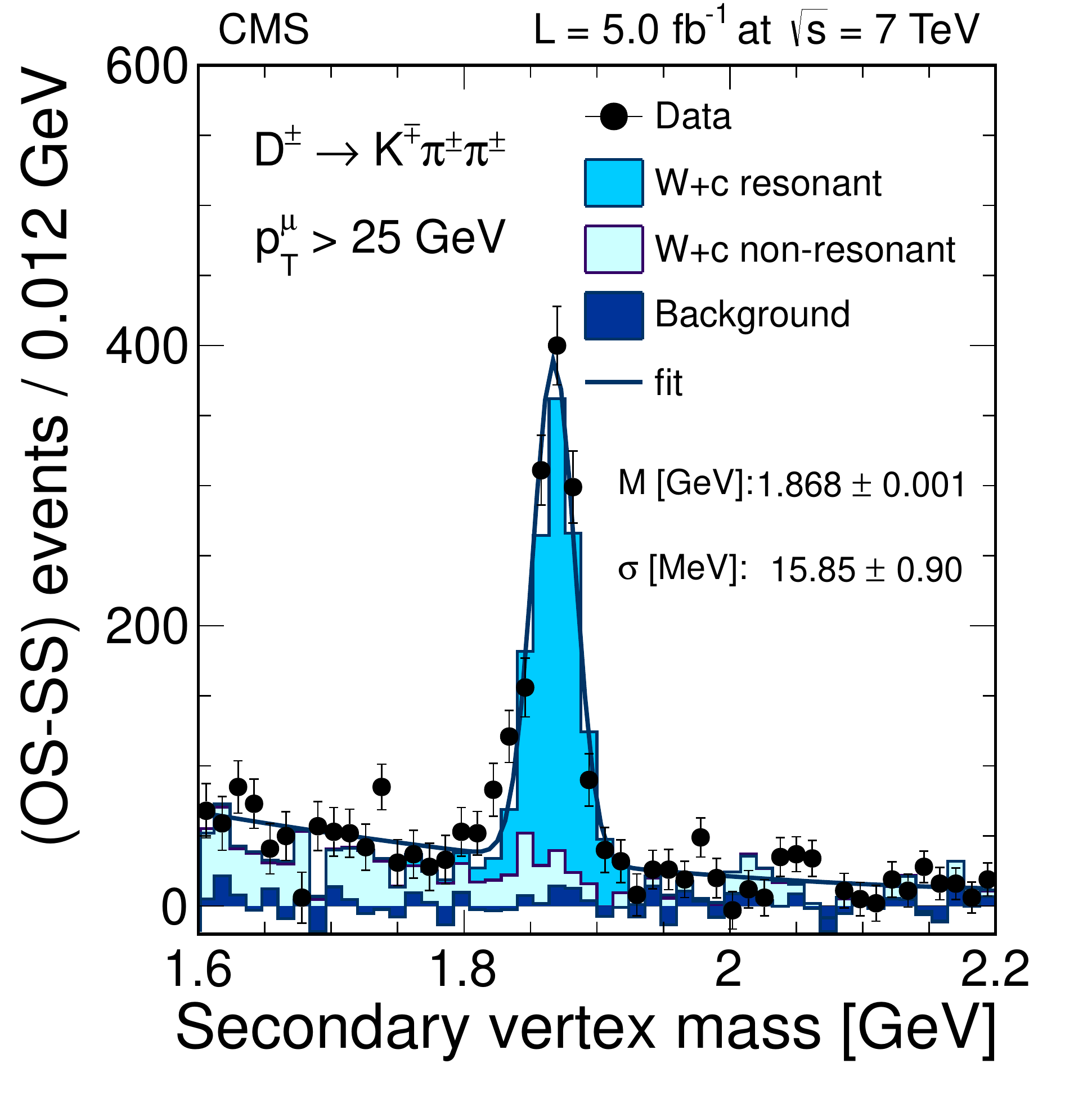}
    \includegraphics[width=0.49\textwidth]{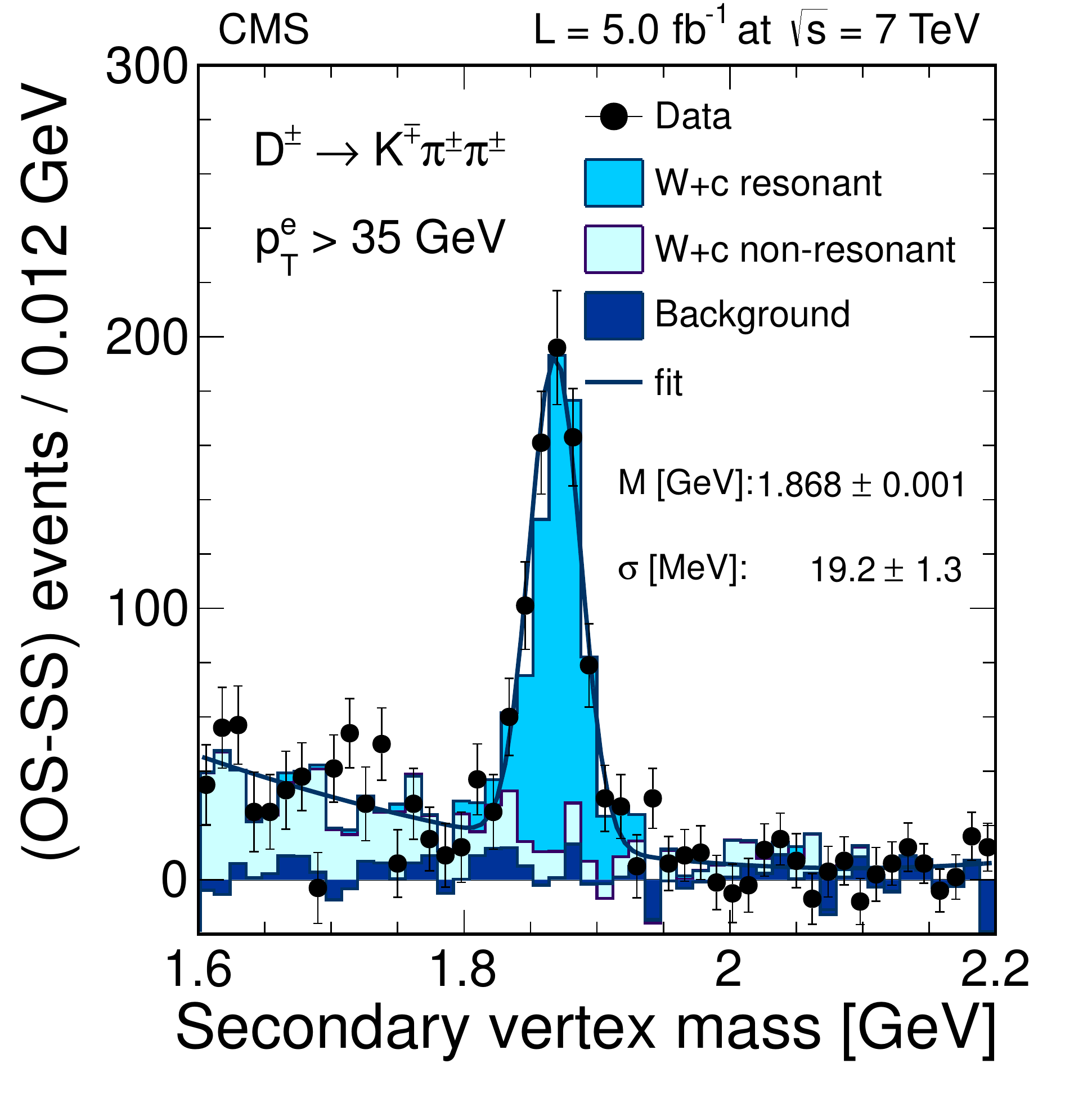}
    \caption{The invariant mass distribution of three-prong secondary
vertices in data, after subtraction of the SS component. The position and width of
the resonance peak are in reasonable agreement with the MC expectations (only
statistical uncertainties are quoted). The channels shown
correspond to muon and electron decay channels of the $\PW$ boson with
$\pt^\mu > 25\GeV$ (left) and $\pt^\Pe>35\GeV$ (right).
The different contributions shown in the plot are described in the text.
Note that the amount of non-($\Wc$) background predicted by the simulation is almost
negligible.}
    \label{fig:Dpm_mass_subtracted}
\end{center}
\end{figure}

Figure~\ref{fig:Dpm_mass_subtracted} shows the $\OSSS$
distributions of the reconstructed invariant mass for $\Dpm$ candidates associated
with $\PWmn$ and $\Wen$ decays. It is compared with the predictions obtained from the simulated MC samples.
We distinguish two different contributions in the $\Wc$ prediction. A resonant $\Wc$ component is composed of those events with a $\Dpm$ meson
decaying into the $\PK^\mp \pi^\pm\pi^\pm$ final state at generator level; it is visible as a clear peak around the $\Dpm$ mass
in Fig.~\ref{fig:Dpm_mass_subtracted}.
A nonresonant component arises from $\Wc$ events where the charm meson decays to any final state other than $\PK^\mp \pi^\pm\pi^\pm$.
The reconstructed invariant mass distribution in this case extends as a continuum over the whole spectrum.
The distribution presented in Fig.~\ref{fig:Dpm_mass_subtracted} is almost exclusively populated by $\Wc$ events.
The contribution from the non-($\Wc$) processes introduced in Section~\ref{sec:samples} is shown as part of the background.

The MC prediction for the $\Dpm$ signal is scaled by the ratio of the branching fractions
${\mathcal{B}}(\cPqc\to\Dpm\to \PK^\mp \pi^\pm\pi^\pm)$ used in the simulation and measured experimentally.
The branching fraction used in the \PYTHIA simulation, $(1.528\pm 0.008)\%$, is about 25\% smaller than
the experimental measurement, $(2.08 \pm 0.10)\%$.
This value is the combination of three measurements performed at LEP~\cite{charm_OPAL_Dpm,charm_ALEPH,charm_DELPHI_Dpm} of
this branching fraction times the relative partial decay width of the Z boson into charm-quark pairs,
$R_\cPqc={\Gamma(\cPZ \to {\cPqc\cPaqc})}/{\Gamma(\cPZ \to \text{hadrons})}$.
The original LEP measurements are divided by the latest experimental value from the PDG~\cite{PDG} of
$R_\cPqc=0.1721\pm 0.0030$.
In the combination of these three experiments, we have assumed that experimental systematic uncertainties are
uncorrelated among the measurements, given the substantially
different sources of uncertainty considered by each experiment, whereas
the experimental uncertainty in $R_\cPqc$ is propagated in a correlated way.
Agreement between data and predictions is reasonable, although a small signal excess over the predictions (of about 10\%)
is visible in Fig.~\ref{fig:Dpm_mass_subtracted}.

For illustration purposes, the sum of a Gaussian function to describe the signal
plus a second-degree polynomial for the nonresonant background is fitted to the data distribution. The
PDG value of the $\Dpm$ mass is reproduced precisely in all cases.

\subsection{Selection of exclusive \texorpdfstring{$\Dstar$}{D*(2010)} decays~\label{sec:Dstar}}

The first step in the identification of
$\Dstarp\to\Dz \pi^+$ ($\Dstarm\to\Dzbar \pi^-$) decays is the
selection of a secondary vertex with two tracks of opposite
charge, as expected from a $\Dz\to \PKm\pi^+$ ($\Dzbar\to \PKp\pi^-$) decay.
This two-track system is combined with a primary track having $\pt>0.3\GeV$
found in a cone of $\Delta  R=0.1$ around the direction of the $\Dz$ candidate
momentum. The secondary track with charge opposite to the charge of the
primary track is assumed to be the kaon in the $\Dz$ decay. Only combinations with
a reconstructed mass differing from the $\Dz$ mass ($1864.86\pm 0.13\MeV$~\cite{PDG})
by less than $70\MeV$ are kept. The $\Dstar$ signal is identified as a
peak in the distribution of the difference
between the reconstructed $\Dstar$ and $\Dz$ masses near the expected value,
$m^{\rec}(\Dstar)-m^{\rec}(\Dz) = 145.421\pm 0.010\MeV$~\cite{PDG}.

\begin{figure}[htbp]
\begin{center}
    \includegraphics[width=0.49\textwidth]{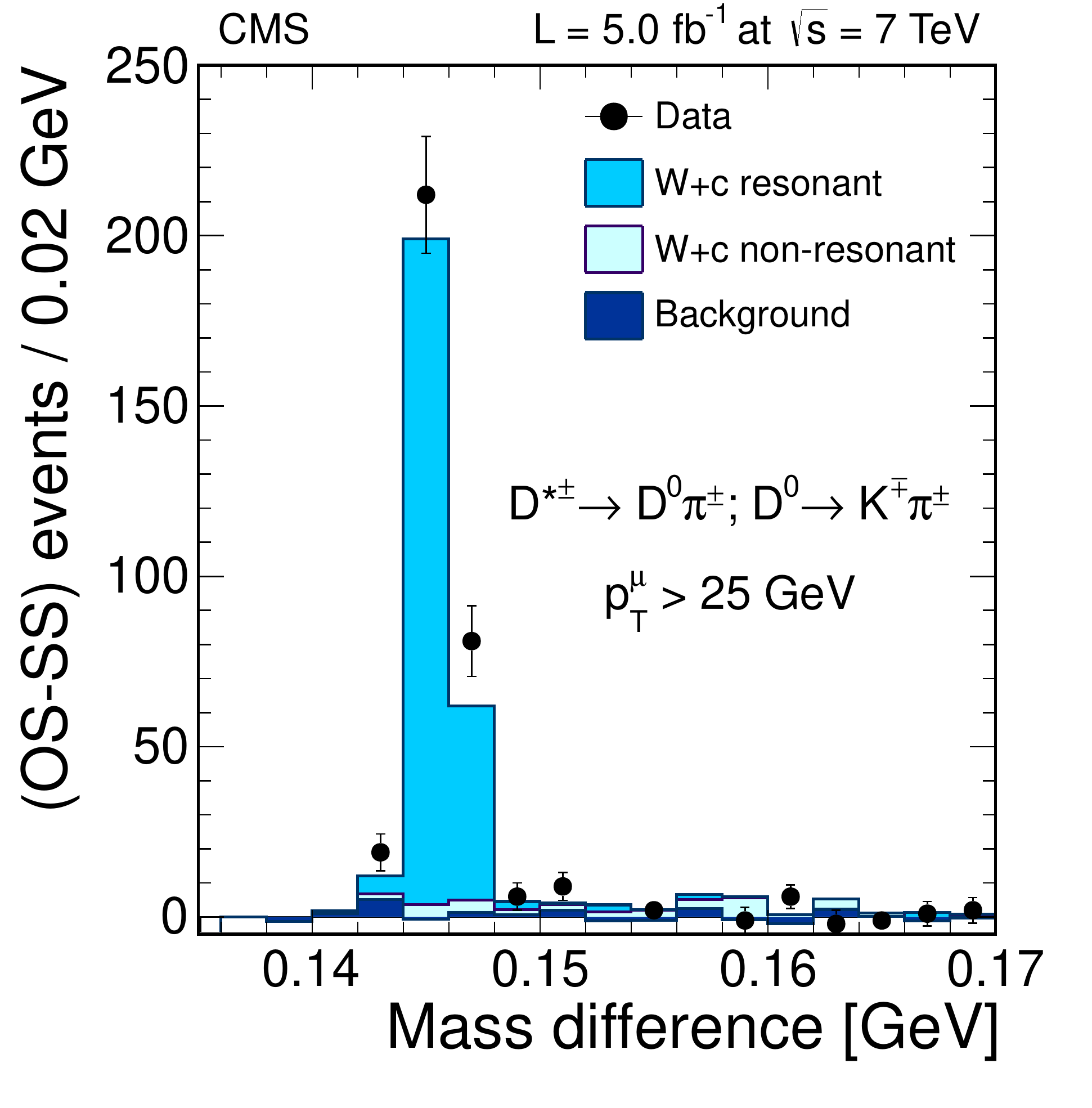}
    \includegraphics[width=0.49\textwidth]{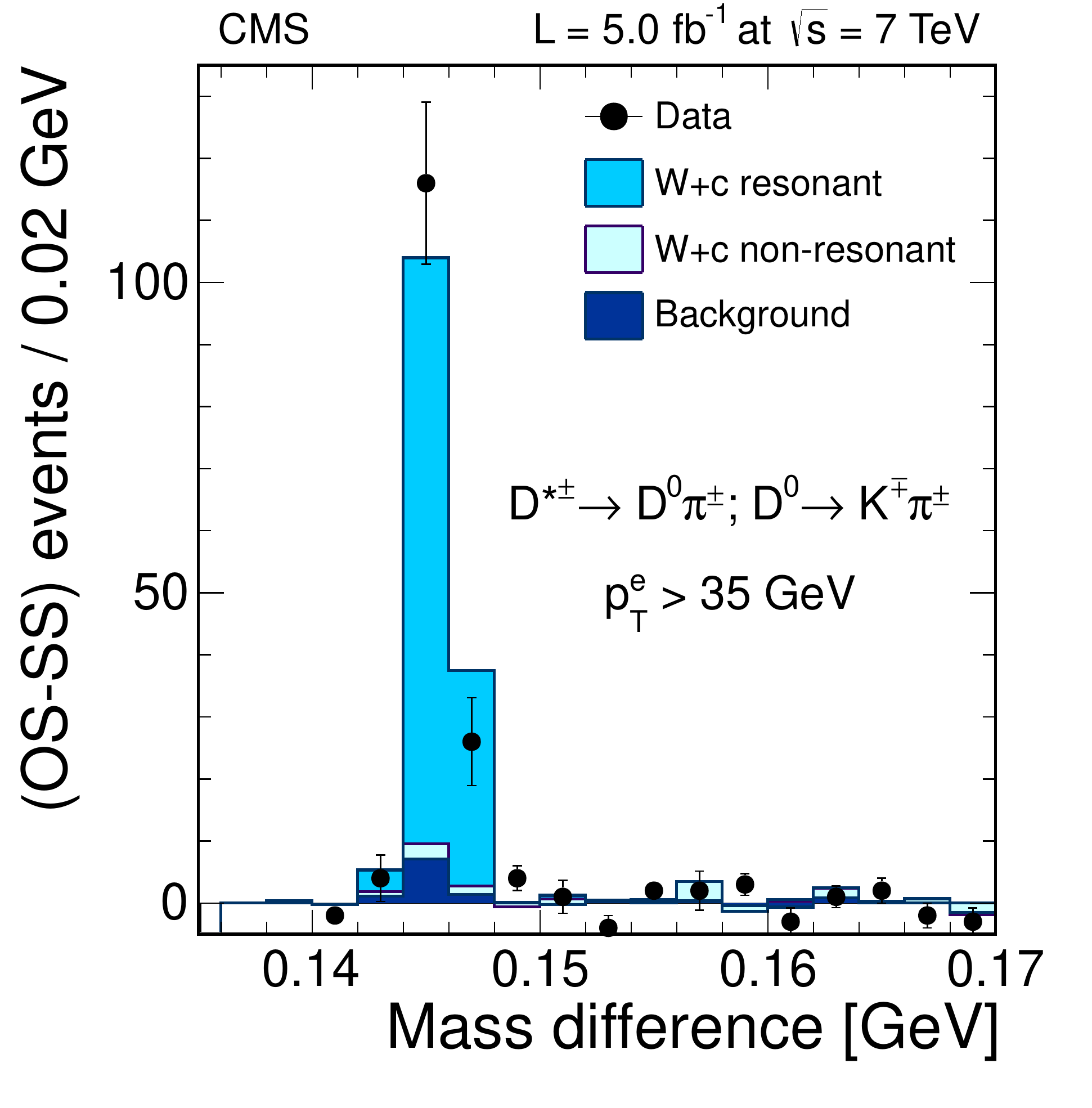}
    \caption{Distribution of the reconstructed mass difference between $\Dstar$ and
$\Dz$ candidates in the selected $\Wc$ sample, after subtraction of the SS
component. The position and width of the peak near $145\MeV$ are in agreement with the
MC expectations.
The different contributions shown in the plot are described in the text.
The channels shown
correspond to muon and electron decay channels of the $\PW$ boson with
$\pt^\mu > 25\GeV$ (left) and $\pt^\Pe>35\GeV$ (right).}
\label{fig:Dstar_massDiff_subtracted}
\end{center}
\end{figure}

The $\OSSS$ distribution of the reconstructed mass difference
$m^{\rec}(\Dstar)-m^{\rec}(\Dz)$ is shown in
Fig.~\ref{fig:Dstar_massDiff_subtracted}.
Both $\PWmn$ and $\Wen$ decays are considered,
with transverse momentum requirements of $\pt^\mu>25\GeV$ and $\pt^\Pe>35\GeV$.
The  resonant $\Wc$ component is composed here of those events with a $\Dstar$ meson
decaying into the $\Dz \pi^+; \Dz\to \PKm\pi^+$ ($\Dzbar \pi^-; \Dzbar\to \PKp\pi^-$) final state at generator level; it is visible as
a clear peak around the nominal mass difference $m^{\rec}(\Dstar)-m^{\rec}(\Dz)$
in Fig.~\ref{fig:Dstar_massDiff_subtracted}.
The nonresonant component comes from $\Wc$ events where the charm meson decays to any final state other than
$\Dz \pi^+; \Dz\to \PKm\pi^+$ ($\Dzbar \pi^-; \Dzbar\to \PKp\pi^-$).
Note that the amount of background predicted by the simulation, and also observed
in data, is extremely small.

The MC prediction for the full $\Dstar$ decay chain is scaled by the ratio between the product of the branching fraction for the decay chain
${\mathcal{B}}(\cPqc\to\Dstarp)\times {\mathcal{B}}(\Dstarp\to\Dz\pi^+)\times {\mathcal{B}}(\Dz\to \PKm \pi^+)$
 used in the simulation and the experimental measurement.
The product of the branching fractions
used in the \PYTHIA simulation is $(0.743\pm 0.005)\%$, which
is about 20\% larger than our estimation of the experimental value,
$(0.622 \pm 0.020)\%$. The latter number is a weighted average that uses as inputs
the dedicated measurements of this product times $R_\cPqc$ by
ALEPH~\cite{charm_ALEPH} and OPAL~\cite{charm_OPAL_Dstar}, as well as the measurement of
${\mathcal{B}}(\cPqc\to\Dstarp)\times {\mathcal{B}}(\Dstarp\to\Dz\pi^+)$ by DELPHI~\cite{charm_DELPHI_Dstar}.
To obtain the charm fractions needed for the $\Wc$ cross section normalization,
the  ALEPH~\cite{charm_ALEPH} and OPAL~\cite{charm_OPAL_Dstar} measurements are divided by the world-average $R_\cPqc$ experimental value
and the DELPHI~\cite{charm_DELPHI_Dstar} measurement is multiplied by the world-average
${\mathcal{B}}(\Dz\to \PKm \pi^+)=0.0388\pm 0.0005$, both taken from the PDG~\cite{PDG}.
Also in this case, experimental systematic uncertainties are
assumed to be uncorrelated among the three LEP measurements
and the experimental uncertainty in $R_\cPqc$ is propagated in a correlated way.
A small excess of data over the theoretical predictions is also observed in this channel.

\subsection{Selection of semileptonic charm decays~\label{sec:dileptons}}

In addition to the previous exclusive channels, we consider the identification
of charm-quark jets via semileptonic decays of the c quark. Only jets containing
semileptonic decays into muons are
considered.
Muons in jets are identified with the same criteria used for muon identification
in $\PW$-boson decays, with the exception that the isolation requirements are not applied. Since
the $\OSSS$ strategy effectively suppresses all backgrounds except Drell--Yan processes,
additional requirements are applied in order to reduce the Drell--Yan contamination
to manageable levels without affecting the signal in an appreciable way.
We require $\pt^{\mu}<25\GeV$, $\pt^{\mu}/\pt^\text{jet}<0.6$, and
$\pt^{\rm rel}<2.5\GeV$, where $\pt^{\mu}$ denotes here the transverse momentum of the
muon identified inside the jet and $\pt^{\rm rel}$ is its transverse momentum with
respect to the jet direction. We also require the invariant mass of the dilepton
system to be above $12\GeV$, in order to avoid the region of low-mass resonances.
Finally, dimuon events with an invariant mass above $85\GeV$ are rejected. The latter requirement
is not applied to the sample with $\PW$-boson decays into electrons, which is
minimally affected by high-mass dilepton contamination.

For the input semileptonic branching fraction of charm-quark jets,
we employ the value
${\mathcal{B}}(\cPqc\to\ell) = 0.091 \pm 0.005$, which is the average of the inclusive
value, $0.096\pm 0.004$~\cite{PDG}, and of the exclusive sum of the individual
contributions from all weakly decaying charm hadrons,
$0.086\pm 0.004$~\cite{charm-fractions,PDG}.
The uncertainty is increased in order to cover both central values within one
standard deviation. This value is consistent with the \PYTHIA value
present in our simulations ($9.3\%$).

\begin{figure}[htbp]
\begin{center}
    \includegraphics[width=0.49\textwidth]{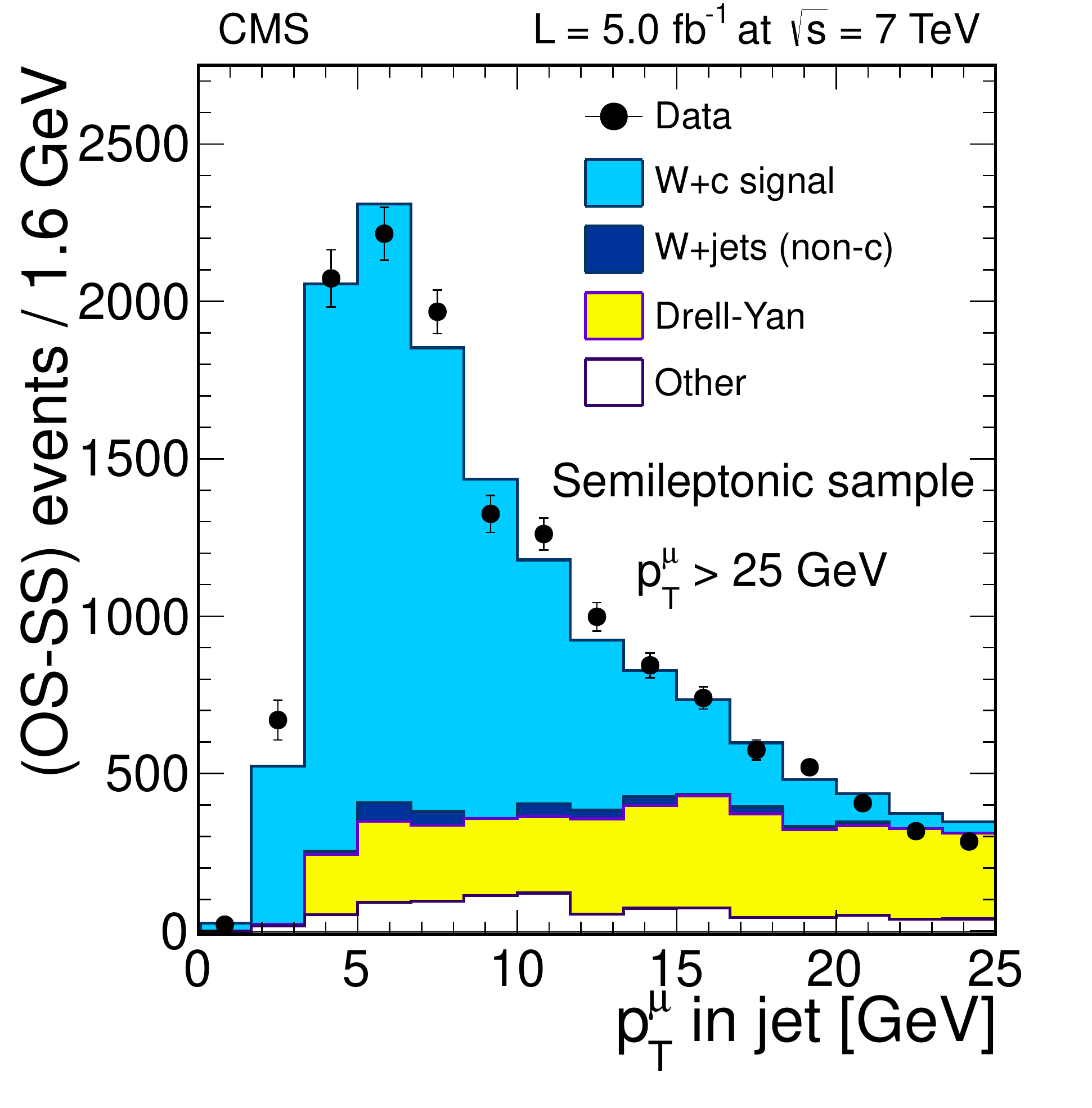}
    \includegraphics[width=0.49\textwidth]{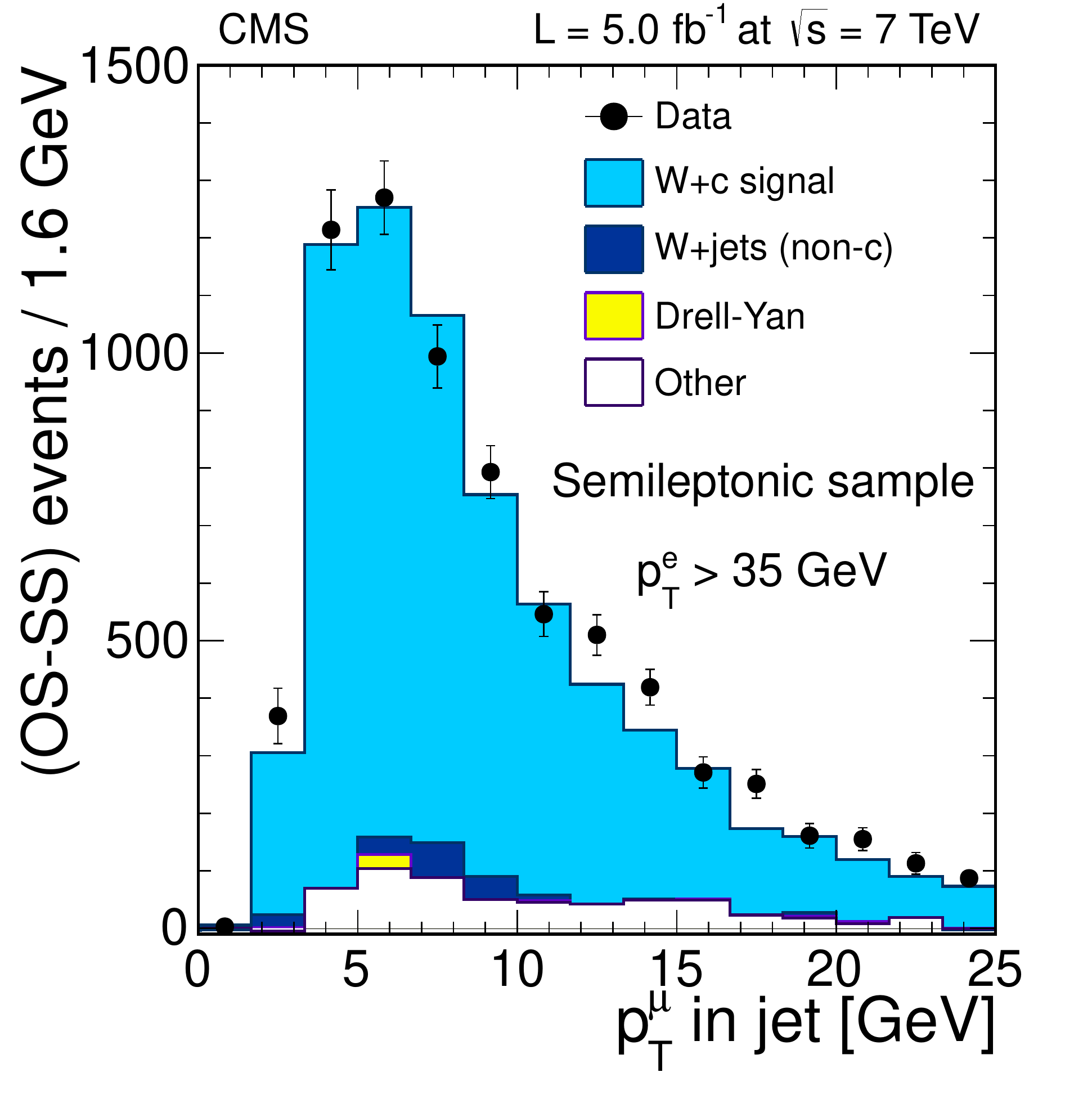}
    \caption{Distributions of the transverse momentum of the muon inside the leading
jet of the event, after subtraction of the SS
component. The channels shown
correspond to muon and electron decay channels of the $\PW$ boson with
$\pt^\mu > 25\GeV$ (left) and $\pt^\Pe>35\GeV$ (right).}
\label{fig:Dileptons_ptmu_subtracted}
\end{center}
\end{figure}

Figure~\ref{fig:Dileptons_ptmu_subtracted} shows the resulting transverse momentum
distribution of the selected muons inside the leading jet after the $\OSSS$ subtraction procedure.
Again, both $\PWmn$ and $\Wen$ decays are considered,
with transverse momentum requirements of $\pt^\mu>25\GeV$ and $\pt^\Pe>35\GeV$ for the leptons
from the $\PW$-boson decay.
The background predicted by the simulation is rather small in the electron
channel, but has a substantial Drell--Yan component in the muon channel.
The visible excess of data over the predictions is consistent with the observations in the $\Dpm$ and $\Dstar$ channels.

\subsection{Selection of inclusive \texorpdfstring{$\Dpm$ and $\Dstar$}{D+/- and D*(2010)} decays~\label{sec:inclusive}}

Enlarged samples of $\Wc$ candidates are selected from the events with secondary vertices with three or two tracks,
in order to increase the size of the samples available for the differential measurements.
We refer to them as inclusive three-prong and two-prong samples, respectively.

\begin{figure}[htb]
\begin{center}
    \includegraphics[width=0.49\textwidth]{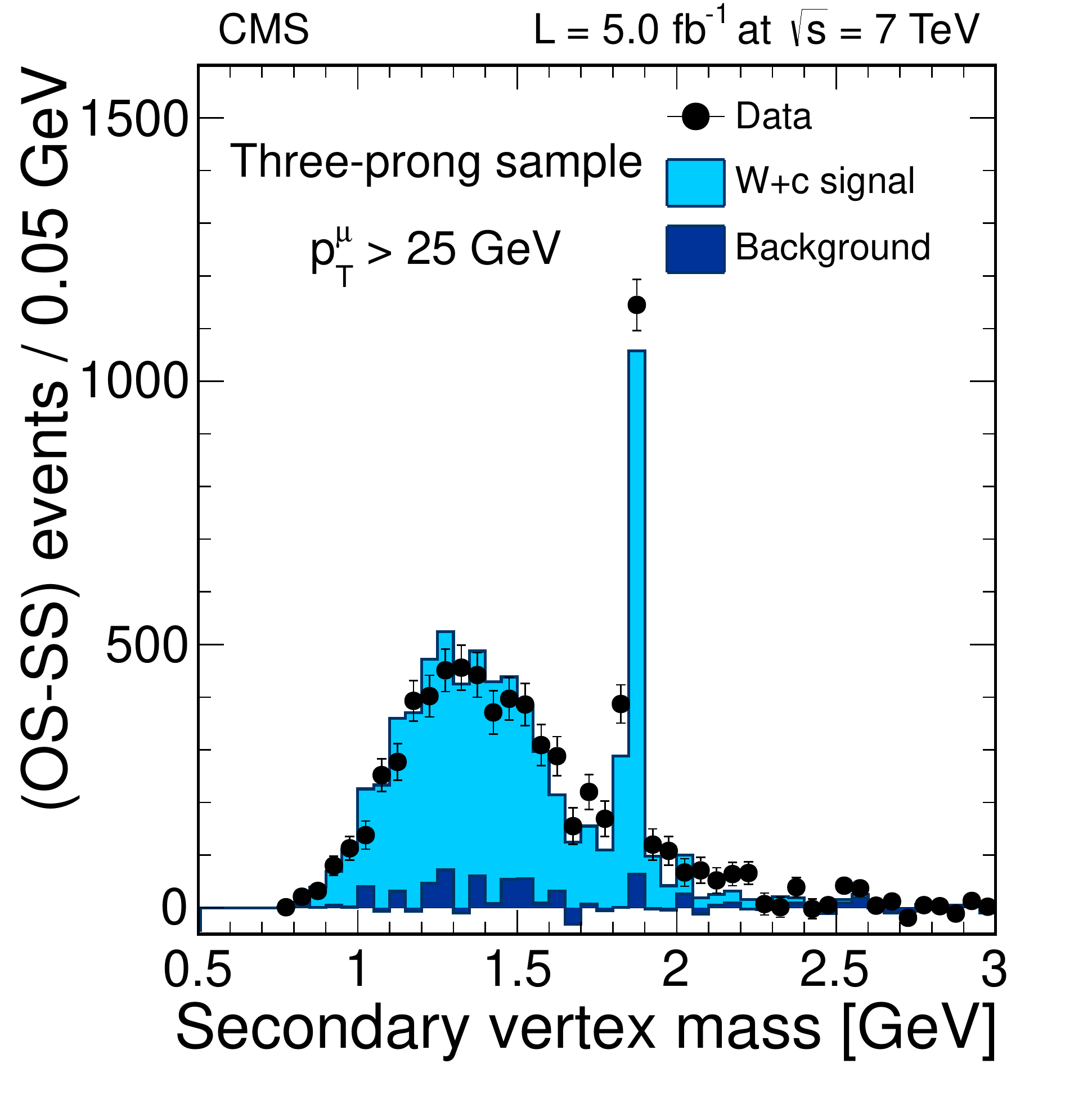}
    \includegraphics[width=0.49\textwidth]{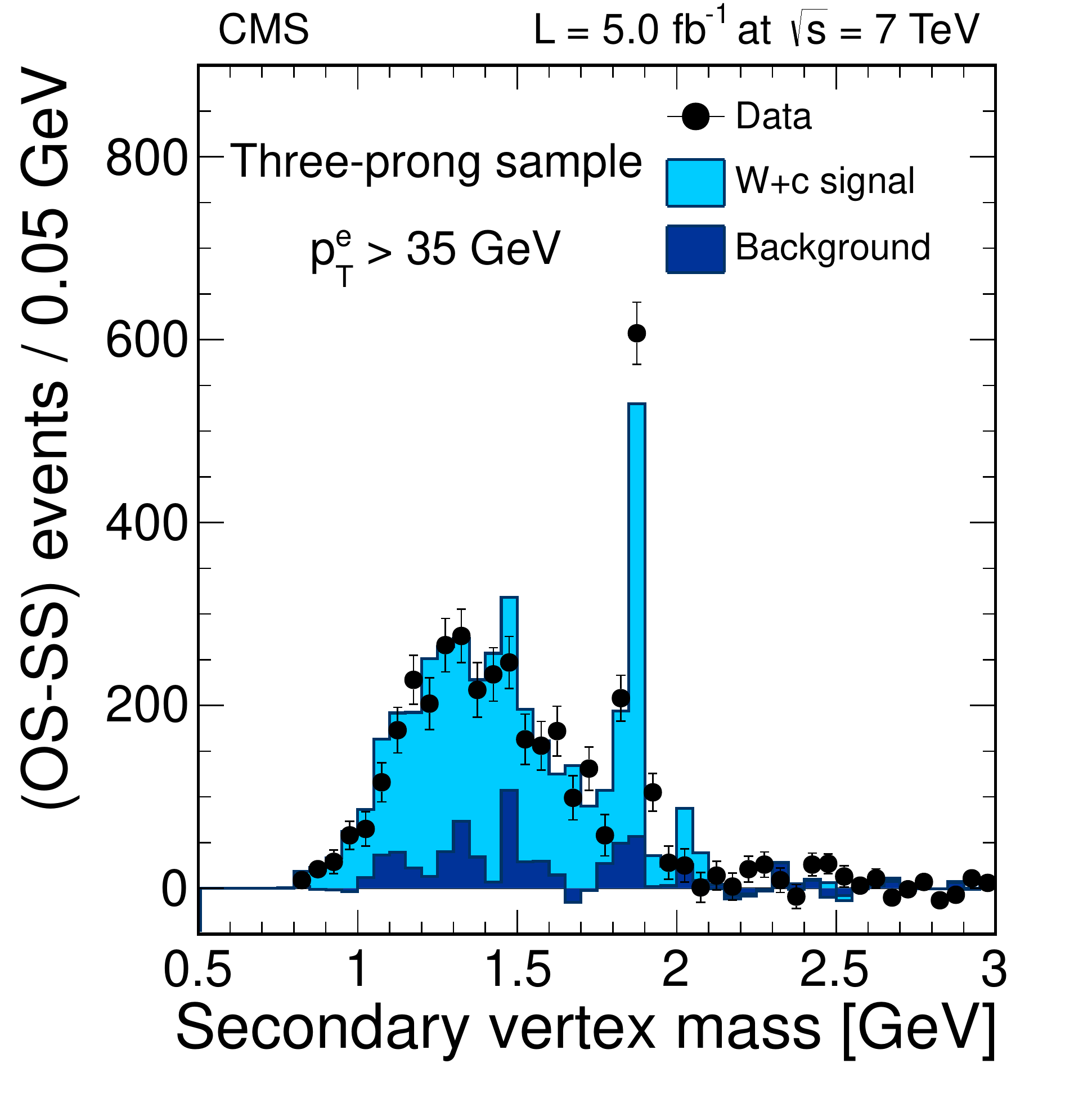}
    \caption{Inclusive three-prong samples: Invariant mass distribution of the three tracks composing a secondary vertex
assuming a $\Dpm\to \PK^\mp \pi^\pm\pi^\pm$ hypothesis.
    The left plot is for $\PWmn$ events, with $\pt^\mu > 25\GeV$. The right plot is for $\Wen$ events, with $\pt^\Pe > 35\GeV$.
    Distributions are presented after subtraction of the SS component.}
    \label{fig:Dpm_mass_inc_subtracted}
\end{center}
\end{figure}

Candidates for charm meson decays in the $\Dpm\to \PK^\mp \pi^\pm\pi^\pm$ decay mode are selected among the events
with a secondary vertex with three tracks and
with a vertex charge equal to ${\pm}1$, which is computed as the sum of the charges associated with the tracks constituting the vertex.
The mass assignment for the secondary tracks follows the procedure described in Section~\ref{sec:Dpm}. However, the constraint that the invariant mass
of the secondary vertex be compatible with the $\Dpm$ nominal mass within $50\MeV$ is not required in this case.
The $\OSSS$ distribution of the reconstructed invariant mass in events with three prongs is presented in Fig.~\ref{fig:Dpm_mass_inc_subtracted}.
In addition to the resonant peak at the $\Dpm$ mass, there is a nonresonant spectrum with lower values of the invariant mass corresponding mainly to $\Dpm$ decays with one or more unaccounted neutral particles in the final state.
For the differential cross section measurement, we consider the region of the invariant mass spectrum $m(\PK^\mp \pi^\pm\pi^\pm) < 2.5\GeV$.
This results in a sample five times larger than the $\Dpm$ exclusive sample.

\begin{figure}[htb]
\begin{center}
    \includegraphics[width=0.49\textwidth]{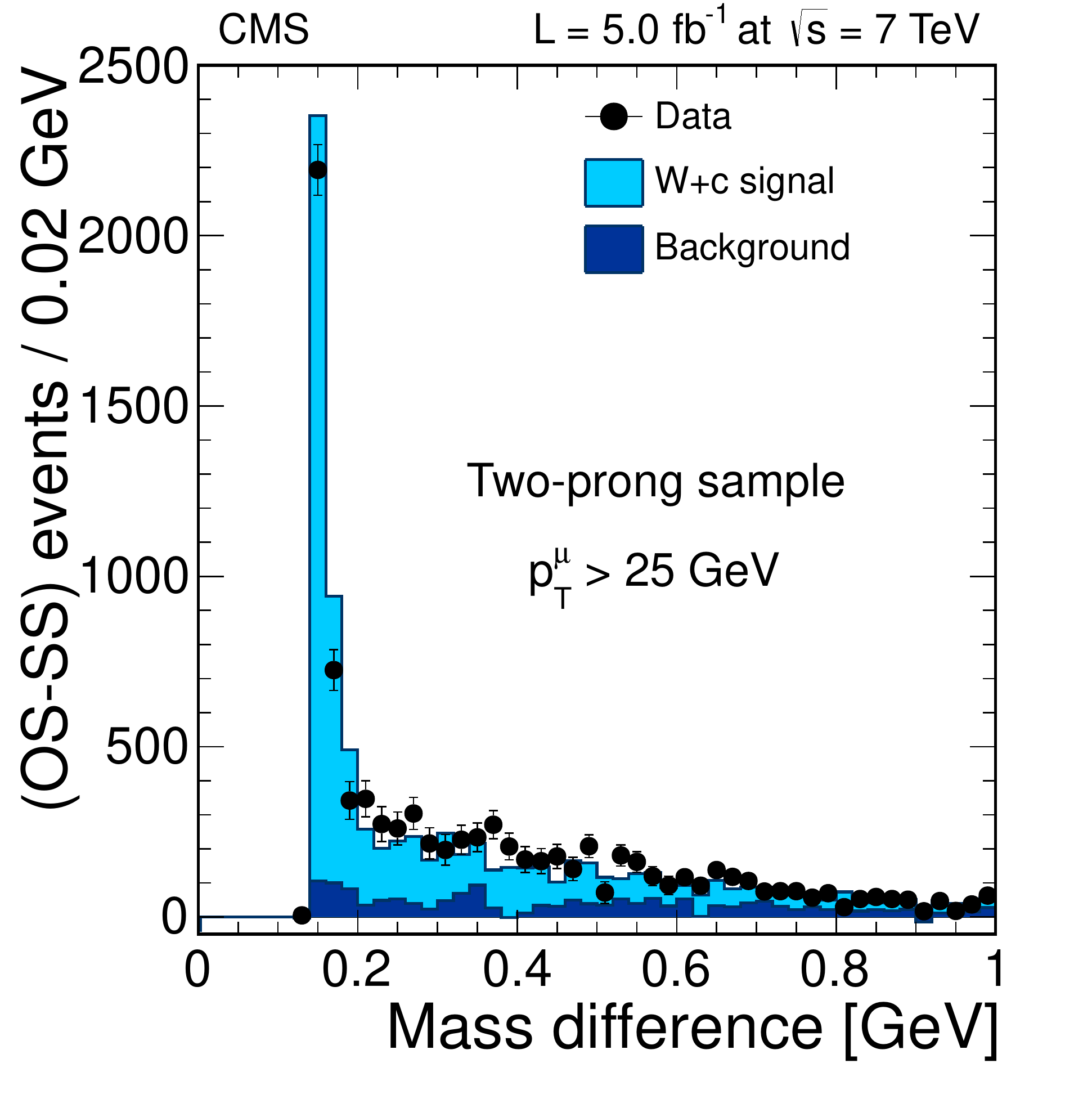}
    \includegraphics[width=0.49\textwidth]{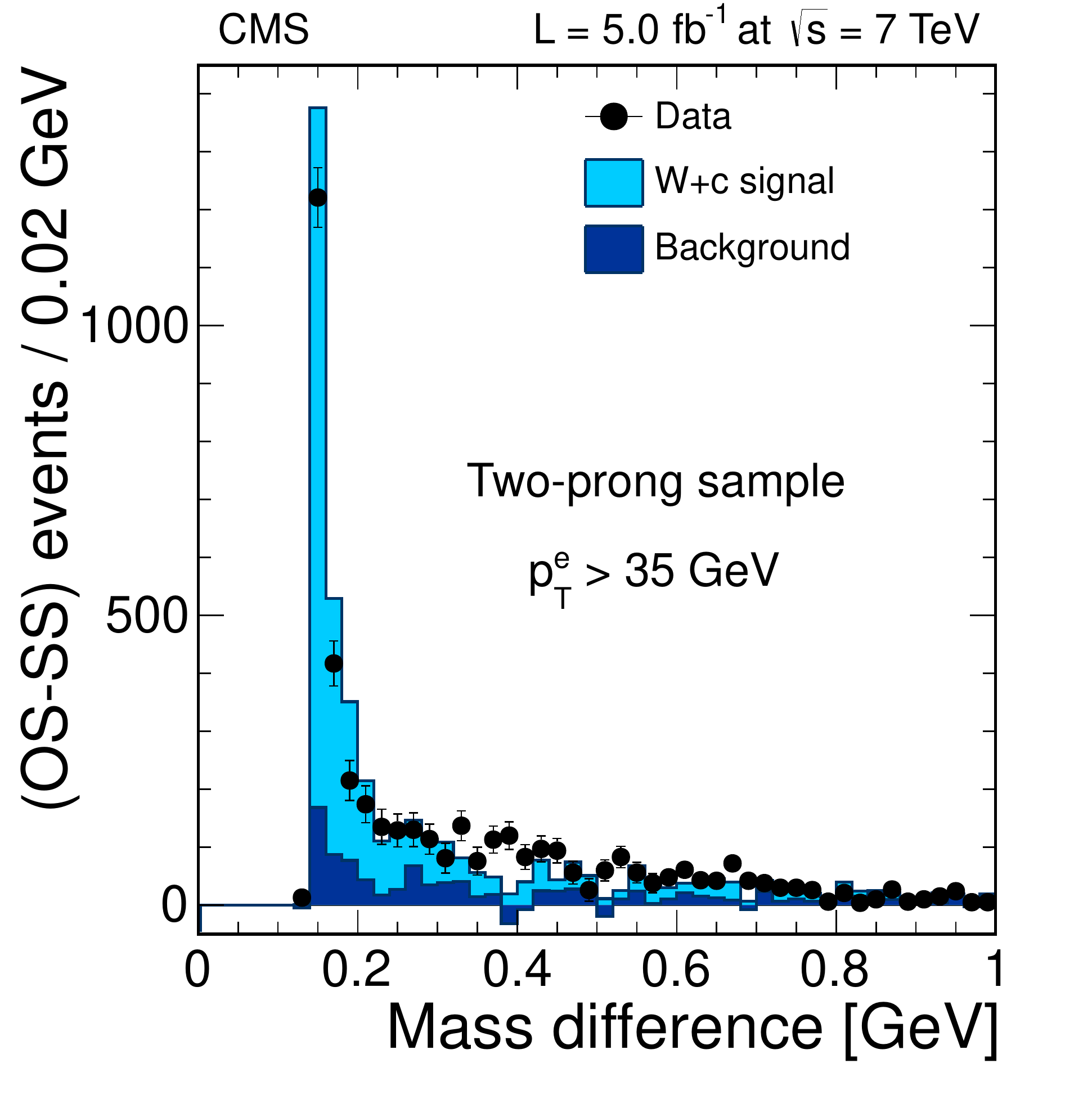}
    \caption{Inclusive two-prong samples: Distribution of the difference between the invariant mass of the two-track system
and the closest track from the primary pp interaction vertex and the invariant mass of the two secondary vertex tracks
($m(\PK^\mp\pi^\pm\pi^\pm)-m(\PK^\mp\pi^\pm)$), assuming the decay chain $\Dstar\to\Dz\pi^\pm \to \PK^\mp\pi^\pm\pi^\pm$.
    The sharp peak at $145\MeV$ reflects the nominal mass difference between the invariant mass of the $\Dz$ and the primary-pion system and
the $\Dz$ mass for the decay $\Dstar\to\Dz\pi^\pm$.
    The left plot is for $\PWmn$ events, with $\pt^\mu > 25\GeV$. The right plot is for $\Wen$ events, with $\pt^\Pe > 35\GeV$.
    The distributions are presented after subtraction of the SS component.
    }
    \label{fig:Dstar_mass_inc_subtracted}
\end{center}
\end{figure}

Similarly, candidates for $\Dz$ charm meson decays are reconstructed in the $\Wj$ events with a displaced secondary
vertex built from two
tracks of opposite curvature.
The two tracks are assumed to correspond to the decay products of a $\Dz$.
The decay chain $\Dstar\to\Dz\pi^{\pm}$, $\Dz\to \PK^\mp\pi^\pm$ is identified according to the procedure described in Section~\ref{sec:Dstar}, but dropping
the $\Dz$ mass constraint $|m(\PK^\mp\pi^\pm)-1864.86\MeV| < 70\MeV$.
Figure~\ref{fig:Dstar_mass_inc_subtracted} shows the $\OSSS$ distributions of the mass difference $m(\PK^\mp\pi^\pm\pi^\pm)-m(\PK^\mp\pi^\pm)$,
where one of the pions is the closest track from the primary pp interaction vertex.
The peak at $m(\PK^\mp\pi^\pm\pi^\pm)-m(\PK^\mp\pi^\pm) {\sim} 145\MeV$ corresponds to the nominal $\Dstar-\Dz$ mass difference~\cite{PDG}.
$\Wc$ events are still the dominant contribution at larger values of the mass difference. The remaining background is small and it is mainly due to residual $\Wuds$-quark jets, $\Wcc$, and $\ttbar$ production.
We select the events with an invariant mass difference $m(\PK^\mp\pi^\pm\pi^\pm)-m(\PK^\mp\pi^\pm) < 0.7\GeV$.
The size of the sample is increased by a factor of ${\sim}25$ with respect to the exclusive $\Dstar$ sample.

\section{Measurement of the \texorpdfstring{$\Wc$}{W + c} cross section~\label{sec:total_xsec}}

The measurement of the $\Wc$ cross section
is performed with several different final states containing
a well-identified $\Wln$ decay plus a leading jet with charm content.
We use the exclusive $\Dpm$ and $\Dstar$ samples and the semileptonic sample,
 described in Section~\ref{sec:selection}.
Two sets of measurements are provided:
one with $\pt^\ell > 25\GeV$ using only $\PWmn$ decays; and a second one, using
both $\PWmn$ and $\Wen$ decays with $\pt^\ell > 35\GeV$.

For all channels under study, the $\Wc$ cross section is determined in the fiducial region
$\pt^{\ell}>25~(35)\GeV$, $\abs{\eta^\ell}<2.1$, $\pt^\text{jet}>25\GeV$, $\abs{\eta^\text{jet}}<2.5$
using the following expression:
\begin{equation*}
      {\sigma(\Wc) = \frac{N_\text{sel}-N_\text{bkg}}{\Lint~\mathcal{B}~\AccEff}},
\end{equation*}
where $N_\text{sel}$ is the number of $\OSSS$ events selected in the defined signal region, $N_\text{bkg}$ is the
estimated number of background events after $\OSSS$ subtraction, $\Lint$ is the integrated luminosity, and
$\mathcal{B}$ is the relevant charm branching fraction, derived in Section~\ref{sec:selection}, for the channel under study, \ie
$\mathcal{B}\equiv \mathcal{B}(\cPqc\to\Dp; \Dp\to \PKm \pi^+\pi^+) = (2.08 \pm 0.10)\%$ in the case of
the $\Dpm$ channel,
$\mathcal{B}\equiv \mathcal{B}(\cPqc\to\Dstarp; \Dstarp\to\Dz\pi^+;\Dz\to \PKm\pi^+) = (0.622 \pm 0.020)\%$
for the $\Dstar$ channel, and $\mathcal{B}\equiv \mathcal{B}(\cPqc\to\ell) = (9.11 \pm 0.49)\%$ for the
semileptonic channel.

The factor $\AccEff$ accounts for limited acceptances and efficiencies.
In $\Wc$ events, less than $20\%$ of the events have a well-identified secondary vertex,
while less than $50\%$ of the muons from semileptonic charm decays have sufficiently high energy
to be reconstructed and identified in the muon spectrometer.
The simulated $\Wj$ sample  generated by $\MADGRAPH$ + \PYTHIA is used to calculate the fraction of events
within the fiducial region that fulfil the criteria for the several charm-quark jet categories.
These simulated samples are corrected for any differences between data and MC description in lepton trigger,
identification and reconstruction efficiencies.
Scaling factors, defined as the ratio $\text{efficiency}_\text{data}/\text{efficiency}_\mathrm{MC}$ as a function of the lepton pseudorapidity,
are determined with samples of $\cPZ \to \ell^+\ell^-$ events.
An invariant mass ($m_{\ell^+\ell^-}$) constraint and tight quality requirements assigned to one of the leptons (``tag'') allow
the other lepton to be used as a probe to test the different steps in lepton identification (``tag-and-probe'' method)~\cite{WZ-paper}.
The precision in the factor $\AccEff$ is limited by the size of the MC sample employed;
its statistical uncertainty is propagated as a systematic
uncertainty to the $\Wc$ cross section.

The signal region for the $\Dpm$ channel is defined by the constraint
$\Delta m(\Dpm) \equiv |m^{\rec}(\Dpm)-1.87\GeV|<0.05\GeV$, where $m^{\rec}(\Dpm)$ is the reconstructed mass of the $\Dpm$
candidate~(Fig.~\ref{fig:Dpm_mass_subtracted}).
The same requirement is applied to the MC simulations in order to determine
the correction factor $\AccEff$. We estimate values of
$\AccEff=0.1114 \pm 0.0033~(\pt^\mu>25\GeV)$
and $\AccEff=0.0834 \pm 0.0032~(\pt^\Pe>35\GeV)$, where the quoted uncertainties are
statistical only. The background is fully dominated by the
nonresonant $\Wc$ component. It is subtracted from the selected number of events in
the data window by using the number of events selected in a control region
away from the resonance, extending up to a difference of $200\MeV$ with respect to the nominal $\Dpm$ mass,
$N[0.05\GeV<\Delta m(\Dpm)<0.20\GeV]$.
This number is scaled by the ratio
$N[\Delta m(\Dpm)<0.05\GeV]/N[0.05\GeV<\Delta m(\Dpm)<0.20\GeV]$
observed in the simulation in order to obtain the number of background events expected in the
reference window. This procedure is largely independent of uncertainties in the charm
fractions present in \PYTHIA.
Systematic biases due to the assumed nonresonant background subtraction
are expected to be negligible compared to the statistical
uncertainty, given the approximate agreement between data and MC distributions.

The signal region for the $\Dstar$ channel is restricted to the interval
$\Delta m(\Dstar) \equiv |m^{\rec}(\Dstar)-m^{\rec}(\Dz)-145\MeV|<5\MeV$, where $m^{\rec}(\Dstar)-m^{\rec}(\Dz)$ is
the reconstructed mass difference between the D mesons
(Fig.~\ref{fig:Dstar_massDiff_subtracted}).
The same procedure is applied to the MC simulations in order to determine the correction factor $\AccEff$. We
estimate values of $\AccEff=0.0849 \pm 0.0040~(\pt^\mu>25\GeV)$ and
$\AccEff=0.0559 \pm 0.0036~(\pt^\Pe>35\GeV)$, where the quoted uncertainties are
statistical only. As in the $\Dpm$ case, the background
is subtracted from the selected number of data events in a sideband sample,
$5~\MeV<\Delta m(\Dstar) <20\MeV$.
This number is scaled by the ratio
$N[\Delta m(\Dstar) < 5\MeV]/N[5\MeV < \Delta m(\Dstar) < 20\MeV]$ observed in the simulation.

For the semileptonic channel, $N_\text{sel}$ is given by
the number of events with a $\PW$-boson candidate decaying into a high-$\pt$ muon or electron and an identified muon inside the jet
passing the requirements described in Section~\ref{sec:dileptons}.
The correction factors $\AccEff$ for the different lepton thresholds are estimated in the MC simulation as
$\AccEff=0.2035 \pm 0.0021~(\pt^\mu>25\GeV)$ and
$\AccEff=0.1706 \pm 0.0021~(\pt^\Pe>35\GeV)$, where the quoted uncertainties are
statistical only.
The number of background events remaining after selection is estimated from the simulated samples.
In the sample with two opposite-sign muons, the residual Drell--Yan background corresponds
to events with significant missing transverse energy and one low-$\pt$ muon inside a jet.
Potential discrepancies between data and MC description in this particular phase space region are evaluated
by analyzing the Drell--Yan-dominated control sample with dimuon invariant masses above
$85\GeV$. A correction factor of $1.2 \pm 0.1$ provides agreement between data and MC simulation in this region
and it is applied to estimate the background in the signal region. The uncertainty in this correction factor is propagated as a systematic
uncertainty in the cross section measurement. This takes into account possible differences in the description
of events below and around the Z-boson peak, where this factor is derived.

Table~\ref{tab:all_cross_sections_25} contains all the relevant inputs used in the measurements
and the resulting cross sections in the different subchannels.
The sources of systematic uncertainties affecting the measurement are discussed in Section~\ref{sec:syst_total}.
\begin{table}[htbp]
\begin{center}
\topcaption{Cross section results for three specific final states.
Here $N_\text{sel}$
is the estimated number of selected events in the signal region (around the resonance in the case
of $\Dpm$ and $\Dstar$ final states). $N_\text{sel}-N_\text{bkg}$ 
is the estimate for the signal events after background subtraction using the method described in the text,
$\AccEff$ is the acceptance and efficiency correction factor, and
$\sigma(\Wc)$ is the measured $\Wc$ cross section after correction
for the charm fractions as discussed in the text.
Results obtained with the sample of $\PW$ bosons decaying into a muon and a neutrino and for the two muon transverse momentum thresholds ($\pt^\mu > 25\GeV$ and $\pt^\mu > 35\GeV$) are shown in the first two blocks of the table. Results obtained when the $\PW$ boson decays into an electron and a neutrino ($\pt^\Pe > 35\GeV$) are given in the lowest block of the table.
All uncertainties quoted in the table are statistical, except for the measured cross sections, which include systematic uncertainties
due to the sources discussed in Section~\ref{sec:syst_total}.}
\label{tab:all_cross_sections_25}
\begin{tabular}{ccccc}
\hline\hline
            & \multicolumn{4}{c}{$\PWmn$, $\pt^\mu > 25\GeV$} \\ \hline
Final state & $N_\text{sel}$ & $N_\text{sel}-N_\text{bkg}$ & $\AccEff$ [\%] & $\sigma(\Wc)$ [pb] \\
\hline
$\Dpm$    & $\phantom{4}1502 \pm \phantom{1}62$          & $1203 \pm \phantom{2}91$            & $11.1 \pm 0.3$           & $103.6 \pm 7.8\stat \pm \phantom{1}8.1\syst$ \\
$\Dstar$     & $\phantom{14}318 \pm \phantom{1}21$ & $\phantom{1}309 \pm \phantom{2}23$ & $\phantom{2}8.5 \pm 0.4$ & $116.9 \pm 8.7\stat \pm 10.0\syst$ \\
$\cPqc\to\mu$ & $14215 \pm 196$                    & $9867 \pm 237$           & $20.4 \pm 0.2$           & $106.5 \pm 2.6\stat \pm \phantom{1}9.6\syst$ \\
\hline\hline
            & \multicolumn{4}{c}{$\PWmn$, $\pt^\mu > 35\GeV$} \\ \hline
Final state & $N_\text{sel}$ & $N_\text{sel}-N_\text{bkg}$ & $\AccEff$ [\%] & $\sigma(\Wc)$ [pb] \\
\hline
$\Dpm$ & $\phantom{1}1209 \pm \phantom{1}55$ & $\phantom{7}981 \pm \phantom{2}79$ & $11.4 \pm 0.4$ & $82.9 \pm 6.7\stat \pm 6.4\syst$ \\
$\Dstar$  & $\phantom{11}260 \pm \phantom{1}19$ & $\phantom{7}248 \pm \phantom{2}21$ &  $\phantom{1}8.6 \pm 0.5$ & $92.3 \pm 7.8\stat \pm 8.2\syst$ \\
$\cPqc\to\mu$ & $11462 \pm 172$ & $7875 \pm 207$ & $21.6 \pm 0.2$ & $79.9 \pm 2.1\stat \pm 6.9\syst$ \\
\hline \hline
            & \multicolumn{4}{c}{$\Wen$, $\pt^\Pe > 35\GeV$} \\ \hline
Final state & $N_\text{sel}$ & $N_\text{sel}-N_\text{bkg}$ & $\AccEff$ [\%] & $\sigma(\Wc)$ [pb] \\
\hline
$\Dpm$ & $\phantom{7}838 \pm \phantom{1}47$ & $\phantom{6}726 \pm \phantom{1}55$ & $\phantom{1}8.3 \pm 0.3$ & $83.5 \pm \phantom{1}6.3\stat \pm 7.1\syst$ \\
$\Dstar$  & $\phantom{7}148 \pm \phantom{1}15$ & $\phantom{6}145 \pm \phantom{1}18$ & $\phantom{1}5.6 \pm 0.4$ & $83.3 \pm 10.4\stat \pm 8.5\syst$ \\
$\cPqc\to\mu$ & $7156 \pm 151$ & $6701 \pm 175$ & $17.1 \pm 0.2$ & $86.5 \pm \phantom{1}2.2\stat \pm 6.9\syst$ \\
\hline
\hline
\end{tabular}
\end{center}
\end{table}

For each $\PW$-boson decay channel and lepton $\pt$ threshold considered,
the cross sections measured from the three charm meson decay samples are consistent and are combined.
Measurements performed in the muon and electron channel with a lepton $\pt$ threshold of $35\GeV$ are also combined.
The combination is a weighted average of the individual measurements taking into account their statistical and systematic uncertainties.
Systematic uncertainties arising from a common source and affecting several measurements are considered to be fully correlated.

For $\pt^\mu>25\GeV$ the average $\Wc$ cross section is
\begin{equation*}
      \sigma(\ppWc) \times \mathcal{B}(\PW \rightarrow \mu\nu) (\pt^\mu>25\GeV)  = 107.7 \pm 3.3\stat \pm 6.9\syst\unit{pb}.
\end{equation*}

For $\pt^\ell>35\GeV$ we obtain
\begin{align*}
      \sigma(\ppWc) \times \mathcal{B}(\PW \rightarrow \mu\nu) (\pt^\mu>35\GeV) & =  82.9 \pm 2.6\stat \pm 5.1\syst\unit{pb}, \\
      \sigma(\ppWc) \times \mathcal{B}(\PW \rightarrow \Pe\nu) (\pt^\Pe>35\GeV) & =  85.3 \pm 2.5\stat \pm 5.7\syst\unit{pb}, \\
           \\
      \sigma(\ppWc) \times \mathcal{B}(\PW \rightarrow \ell\nu) (\pt^\ell>35\GeV) & =  84.1 \pm 2.0\stat \pm 4.9\syst\unit{pb}.
\end{align*}
The average cross sections are dominated by the measurements in the semileptonic channel (${\sim}50\%$), followed by the $\Dpm$ channel (${\sim}30\%$) and the $\Dstar$
channel (${\sim}20\%$). The weight of the $\PWmn$ channel in the cross section measurement with a lepton $\pt$ threshold of $35\GeV$ is ${\sim}30\%$ higher than
the contribution from the $\Wen$ channel.

These measurements are largely background-free. The overall relative
uncertainty, 6--7\%, is dominated by systematic uncertainties in the theoretical
modeling of the signal and by experimental uncertainties in the efficiency of the selection criteria.
A detailed comparison with theoretical predictions is provided in
Section~\ref{sec:results}.

\subsection{Systematic uncertainties in the \texorpdfstring{$\Wc$}{W c} cross section measurement~\label{sec:syst_total}}

The various sources of systematic uncertainties are presented in Table~\ref{tab:xsection_systematics}.
The limited precision in the branching fractions of the charm decays
is one of the dominant sources of uncertainties.

Tracking reconstruction inefficiencies are intrinsically small
($<1\%$~\cite{CMS-PAS-TRK-10-002}). Given the nature of the method
used to build secondary vertices, tracks are assigned
to either the primary or secondary vertex in a way that may be
different in data and MC simulation. In order to estimate the size of a potential
discrepancy, the set of secondary tracks is either increased by adding a nearby
primary track or decreased by dropping one of the original secondary tracks.
A systematic uncertainty of 3.3\% in the measured cross sections is estimated from
the observed differences at the resonant $\Dz$ and $\Dpm$ peaks between data and
simulation.
Its impact on the final cross sections is reduced after combination with the results from
the semileptonic channel, which is free of this uncertainty.

Uncertainties due to the pileup modeling are calculated using a modified pileup profile obtained with a
minimum bias cross section increased by its estimated uncertainty, ${\approx}6\%$. Jet energy
scale uncertainties are extracted from dedicated CMS
studies~\cite{CMS-PAPER-JME-10-011}, which also take into account possible variations
in the jet flavour composition. Additional $\ETmiss$ effects are estimated
by smearing the $\MT$ distribution in simulation in order to match the $\MT$ shape
observed in data.  Their impact is ${\approx}2\%$ on the final measurement.

Lepton trigger and selection inefficiencies are included in the simulation by applying the corresponding data/MC scale factors determined
in dedicated ``tag-and-probe'' studies as a function of the lepton pseudorapidity.
For muons we estimate a 0.7\% uncertainty according to CMS studies on dimuon events in the Z-boson mass peak.
In the electron case we consider the difference between switching on and off the efficiency scale factors, because of the
presence of missing transverse energy requirements at the trigger level that
cannot be fully accounted by using ``tag-and-probe'' techniques.
The effect of momentum and energy resolution corrections determined at the Z-boson mass peak
is also propagated as an additional uncertainty.
We combine the uncertainties due to lepton identification, isolation, and trigger efficiencies with the
uncertainty in the lepton momentum and energy resolution in a single entry in Table~\ref{tab:xsection_systematics}.

The efficiency uncertainty for muons inside jets is taken
to be 3.0\% according to dedicated studies in multijet events.
The systematic uncertainty arising from the Drell--Yan background subtraction in the semileptonic channel is determined as the change
in the cross section when the correction factor to the MC simulation is varied within its uncertainties.

The propagation of the statistical uncertainty in the factor $\AccEff$ to the cross section
is not negligible due to the limited size of the MC samples used.
The uncertainties related to initial-state radiation (ISR) are estimated by recalculating the factor $\AccEff$
from samples generated with different renormalization and factorization scales (half and twice the default scale $Q^2$ used
in the generation). The average value of the meson energy fraction in charm
decays is varied by 4\%, which is about twice the uncertainty in the $\Dstar$ fragmentation
determined at LEP~\cite{charm_ALEPH,charm_OPAL_Dstar}, in order to cover possible uncertainties in the assumed shape.
Other theoretical uncertainties in $\AccEff$ include PDF effects
and potential biases due to the adoption of the $\MADGRAPH$ jet-parton matching scheme
as the reference to be compared with the \MCFM calculations ($\approx 1\%$).

The integrated luminosity measurement has a 2.2\% uncertainty~\cite{CMS-PAS-SMP-12-008}. Physics
backgrounds, including the gluon-splitting $\Wcc$ component, have a negligible
contribution to the systematics compared with the statistical uncertainties in the
background subtraction.
\begin{table}[htbp]
\begin{center}
 \topcaption{Breakdown of the different contributions to the total systematic uncertainty ($\Delta_{\text {syst}}$)
in the combined $\sigma(\Wc)$ measurements in the fiducial region given by
$\pt^\text{jet}>25\GeV$, $\abs{\eta^\text{jet}}<2.5$, $\abs{\eta^\ell}<2.1$ for two different
thresholds of the transverse momentum of the lepton from the $\PW$-boson decay:
$\pt^\ell > 25\GeV$ (muon channel only) and $\pt^\ell > 35\GeV$ (muon and electron channels combined).}
\label{tab:xsection_systematics}
\begin{tabular}{ccc}
\hline\hline
       & $\pt^\mu>25\GeV$ & $\pt^\ell>35\GeV$ \\
Source & $\Delta_{\text{syst}} [\%]$ & $\Delta_{\text{syst}} [\%]$ \\
\hline
${\mathcal{B}}(\cPqc\to\Dpm\to \PK^\mp \pi^\pm\pi^\pm)$ & 1.5 & 1.5 \\
${\mathcal{B}}(\cPqc\to\Dstar\to\Dz\to \PK^\mp\pi^\pm)$ & 0.7 & 0.6 \\
${\mathcal{B}}(\cPqc\to\mu)$                            & 2.6 & 2.7 \\
Vertex reconstruction                   & 1.8   & 1.7   \\
Pileup                                  & 0.9   & 0.8   \\
Jet energy scale                        & 3.0   & 1.7   \\
$\ETmiss$                                  & 2.0   & 2.0   \\
Lepton efficiency, resolution           & 0.8   & 1.5   \\
Muon efficiency in charm decay          & 1.4   & 1.5   \\
Drell--Yan background                   & 1.4   & 0.9   \\
MC statistics ($\AccEff$ stat. uncert.) & 1.6   & 1.3   \\
ISR and renormalization/                & \multirow{2}{*}{0.2} & \multirow{2}{*}{0.2}  \\
factorization scales                    &       &       \\
Fragmentation function                  & 0.8   & 0.6   \\
Other theoretical uncertainties         & 0.8   & 0.7   \\
Luminosity                              & 2.2   & 2.2   \\
\hline
Total                                   & 6.3   & 5.7   \\
\hline\hline
\end{tabular}
\end{center}
\end{table}

\subsection{Characterization of \texorpdfstring{$\Wc$}{W c} kinematics~\label{sec:kinematics}}

\begin{figure}[htbp]
\begin{center}
    \includegraphics[width=0.4\textwidth]{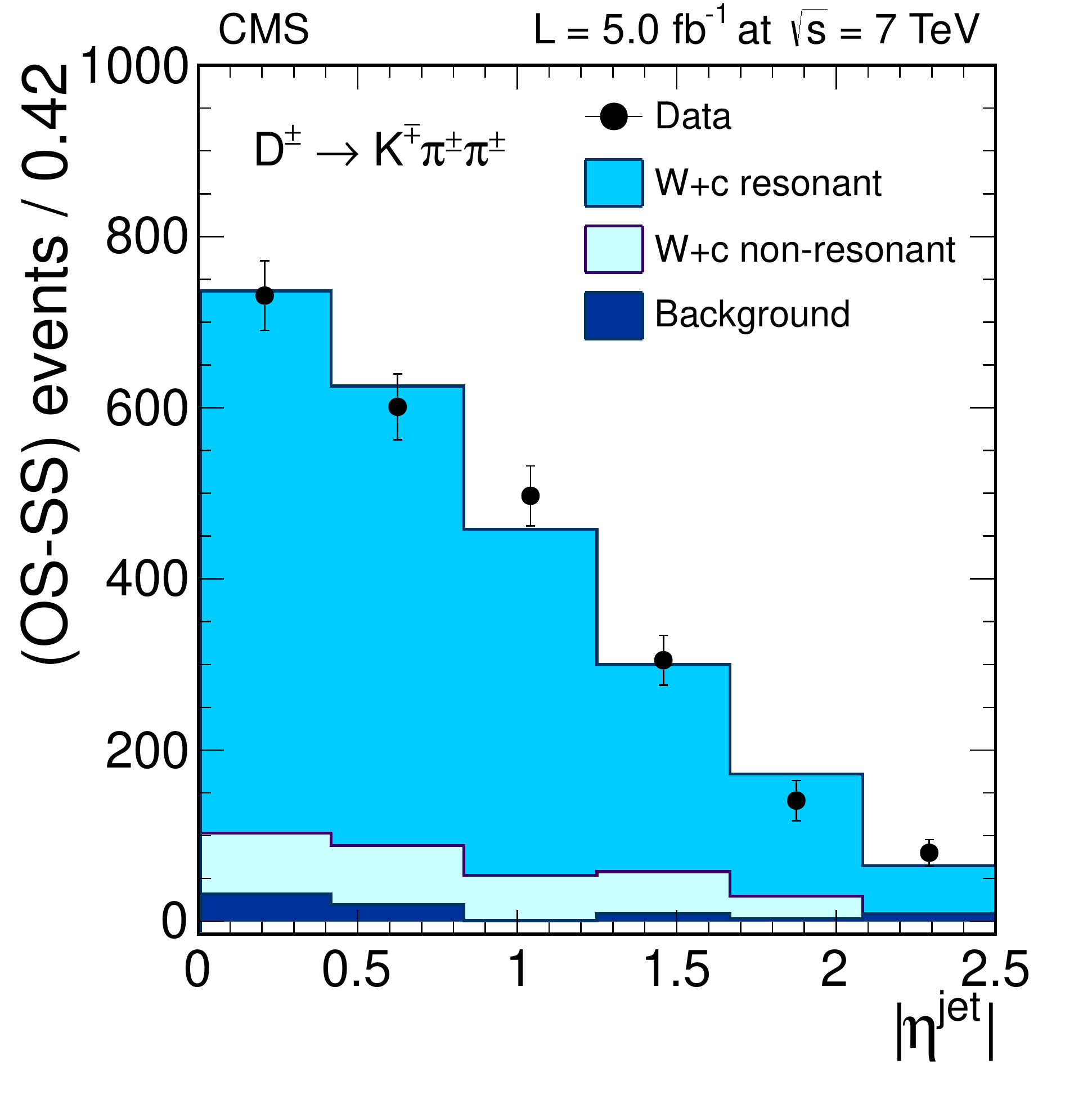}
    \includegraphics[width=0.4\textwidth]{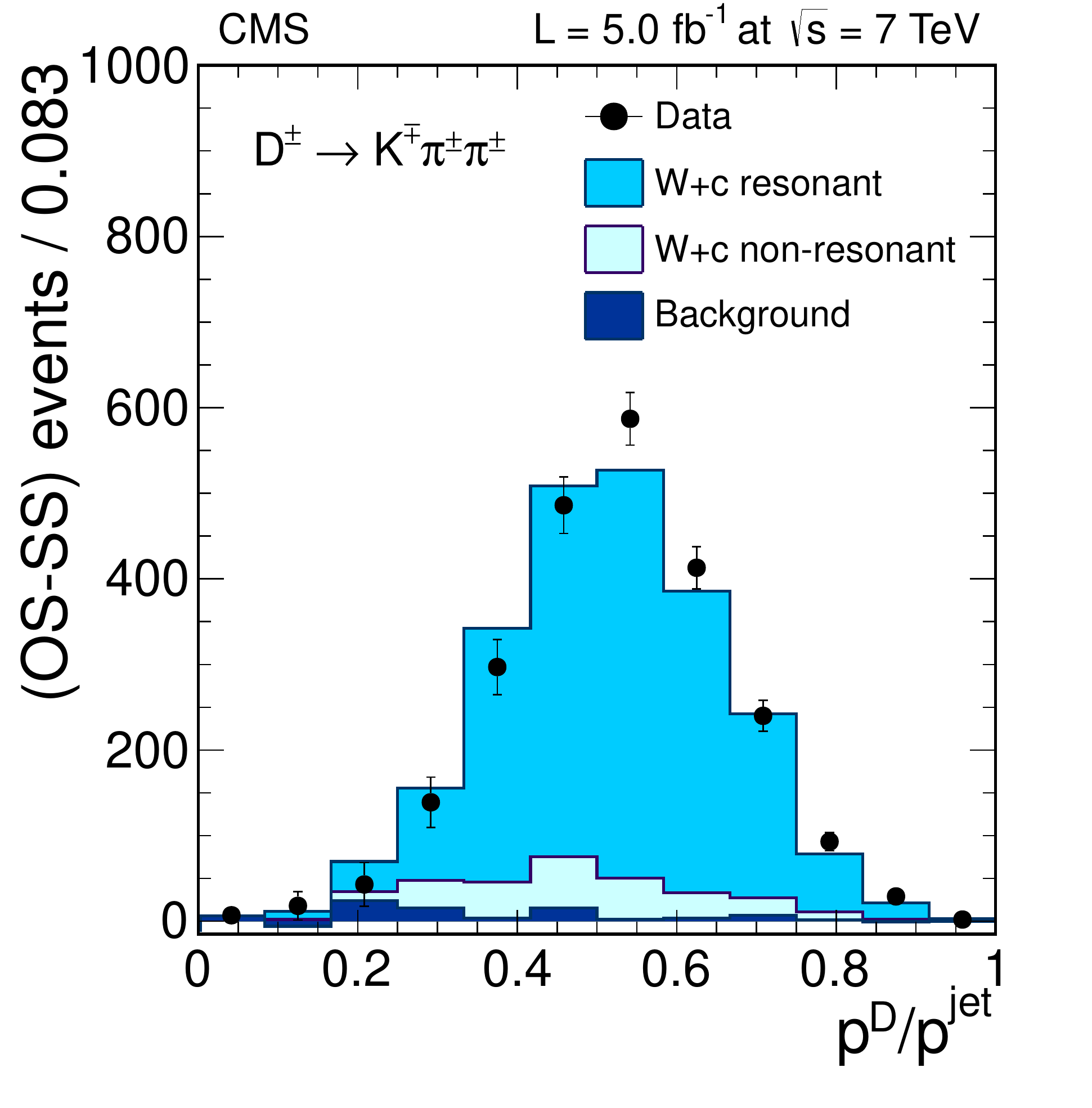}\\
    \includegraphics[width=0.4\textwidth]{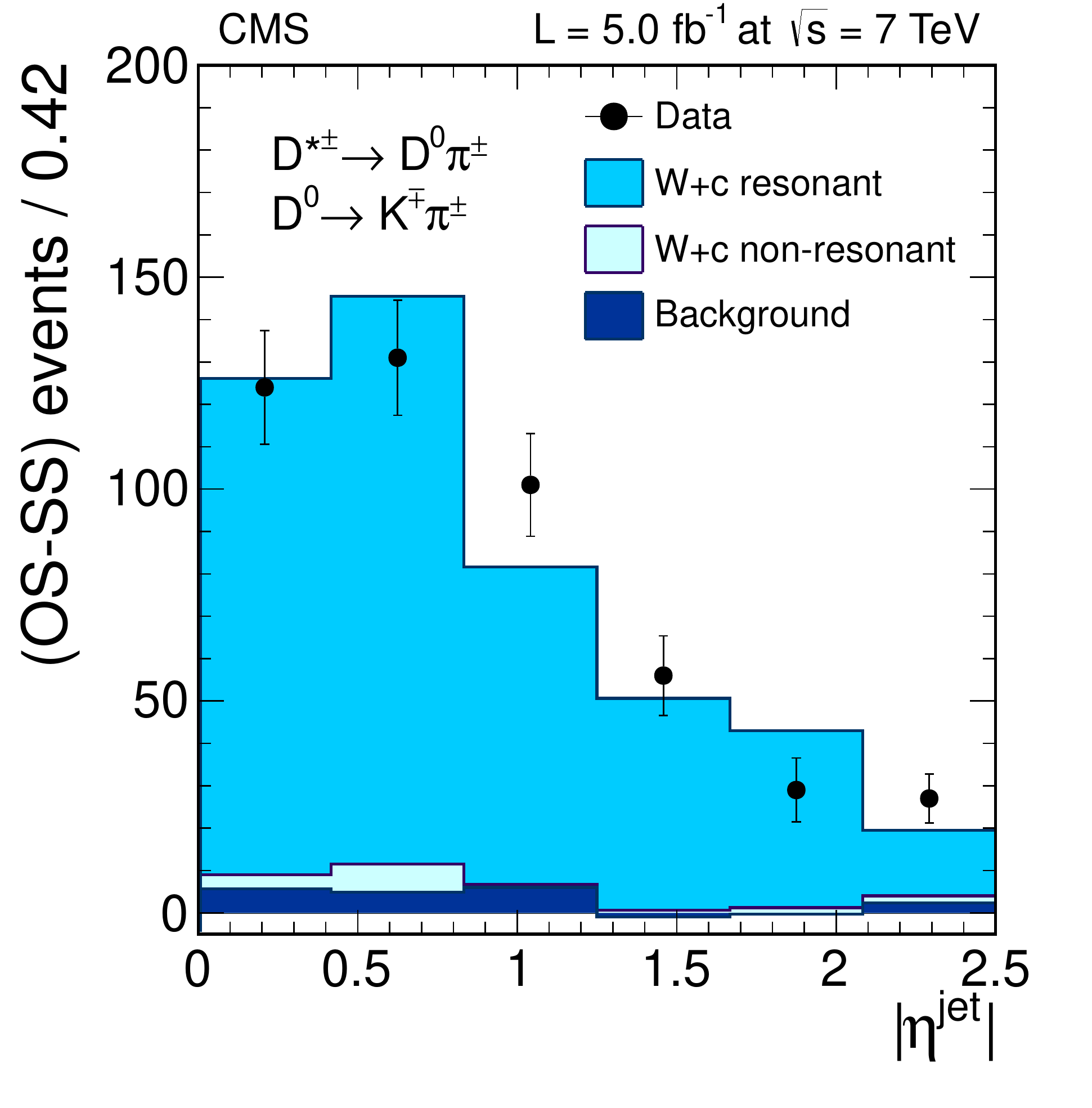}
    \includegraphics[width=0.4\textwidth]{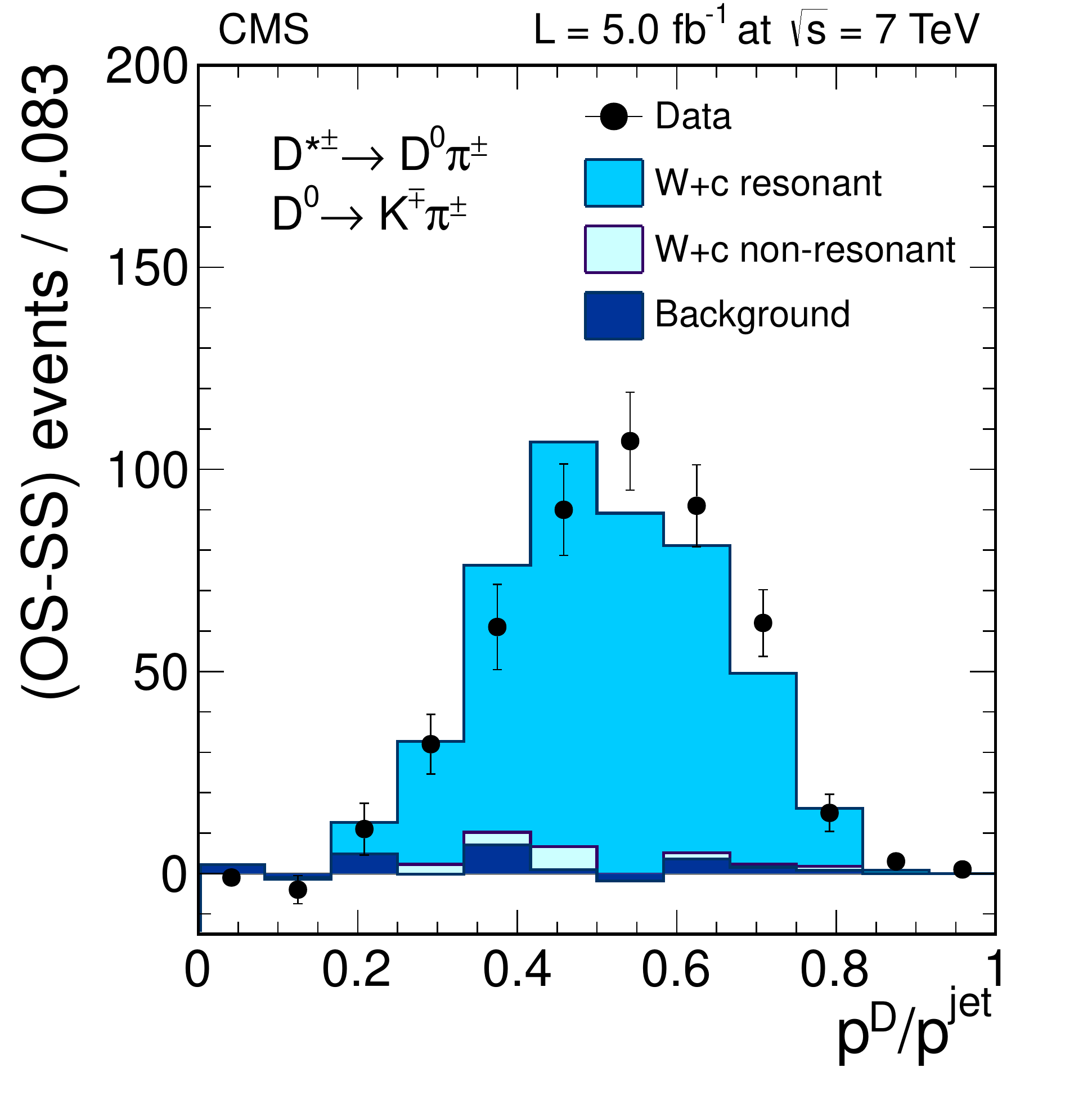}\\
    \includegraphics[width=0.4\textwidth]{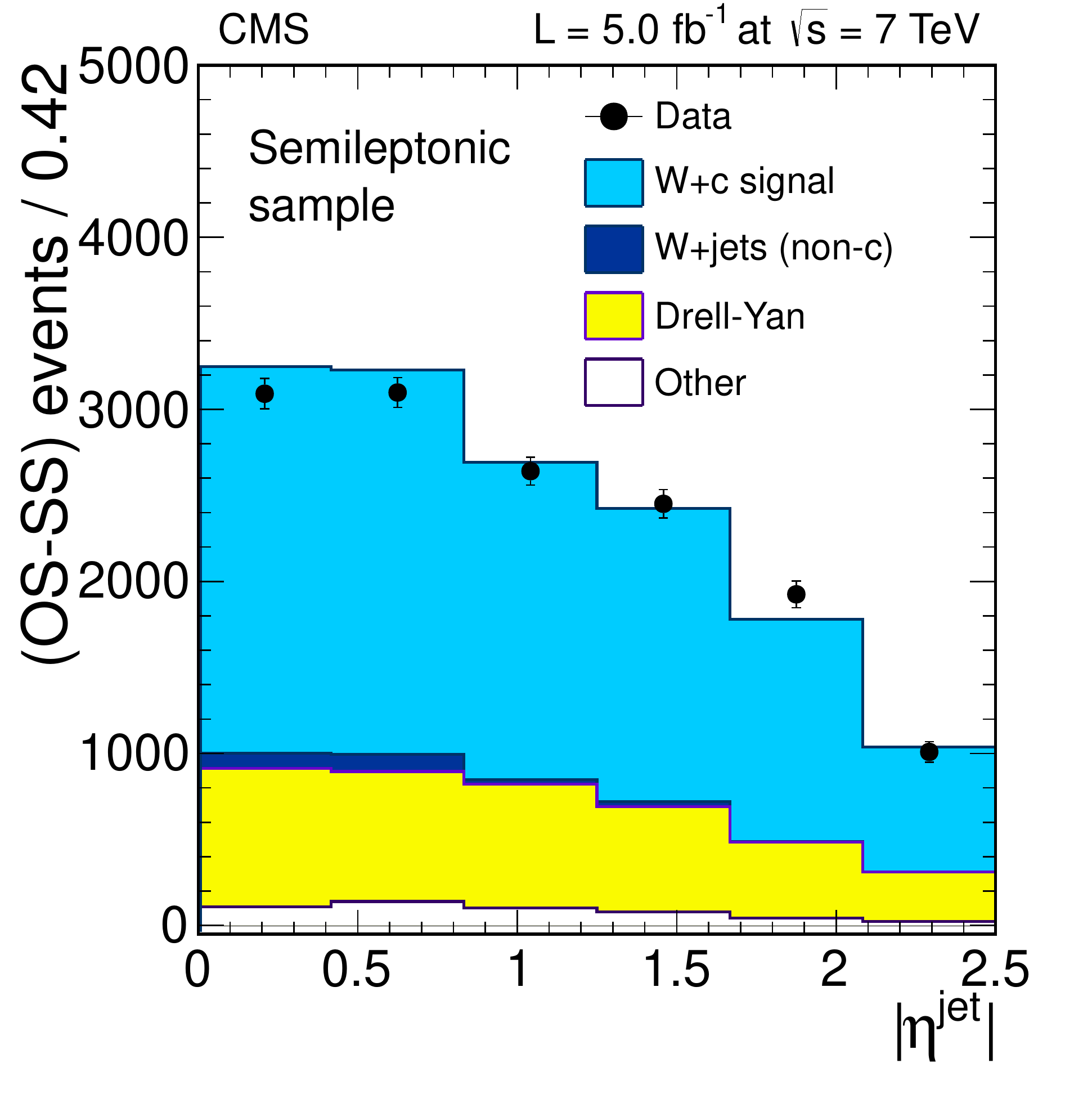}
    \includegraphics[width=0.4\textwidth]{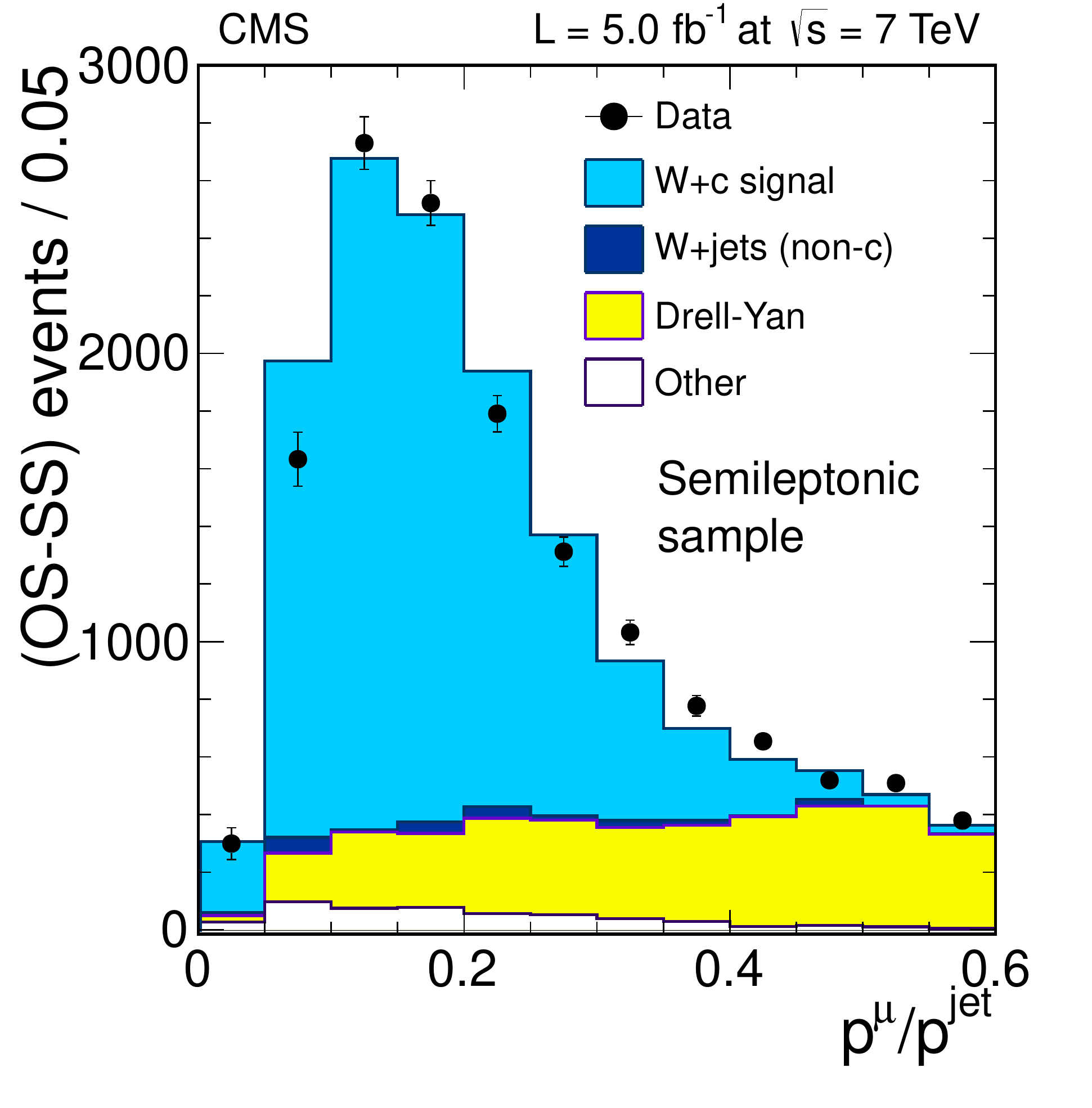}
    \caption{Distributions of $\Wc$ selected events in the different charm
decay channels as a function of the jet pseudorapidity (left) and the
jet momentum fraction (right) carried by the D meson or by the muon inside the jet.
The top row corresponds to the $\Dpm$ decay channel, the middle row corresponds to the
$\Dstar$ decay channel, and the bottom row corresponds to semileptonic charm decays into muons.
Only events in the signal region used to determine the cross section are
used. The Monte Carlo predictions
have been scaled by a factor of $1.1$ in order to approximately match the $\Wc$ yield
measured in data.}
    \label{fig:distributions}
\end{center}
\end{figure}

The high signal purity of the selected samples allows a deeper study of the properties
of $\Wc$ events.
Figure~\ref{fig:distributions} shows the distributions of the jet pseudorapidity
and the jet momentum fraction carried by the $\Dpm$ candidates (top row of plots) and the $\Dstar$ candidates (middle row of plots), while the
jet pseudorapidity and the jet momentum fraction carried by the muon is shown for the semileptonic candidates (bottom row of plots).
The latter observable is directly related to the charm fragmentation function.
The normalization of the $\Wc$ component in
the simulation has been scaled by a factor of 1.1 in order to match approximately
the experimental rate measured in data.
Electron and muon channels are added
in order to enhance the statistical power of the comparison.
All distributions show reasonable agreement with the predictions of $\MADGRAPH+\PYTHIA$, although
the experimental charm fragmentation spectra are slightly harder than the predicted ones.

\section{Measurement of the differential cross section as a function of the
lepton pseudorapidity~\label{sec:differential_xec}}

The $\Wc$ cross section is also measured differentially with respect to the absolute value of the pseudorapidity of the lepton from the $\PW$-boson decay.
We first determine the normalized differential cross section, $\SWcdiffline$. The absolute differential cross section is
derived from the normalized one just by scaling to the average cross section presented in the previous section.

For this measurement, the inclusive three-prong and two-prong samples of $\Wc$ candidates are used.
In addition, the semileptonic sample is employed.
Five bins in the absolute value of the lepton pseudorapidity are considered: [0, 0.35],~[0.35, 0.7],~[0.7, 1.1],~[1.1, 1.6],~[1.6, 2.1];
this binning is chosen in order to have a uniform distribution of the events among the five bins.

The normalized differential cross section is computed from the observed number of $\OSSS$ events with the lepton from the $\PW$-boson emitted in
a given pseudorapidity bin ($N_{\text{sel},i}$), after subtraction of the residual background ($N_{\text{bkg},i}$), which is evaluated with the simulated samples.
A bin-by-bin correction ($\AccEffnorm_i$) is used to correct $(N_{\text{sel},i}-N_{\text{bkg},i})$ for detector inefficiencies.
For this differential cross section only the differences among rapidity bins are relevant.
Hence we define the lowest rapidity bin [0, 0.35] as $\AccEffnorm_1=1.0$ and compute the correction factors relative to this bin.
These correction factors are displayed in Table~\ref{tab:acc_eff_diff}.
For $\AccEffnorm_i$ only selection requirements related to the $\PW$-boson identification and jet selection
are applied; these will be used to correct the observed events in the semileptonic sample.
This procedure is done separately for events with a secondary vertex using the correction factors $\AccEffnorm_\mathrm{ SV}$,
which are applied to the events in the inclusive three- and two-prong samples.
Global factors correcting for effects independent of the pseudorapidity of the lepton from the $\PW$-boson decay affect equally all bins
and cancel in the normalization.
The statistical uncertainty in the $\AccEffnorm_i$ factors is propagated as a systematic uncertainty to the normalized differential cross section.
\begin{table}[htbp]
\begin{center}
    \topcaption{Correction factors $\AccEffnorm$ used for the calculation of the differential measurements.
Statistical uncertainties in $\AccEffnorm$ are typically 0.3\% while in $\AccEffnorm_{\rm SV}$ they are roughly 1\%. }
    \label{tab:acc_eff_diff}
\begin{tabular}{ c  c  c  c  c  c  c }
\hline\hline
              & \multicolumn{4}{c}{$\PWmn$} & \multicolumn{2}{c}{$\Wen$} \\ 
              & \multicolumn{2}{c}{$\pt^\mu > 25\GeV$} & \multicolumn{2}{c}{$\pt^\mu > 35\GeV$} & \multicolumn{2}{c}{$\pt^\Pe > 35\GeV$} \\ \cline{2-7}
$[{\abs{\eta}}_\text{min},{\abs{\eta}}_\text{max}]$ & $\AccEffnorm$ & $\AccEffnorm_{\rm SV}$ & $\AccEffnorm$ & $\AccEffnorm_{\rm SV}$ & $\AccEffnorm$  & $\AccEffnorm_{\rm SV}$ \\
\hline
$[0, 0.35]$   & $1.00$ &  $1.00$ & $1.00$ & $1.00$ & $1.00$ & $1.00$\\
$[0.35, 0.7]$ & $1.07$ &  $1.07$ & $1.06$ & $1.06$ & $1.01$ & $0.99$\\
$[0.7, 1.1]$  & $0.98$ &  $0.97$ & $0.98$ & $0.96$ & $1.01$ & $1.01$\\
$[1.1, 1.6]$  & $0.96$ &  $0.94$ & $0.97$ & $0.95$ & $0.73$ & $0.69$\\
$[1.6, 2.1]$  & $0.90$ &  $0.86$ & $0.91$ & $0.87$ & $0.72$ & $0.65$\\
\hline\hline
\end{tabular}
\end{center}
\end{table}

The number of events selected, $N_{\text{sel},i}$, in the inclusive three-prong sample is subject to the constraint
that the invariant mass of the three tracks from the vertex, $m(\PK^\mp \pi^\pm\pi^\pm)$ is smaller than $2.5\GeV$.
The events included in the inclusive two-prong sample have a mass difference
of less than $0.7\GeV$ between (1)~the invariant mass of the two-track system plus the closest track from the primary pp interaction $m(\PK^\mp\pi^\pm\pi^\pm)$,
and (2)~the invariant mass of the two-track system $m(\PK^\mp\pi^\pm)$.
For the semileptonic channel $N_{\text{sel},i}$ is given by the number of events with a $\PW$-boson candidate decaying into a high-$\pt$ lepton and
an identified muon inside the jet passing the requirements described in Section~\ref{sec:dileptons}.
The assignment to the corresponding $i${th} bin in the differential distribution is determined by the absolute value of the pseudorapidity of the lepton
from the $\PW$-boson decay.

\begin{figure}[htbp]
\begin{center}
    \includegraphics[width=0.32\textwidth]{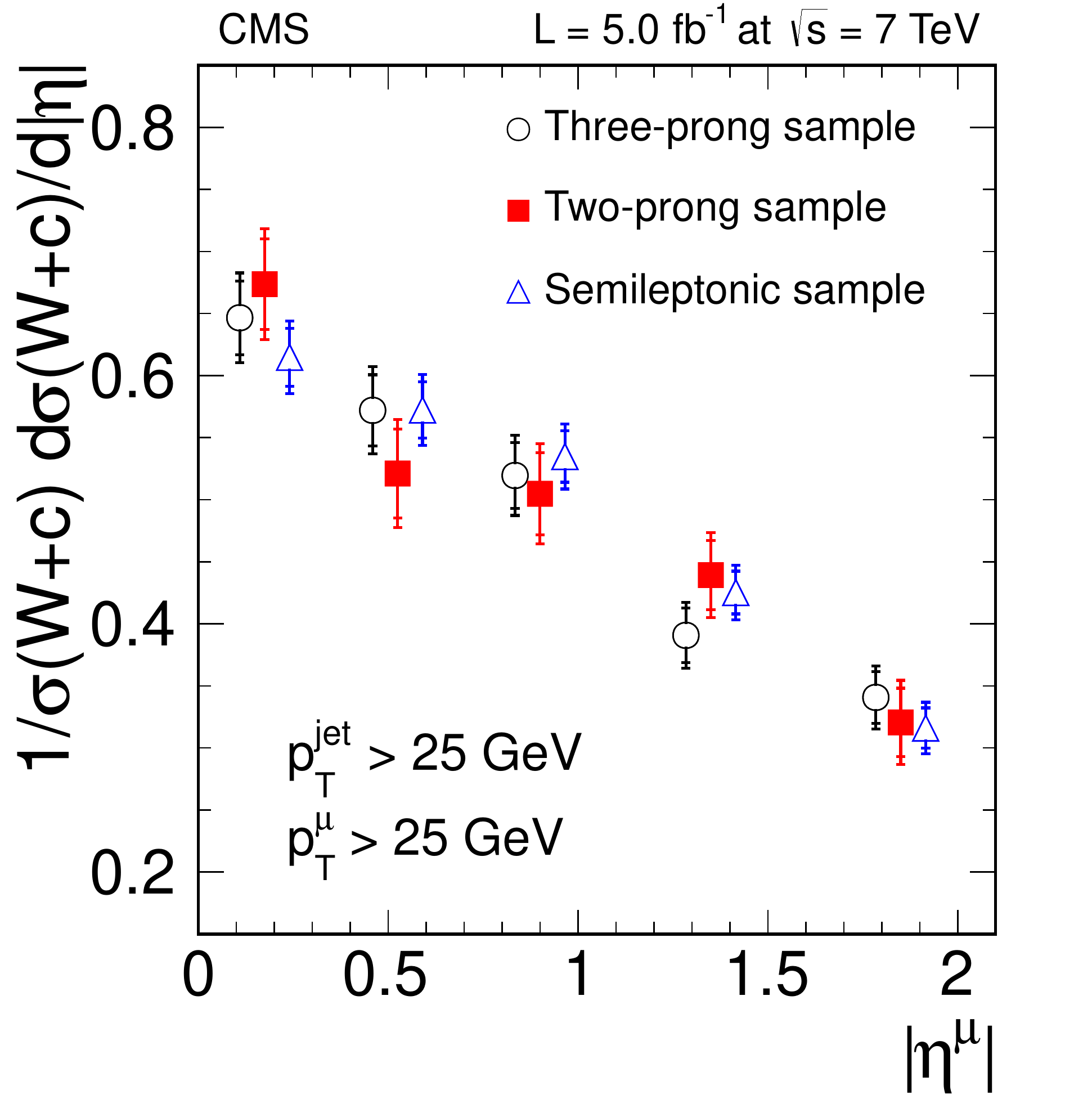}
    \includegraphics[width=0.32\textwidth]{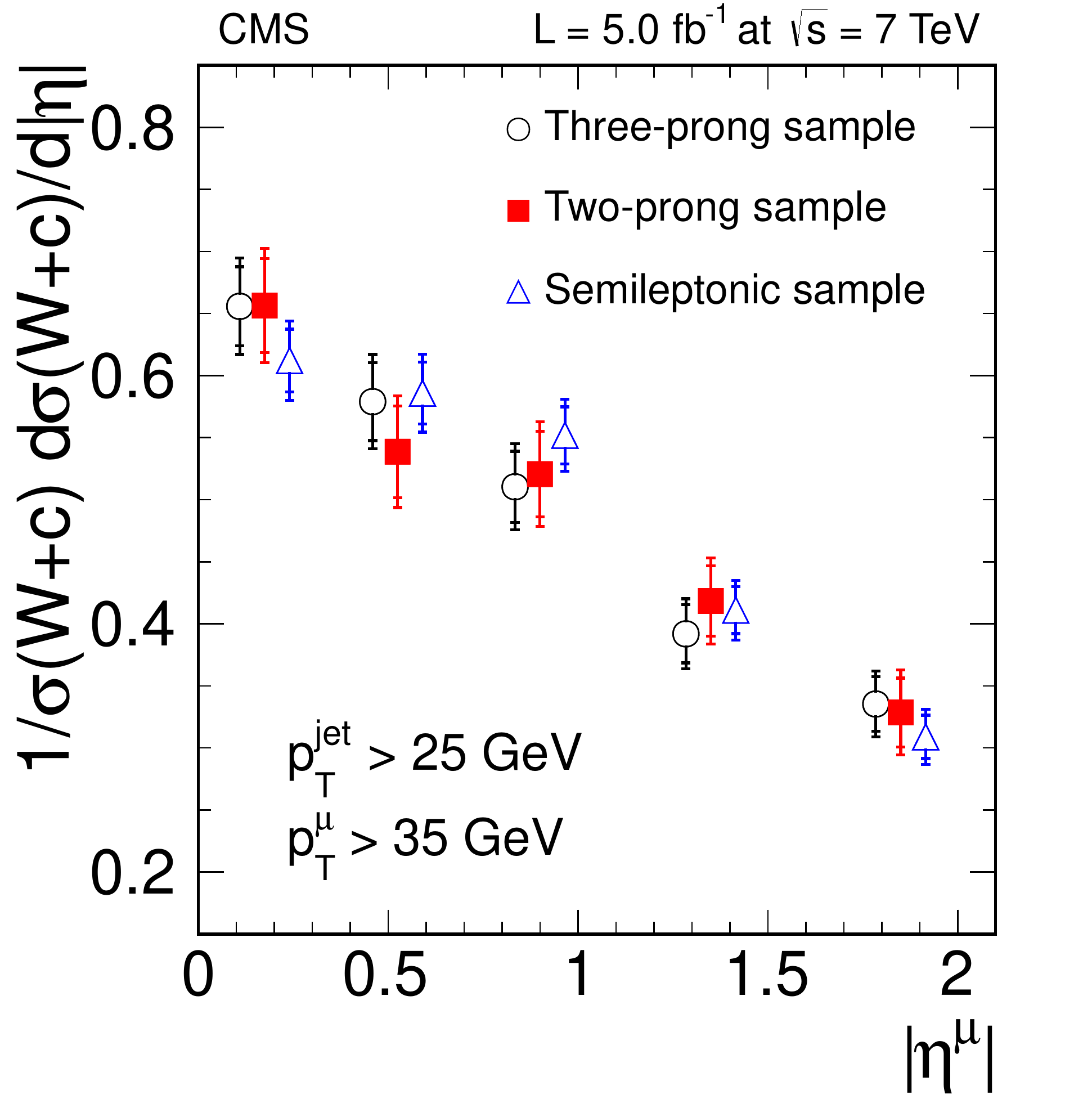}
    \includegraphics[width=0.32\textwidth]{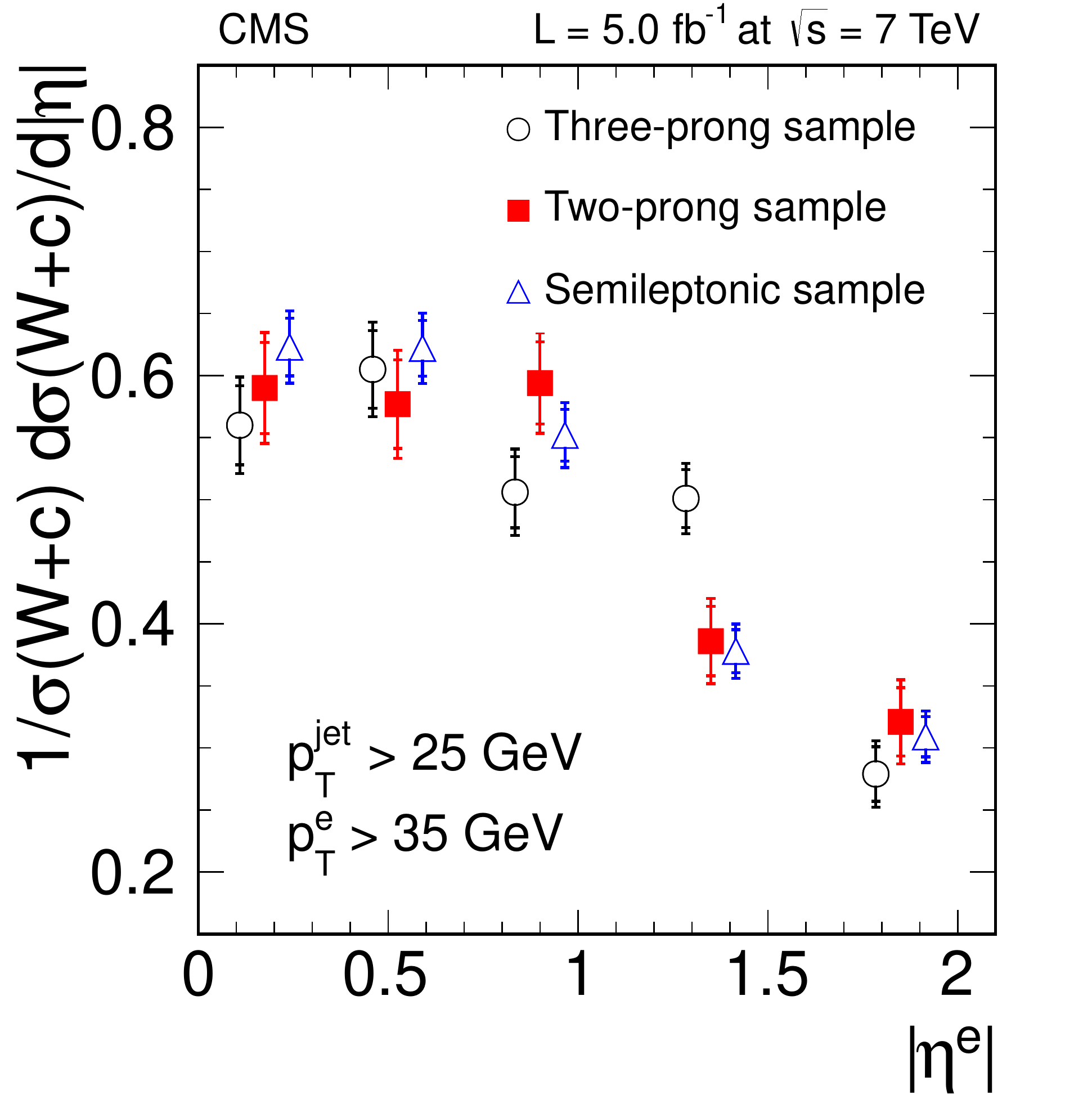}
    \caption{Normalized differential cross section distribution of $\Wc$ ($\Wln$) events as a function of the absolute value of the pseudorapidity of the lepton
from the $\PW$-boson decay. The first two plots show the results from the $\PWmn$ sample, with $\pt^\mu > 25\GeV$ (left plot) and
$\pt^\mu > 35\GeV$ (middle plot). The right plot shows the results from the $\Wen$ sample, with $\pt^\Pe > 35\GeV$.
    The results obtained with the inclusive three-prong sample are shown as open points.
    Solid squares represent the results obtained with the inclusive two-prong sample
    and the open triangles give the result from the semileptonic sample.
    Data points showing the results from the three-prong and the semileptonic samples are slightly
displaced in the horizontal axis for better visibility of the results.}
    \label{fig:Sc_diff}
\end{center}
\end{figure}

The normalized differential cross sections are presented graphically in Fig.~\ref{fig:Sc_diff}.
The number of $\OSSS$ events in each lepton pseudorapidity bin for the three charm meson decay samples are detailed in
Tables~\ref{tab:Dpm_differential}, ~\ref{tab:Dstar_differential}, and~\ref{tab:dilepton_differential} of Appendix~\ref{app:tables_diff},
together with the expected residual background $N_{\text{bkg},i}$
and the numerical values of the normalized cross sections.
The estimation of this background contamination has large statistical uncertainties due to the limited size of the MC samples, mainly for the data
with a displaced secondary vertex. This uncertainty is propagated to the differential cross sections as a systematic uncertainty in the measurement.
Unlike the $\Wen$ sample, there is a sizable background contribution in the $\PWmn$ sample arising from Drell--Yan events.

The normalized differential cross sections measured with the different $\Wc$ subsamples and for the two $\Wln$ decay channels are consistent.
Therefore, the results obtained in the $\PWmn$ channel with $\pt^\mu > 25\GeV$ are averaged, as are the results for the $\PWmn$ and $\Wen$ channels with $\pt^\ell > 35\GeV$.
These combinations are a weighted average of the individual measurements taking into account their statistical and systematic uncertainties.
Systematic uncertainties arising from a common source and affecting several measurements are considered to be fully correlated among them.
The existing statistical correlations among the normalized cross section in the five pseudorapidity bins are included in the combination.
These averaged values are given in Table~\ref{tab:averaged_norm_diff_xsec}.
The corresponding correlation matrices are presented in Table~\ref{tab:correlation_norm_diff_xsec}.

The normalized differential cross sections obtained for  $\pt^\mu > 25\GeV$ and  $\pt^{\ell} > 35\GeV$ are combined with the respective $\Wc$ cross sections presented in Section~\ref{sec:total_xsec} to obtain the absolute differential cross sections, $\SWcdifftline$.
Results are shown in Table~\ref{tab:averaged_diff_xsec}.
Normalized differential cross section and total cross section measurements are essentially uncorrelated and the full covariance matrices for the absolute
differential cross sections can be obtained by propagating the information contained in
Tables~\ref{tab:averaged_norm_diff_xsec} and~\ref{tab:correlation_norm_diff_xsec} and the total uncertainty in the $\Wc$ cross sections.
\begin{table}[htbp]
\begin{center}
    \topcaption{The normalized differential cross section as a function of the absolute value of the lepton pseudorapidity. These results are the average
of the three samples (inclusive three-prong, inclusive two-prong, and semileptonic). The left column shows the results obtained with the
$\PWmn$ sample for muons with $\pt>25\GeV$, while
the right column combines the results obtained with the $\PWmn$ and $\Wen$ samples for leptons with  $\pt>35\GeV$.
    }
    \label{tab:averaged_norm_diff_xsec}
\begin{tabular}{ c  c  c  }
\hline\hline
          & \multicolumn{2}{c}{Normalized differential cross section, $\SWcdiffline$}  \\ \cline{2-3}
$[{\abs{\eta}}_\text{min},{\abs{\eta}}_\text{max}]$ & $\pt^{\ell} > 25\GeV$  & $\pt^{\ell} > 35\GeV$ \\
\hline
$[0, 0.35]$    & $ 0.638 \pm 0.016\stat \pm 0.012\syst $ & $ 0.622 \pm 0.013\stat \pm 0.010\syst $ \\
$[0.35, 0.7]$  & $ 0.556 \pm 0.016\stat \pm 0.012\syst $ & $ 0.585 \pm 0.014\stat \pm 0.010\syst $ \\
$[0.7, 1.1]$   & $ 0.527 \pm 0.015\stat \pm 0.011\syst $ & $ 0.541 \pm 0.012\stat \pm 0.009\syst $ \\
$[1.1, 1.6]$   & $ 0.416 \pm 0.012\stat \pm 0.009\syst $ & $ 0.407 \pm 0.010\stat \pm 0.008\syst $ \\
$[1.6, 2.1]$   & $ 0.326 \pm 0.012\stat \pm 0.009\syst $ & $ 0.316 \pm 0.010\stat \pm 0.007\syst $ \\
\hline\hline
\end{tabular}
\end{center}
\end{table}
\begin{table}[htbp]
\begin{center}
    \topcaption{Correlation matrices for the averaged normalized differential cross sections $\SWcdiffline$.
Matrices are symmetric and only the lower part of them is shown.
The top matrix is for the normalized differential cross section requiring that the $\pt$ of the lepton be larger than $25\GeV$ ($\PWmn$ sample only).
The bottom one refers to the combination of results obtained with the $\PWmn$ and $\Wen$ samples for leptons with  $\pt>35\GeV$.
    }
    \label{tab:correlation_norm_diff_xsec}
\begin{tabular}{ c | ccccc } \hline \hline
\multicolumn{6}{c}{$\pt^\ell > 25\GeV$} \\ \hline
$[{\abs{\eta}}_\text{min},{\abs{\eta}}_\text{max}]$ & $[0, 0.35]$ & $[0.35, 0.7]$ & $[0.7, 1.1]$ & $[1.1, 1.6]$ & $[1.6, 2.1]$ \\ \hline
$[0, 0.35]$   &  $ 1.00$ &          &         &         &        \\
$[0.35, 0.7]$ &  $-0.22$ &  $ 1.00$ &         &         &        \\
$[0.7, 1.1]$  &  $-0.24$ &  $-0.22$ & $ 1.00$ &         &        \\
$[1.1, 1.6]$  &  $-0.26$ &  $-0.26$ & $-0.28$ & $ 1.00$ &        \\
$[1.6, 2.1]$  &  $-0.24$ &  $-0.24$ & $-0.26$ & $-0.26$ & $1.00$ \\
\hline\hline
\multicolumn{6}{c}{$\pt^\ell > 35\GeV$} \\ \hline
$[{\abs{\eta}}_\text{min},{\abs{\eta}}_\text{max}]$ & $[0, 0.35]$ & $[0.35, 0.7]$ & $[0.7, 1.1]$ & $[1.1, 1.6]$ & $[1.6, 2.1]$ \\ \hline
$[0, 0.35]$   & $ 1.00$ &          &         &         &        \\
$[0.35, 0.7]$ & $-0.20$ &  $ 1.00$ &         &         &        \\
$[0.7, 1.1]$  & $-0.22$ &  $-0.21$ & $ 1.00$ &         &        \\
$[1.1, 1.6]$  & $-0.26$ &  $-0.26$ & $-0.28$ & $ 1.00$ &        \\
$[1.6, 2.1]$  & $-0.24$ &  $-0.24$ & $-0.25$ & $-0.27$ &  $1.00$ \\
\hline\hline
\end{tabular}
\end{center}
\end{table}
\begin{table}[htbp]
\begin{center}
    \topcaption{The differential cross section as a function of the absolute value of the lepton pseudorapidity.
These results are the average of the three samples (inclusive three-prong, inclusive two-prong, and semileptonic).
The left column shows the results obtained with the $\PWmn$ sample for muons with $\pt>25\GeV$, while
the right column combines the results obtained with the $\PWmn$ and $\Wen$ samples for leptons with  $\pt>35\GeV$.
    }
    \label{tab:averaged_diff_xsec}
\begin{tabular}{ c  c  c  }
\hline\hline
      & \multicolumn{2}{c}{Differential cross section, $\SWcdifftline$ [pb]}  \\ \cline{2-3}
$[{\abs{\eta}}_\text{min},{\abs{\eta}}_\text{max}]$ & $\pt^{\ell} > 25\GeV$  & $\pt^{\ell} > 35\GeV$ \\
\hline
$[0, 0.35]$    & $68.7 \pm 2.7\stat \pm4.6 \syst $\unit{pb} & $52.3 \pm 1.7\stat \pm 3.2 \syst$\unit{pb} \\
$[0.35, 0.7]$  & $59.9 \pm 2.5\stat \pm4.0 \syst $\unit{pb} & $49.2 \pm 1.6\stat \pm 3.0 \syst$\unit{pb} \\
$[0.7, 1.1]$   & $56.7 \pm 2.4\stat \pm3.8 \syst $\unit{pb} & $45.5 \pm 1.5\stat \pm 2.7 \syst$\unit{pb} \\
$[1.1, 1.6]$   & $44.8 \pm 1.9\stat \pm3.2 \syst $\unit{pb} & $34.2 \pm 1.2\stat \pm 2.1 \syst$\unit{pb} \\
$[1.6, 2.1]$   & $35.1 \pm 1.7\stat \pm2.4 \syst $\unit{pb} & $26.6 \pm 1.0\stat \pm 1.7 \syst$\unit{pb} \\
\hline\hline
\end{tabular}
\end{center}
\end{table}

\subsection{Systematic uncertainties in the normalized differential cross section measurement~\label{sec:syst_diff}}

The dominant source of systematic uncertainty in the normalized differential cross sections from the three samples is the limited size of the MC samples.
It impacts the statistical accuracy in the estimation of the residual background after the SS subtraction, and to a lesser extent, 
in the determination of the correction factors $\AccEffnorm_i$.
As summarized below, most of the other sources that have been discussed in Section~\ref{sec:total_xsec} have a negligible impact in the differential distributions since their effects largely cancel out in the ratios.

Differential distributions are mostly independent of jet energy scale effects since they are measured as a
function of the pseudorapidity of the lepton from the $\PW$-boson decay and the spanned jet kinematic region is similar in all cases, independently of the pseudorapidity of the lepton.
Possible effects due to jet energy scale uncertainties are evaluated by changing the jet energy scale in the simulated $\Wc$ sample in accord with the results of
dedicated studies by CMS~\cite{CMS-PAPER-JME-10-011}. The variations observed in the resulting differential distribution can be largely explained by statistical fluctuations in the MC sample.

The calibration factors for lepton momentum scale and resolution have been derived from detailed studies of the position and width of the
Z-boson peak~\cite{mu_scale,CMS-PAPER-EGM-11-001}.
The systematic uncertainty in the normalized differential cross section is estimated in the $\Wen$ channel by comparing the resulting distributions with and without calibration corrections. Variations are smaller than 1\% in the barrel, and of the order of 1.5\% in the endcap region.
In the $\PWmn$ channel the measurement is repeated many times, varying the muon calibration factors
within their uncertainties and comparing to the values obtained when applying the central value of the correcting factors.  The width of the resulting distribution is taken as the systematic uncertainty arising from limited knowledge of the muon momentum scale and resolution. Uncertainties between $0.2\%$ and $0.4\%$ in the normalized differential distributions are obtained, depending on the particular muon pseudorapidity bin, the sample selection, and the $\pt^\mu$ threshold.

We estimate a residual ${\sim}0.35\%$ systematic uncertainty in the muon efficiency scaling factors, which are treated as uncorrelated among the different pseudorapidity bins.
For the $\Wen$ channel, the effect of the efficiency corrections in the measured ratios (${\sim}0.25\%$) is computed and taken as an estimation of the systematic uncertainty.

In the modeling of the background remaining after the SS subtraction, the only physical process with a visible contribution to the final sample is 
Drell--Yan production, which, when one of the two muons is inside a jet, mimics the semileptonic sample in the $\PWmn$ channel.
The correction factor ($1.2 \pm 0.1$) applied to the Drell--Yan prediction is varied by one sigma and the differential distribution is reevaluated.
Variations smaller than 0.3\% are observed and taken as the associated systematic uncertainty.
Top-quark contributions have also been varied by $6\%$ for $\ttbar$ production and by $15\%$ for single-top-quark production. Variations in the differential
distributions are smaller than $0.2\%$. A total systematic uncertainty of $0.3\%$ is assumed to account for the background subtraction.

It is observed that the uncertainties related to the parton distribution function of the strange quark within the same PDF set are smaller than,
or equal to, the differences between the central values obtained with
MSTW08~\cite{MSTW08}, CT10~\cite{CT10}, and NNPDF23~\cite{Ball:2011uy}.
However, no variation in the $\AccEff$ correction factors computed with these sets of PDFs is observed and therefore no change is expected in the final result.

Systematic uncertainties arising from other sources, such as knowledge of the event pileup or the average energy fraction in charm fragmentation have been
evaluated with the $\Wc$ MC sample and are found to be negligible.

The systematic uncertainties in the absolute differential cross sections given in Table~\ref{tab:averaged_diff_xsec} are dominated by the 
uncertainties in the total $\Wc$ cross section. The relative importance of the different sources essentially follows the breakdown 
of the contributions presented in Table~\ref{tab:xsection_systematics}. The effect of the limited MC statistics is increased because both measurements, total and normalized differential cross sections, are affected.

\section{Measurement of the cross section ratio \texorpdfstring{$\SWpc/\SWmc$}{of cross sections for W+ c bar to W- c}~\label{sec:ratio}}

Cross section ratios $\SWpc/\SWmc$ are also measured for the three specific final states discussed in the previous section.
They are determined as the ratio of the $\OSSS$ samples in which the lepton from the $\PW$-boson decay is positively or negatively charged:
\begin{equation*}\label{eqn:chratio}
\Rcpm = \frac{\SWpc}{\SWmc} = \frac{(N^+_\text{sel}-N^+_\text{bkg})}{(N^-_\text{sel}-N^-_\text{bkg})}.
\end{equation*}
The total cross section ratio and the ratio as a function of the absolute value of the pseudorapidity of the lepton from the $\PW$-boson decay are determined.

The numbers for $N^+_\text{sel}$ and $N^-_\text{sel}$ are extracted from the
same subsamples used for the differential cross section measurement presented in the previous section and by separating the events
according to the sign of the lepton from the $\PW$-boson decay.
The background contributions $N^+_\text{bkg}$ and $N^-_\text{bkg}$ to $N^+_\text{sel}$ and $N^-_\text{sel}$
have a small effect in the ratio and are neglected in the calculation. The largest
effect is due to the Drell--Yan contamination in the $\PWmn$ channel and that is reduced by requiring that
the transverse momentum of the muon inside the jet be less than $12\GeV$.
No efficiency corrections
are applied since they affect the positively and negatively charged samples equally and cancel in the ratio.

\begin{figure}[htbp]
\begin{center}
     \includegraphics[width=0.32\textwidth]{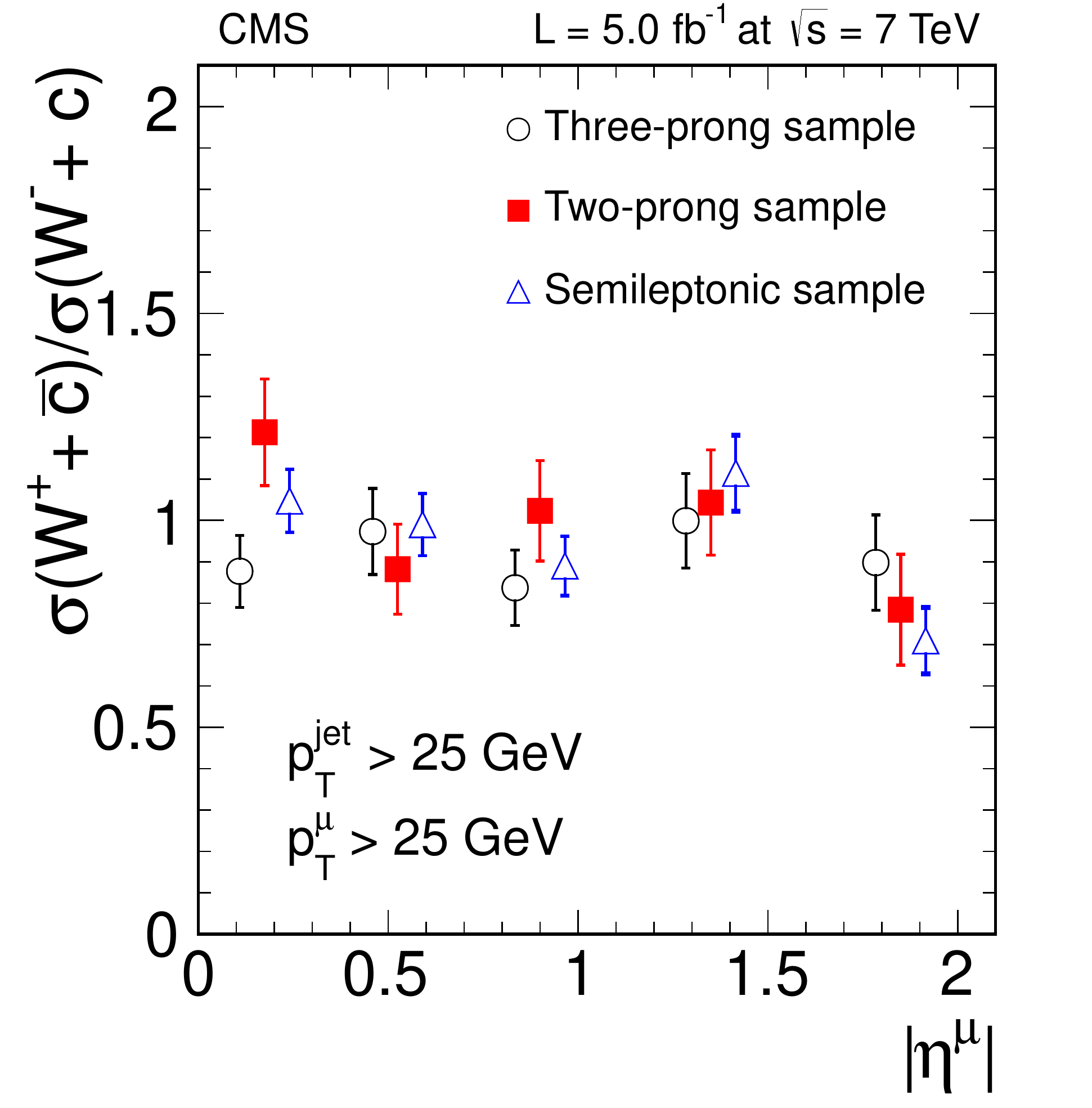}
     \includegraphics[width=0.32\textwidth]{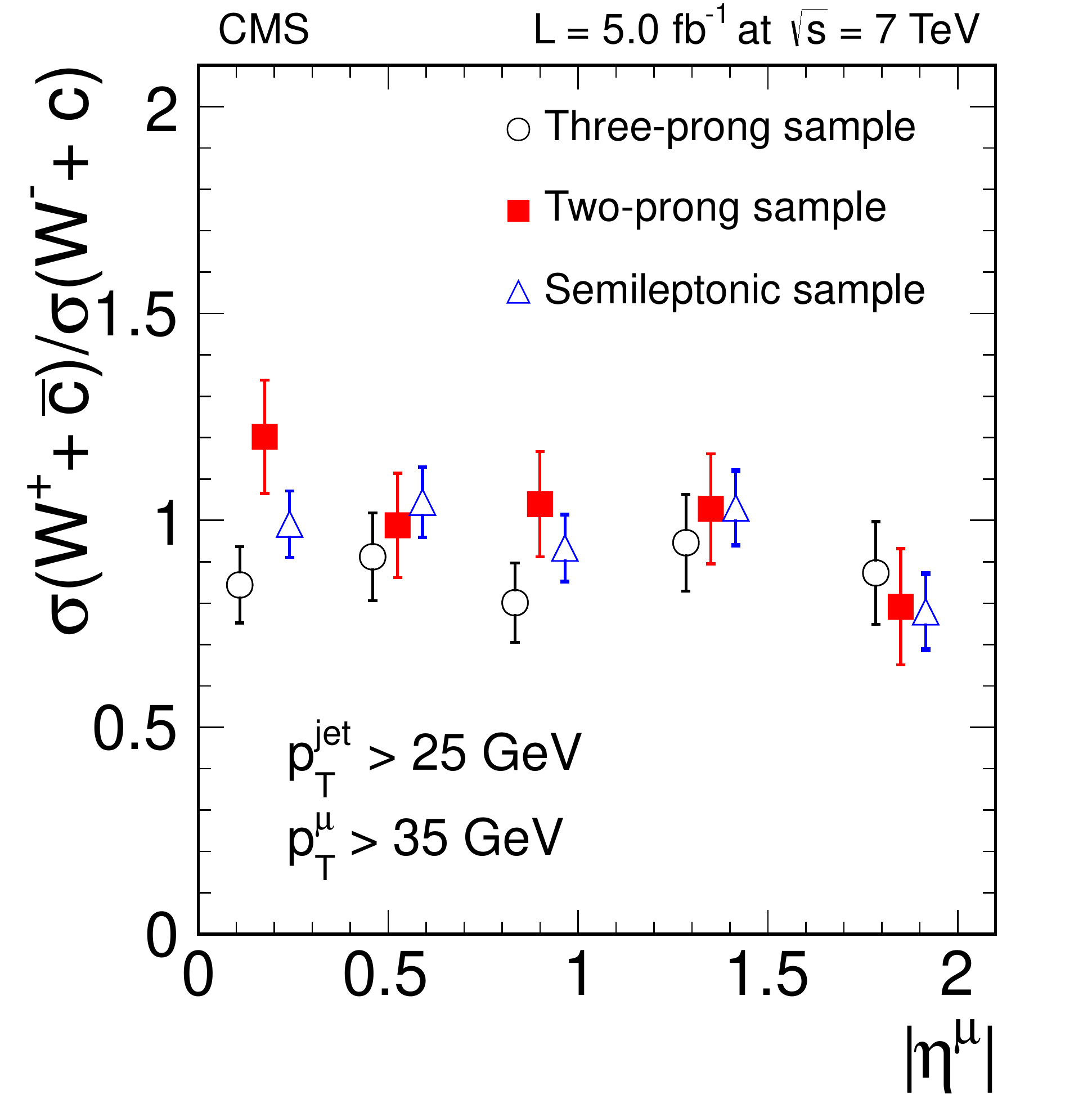}
     \includegraphics[width=0.32\textwidth]{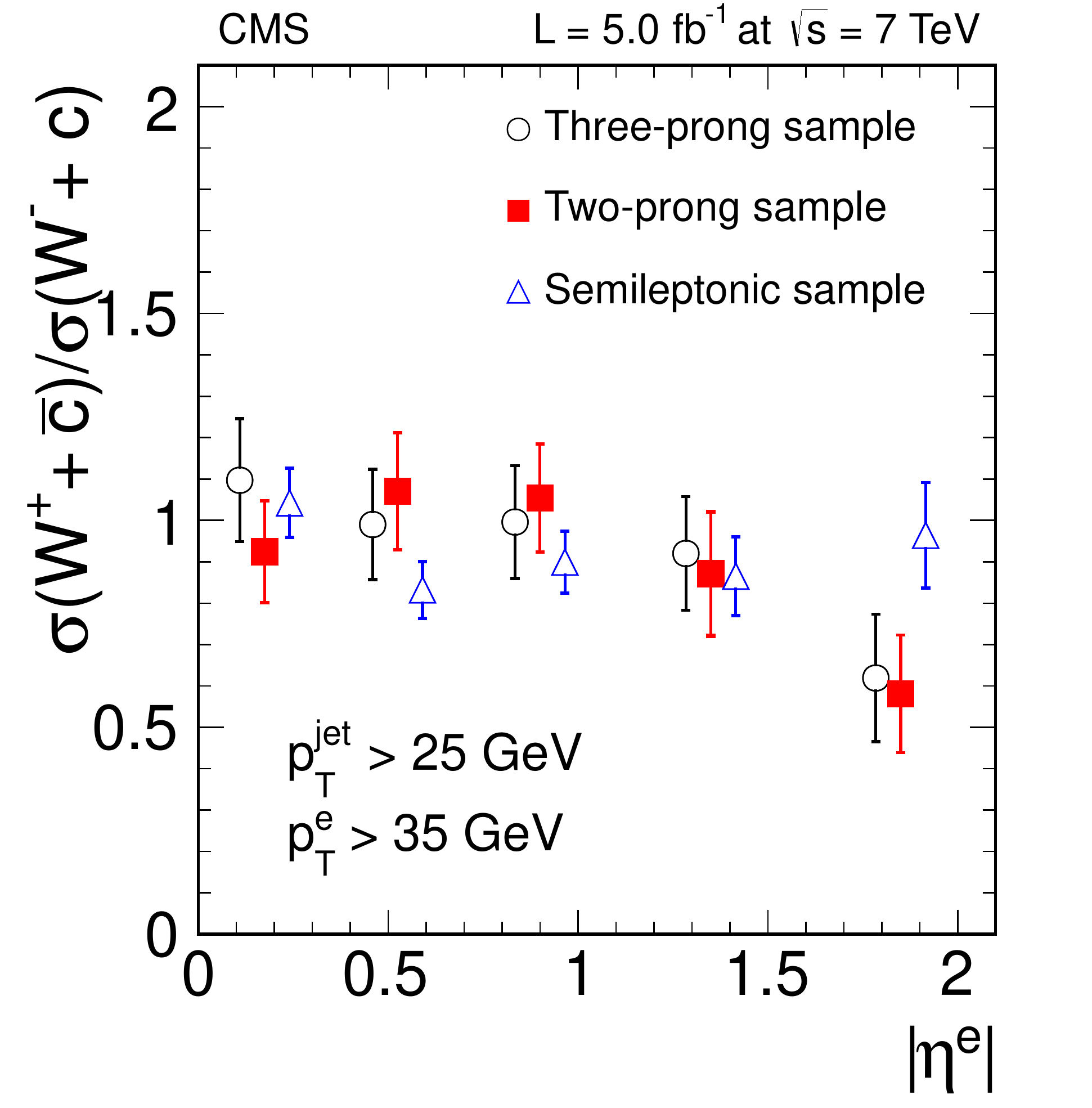}
    \caption{Measured ratios $\SWpc/\SWmc$ as a function of the absolute value of the lepton pseudorapidity from the $\PW$-boson decay.
    The first two plots show the results from the $\PWmn$ sample, with $\pt^\mu > 25\GeV$ (left plot) and  $\pt^\mu > 35\GeV$ (middle plot). The right plot shows the results from the $\Wen$ sample, with $\pt^\Pe > 35\GeV$.
    The results obtained with the inclusive three-prong sample are shown as open points.
    Solid squares represent the results obtained with the inclusive two-prong sample and the open triangles give the result from the semileptonic sample.
    Data points showing the results from the three-prong and the semileptonic samples are slightly
displaced in the horizontal axis for better visibility of the results.}
    \label{fig:Rpm}
\end{center}
\end{figure}

Figure~\ref{fig:Rpm} presents the cross section ratios $\Rcpm(\abs{\eta^\ell})$ obtained from the three samples.
The numerical values of the cross section ratio are detailed in Table~\ref{tab:Rpm} in Appendix~\ref{app:tables_diff}.
The last row of each set of results in the table gives the cross section ratio for the full lepton absolute pseudorapidity range [0., 2.1].

The effect of neglecting the background is estimated to be of the order of $0.3\%$ and $0.2\%$
for the inclusive cross section ratio in the inclusive three- and two-prong samples, respectively.
It is $1\%$ ($0.3\%$) in the semileptonic sample in the $\PWmn$ ($\Wen$) channel.
In the ratios as a function of the absolute value of the pseudorapidity, the largest effect is for
the highest $\abs{\eta}$ bin for all samples (${\sim} 1\%$) except for the semileptonic sample in the $\PWmn$ channel
where it reaches ${\sim} 4\%$.
Other sources of systematic uncertainties in the cross section ratios are those related to lepton reconstruction, identification, and, in particular, any lepton-charge-dependent effect that may affect the $\PWp$ and $\PWm$ candidate samples differently.
The systematic uncertainty in the cross section ratio due to lepton momentum scale and resolution is estimated  following the same technique used for the normalized
differential cross section. The uncertainties in the $\Wen$ channel are smaller than 1\% in the barrel, and approximately 1.5\% in the endcap region.
They vary in the range 0.4--0.8\% in the $\PWmn$ channel, depending again on the muon pseudorapidity bin, the sample, and the muon $\pt$ threshold.
They reduce to ${\sim}0.2$--$0.3\%$ for the inclusive cross section ratios since the effect of muon momentum correction factors for the muon pseudorapidity bins cancels to a large extent, thus decreasing the final uncertainty.
The correction factors to the lepton reconstruction efficiencies for positively and negatively charged leptons are the
same within their statistical uncertainty and thus no additional systematic uncertainties are assigned to this source.

The lepton charge misassignment in CMS is smaller than $0.3\%$ for electrons~\cite{CMS-PAPER-SMP-12-001} and of the order of $10^{-4}$ for muons~\cite{CMS-PAPER-MUO-10-001}.
The associated systematic uncertainty in the cross section ratio is proportional to the relative difference between $\PWpc$ and $\PWmc$ production.
Since this is small because the measured cross section ratios are close to 1, the total effect is neglected.

The cross section ratios, both total and as a function of the lepton pseudorapidity, measured with the different $\Wc$ samples
and for the two $\Wln$ decay channels are consistent.
The results obtained in the $\PWmn$ channel with $\pt^\mu > 25\GeV$ are
averaged, as are the results for the $\PWmn$ and $\Wen$ channels with $\pt^\ell > 35\GeV$.
Statistical and systematic uncertainties of the individual measurements are taken into account in the combination process.
Systematic uncertainties arising from a common source and affecting several measurements are considered to be fully correlated.

The following averaged $\Rcpm$ ratios in the full pseudorapidity interval are derived:
\begin{align*}
\frac{\sigma(\Pp\Pp \rightarrow \mathrm{\PWp + \cPaqc +X})}{\sigma(\Pp\Pp \rightarrow \mathrm{\PWm + c + X})} (\pt^\mu>25\GeV) & = & 0.954 \pm 0.025\stat \pm 0.004\syst, \\
 & & \\
 & & \\
\frac{\sigma(\Pp\Pp \rightarrow \mathrm{\PWp + \cPaqc +X})}{\sigma(\Pp\Pp \rightarrow \mathrm{\PWm + c + X})} (\pt^\mu>35\GeV) & = & 0.947 \pm 0.026\stat \pm 0.005\syst, \\
\frac{\sigma(\Pp\Pp \rightarrow \mathrm{\PWp + \cPaqc +X})}{\sigma(\Pp\Pp \rightarrow \mathrm{\PWm + c + X})} (\pt^\Pe>35\GeV) & = & 0.927 \pm 0.029\stat \pm 0.012\syst, \\
 & & \\
 & & \\
\frac{\sigma(\Pp\Pp \rightarrow \mathrm{\PWp + \cPaqc +X})}{\sigma(\Pp\Pp \rightarrow \mathrm{\PWm + c + X})} (\pt^\ell>35\GeV) & = & 0.938 \pm 0.019\stat \pm 0.006\syst.\\
\end{align*}
and the corresponding averaged values as a function of the absolute value of the pseudorapidity are presented in Table~\ref{tab:averaged_chargeratio}.
\begin{table}[htbp]
\begin{center}
    \topcaption{Measured ratios $\SWpc/\SWmc$ as a function of the absolute value of the pseudorapidity of the lepton from the $\PW$-boson decay.
The results are the average of the three different samples (inclusive three-prong and two-prong and semileptonic).
The left column shows the results obtained with the $\PWmn$ sample for muons with $\pt^\mu>25\GeV$, while
the right column combines the results obtained with the $\PWmn$ and $\Wen$ samples for leptons with  $\pt^\ell>35\GeV$.
    }
    \label{tab:averaged_chargeratio}
\begin{tabular}{ c  c  c  }
\hline\hline
           & \multicolumn{2}{c}{Charged cross section ratio, $\SWpc/\SWmc$ }  \\ \cline{2-3}
$[{\abs{\eta}}_\text{min},{\abs{\eta}}_\text{max}]$ & $\pt^{\ell} > 25\GeV$  & $\pt^{\ell} > 35\GeV$ \\
\hline
$[0, 0.35]$    & $ 1.013 \pm 0.052\stat \pm 0.005\syst $ & $ 0.993 \pm 0.041\stat \pm 0.007\syst $ \\
$[0.35, 0.7]$  & $ 0.960 \pm 0.053\stat \pm 0.005\syst $ & $ 0.977 \pm 0.039\stat \pm 0.007\syst $ \\
$[0.7, 1.1]$   & $ 0.897 \pm 0.051\stat \pm 0.008\syst $ & $ 0.927 \pm 0.040\stat \pm 0.008\syst $ \\
$[1.1, 1.6]$   & $ 1.062 \pm 0.061\stat \pm 0.014\syst $ & $ 0.948 \pm 0.046\stat \pm 0.010\syst $ \\
$[1.6, 2.1]$   & $ 0.776 \pm 0.058\stat \pm 0.016\syst $ & $ 0.784 \pm 0.050\stat \pm 0.011\syst $ \\
\hline\hline
\end{tabular}
\end{center}
\end{table}

A larger production yield of $\PWmc$ than of $\PWpc$ is expected because the former process involves a d quark whereas the latter
involves a $\cPaqd$ (sea) antiquark.
This prediction is confirmed since the measured cross section ratio $\SWpc/\SWmc$ is smaller than 1.0.
The difference in production between $\PWpc$ and $\PWmc$ is not constant over the full pseudorapidity range.
Production cross sections are similar in the central region, $\Rcpm {\sim} 1$, for absolute values of the pseudorapidity
of the lepton smaller than 0.35. The ratio reduces to about 0.8 for the most forward
lepton pseudorapidity.  A decrease of the cross section ratio with the lepton pseudorapidity is expected,
since in this case we are probing a region of Bjorken $x$ where the difference between the d and $\cPaqd$ contributions is larger.

\section{Results and comparisons with theoretical predictions~\label{sec:results}}

The measured total and differential cross sections and cross section ratios can be compared to analytical calculations from the \MCFM program.
The $\Wc$ process is available in \MCFM up to $\mathcal{O}({\alpha_\mathrm{s}}^2)$
with a massive charm quark ($m(\cPqc) =1.5\GeV$).
The \MCFM predictions for this process do not include contributions
from gluon splitting into a $\cPqc{\cPaqc}$ pair, but only contributions
where the strange (or the down) quark couples to the $\PW$ boson.
The implementation of $\Wc$ follows the calculation for the similar $\PW$+top-quark process~\cite{MCFM_WplusC}.

The parameters of the calculation have been adjusted to match the experimental measurement: $\pt^\text{jet}>25\GeV$ and $\abs{\eta^\text{jet}}<2.5$.
Two sets of predictions are computed, utilizing the different lepton $\pt$ thresholds used in the analysis: $\pt^{\ell}>25\GeV$
in the $\PWmn$ channel and $\pt^\ell>35\GeV$ in the $\PWmn$ and in the $\Wen$ channel.

We show predictions for three NNLO PDF sets:
MSTW2008, CT10, and NNPDF2.3.
These three PDF sets have in common the use of a global data set with a wide variety of observables to constrain PDFs, and,
in particular, they include neutrino charm production data to provide information on the strange-quark content of the proton.
In addition, we compare with predictions using the
NNPDF2.3$_\text{coll}$ NNLO set~\cite{NNPDF23},
which is based on  high energy collider data only, and
thus does not rely on the neutrino DIS charm information. In particular,
it includes $\PW$ and $\cPZ$ production data from ATLAS, CMS,
and LHCb, and leads to a larger strangeness content of the proton than that of
global PDF sets.
These four sets span a wide range of values for the strange-quark PDF, and the strangeness content from other PDF analyses falls within this interval.
NNPDF2.3 has the smallest strangeness, and NNPDF2.3$_\text{coll}$ the largest one.
We have also computed the theoretical predictions for the ABM11~\cite{ABM}, JR09~\cite{JR}, and HERAPDF1.5~\cite{CooperSarkar,Radescu}
PDF sets and we discuss these results below as well.

Both the factorization and the renormalization scales are set to the value of the $\PW$-boson mass.
To estimate the uncertainty from missing higher
perturbative orders, cross section predictions are computed by varying
independently the factorization and renormalization scales to twice and half the nominal value (with the constraint that the ratio of scales is never larger than two).
The envelope of the cross sections with these scale variation defines the theoretical scale uncertainty.

The value of $\alpha_\mathrm{s}(M_\cPZ)$ in the calculation is set to the central value given by the respective PDF groups.
Uncertainties in the predicted cross sections associated with $\alpha_\mathrm{s}(M_\cPZ)$ are smaller than the
uncertainties from the PDFs, and have been neglected in the following comparisons.

\subsection{Total cross section~\label{sec:theo_total}}

The measured total cross sections are consistent with theoretical
expectations.
However, there are significant variations depending on the PDF set used in the prediction.
The detailed theoretical predictions
are summarized in Table~\ref{tab:MCFM_Sc} where the central value of the prediction
is given, together with the uncertainty due to the
PDF variations within each set. The experimental results reported in this document are also included in the table.
The size of the PDF uncertainties depends on the different methodology
used by the various groups. In particular, they depend on the parametrization of the strange-quark PDF and on the definition of the one-standard-deviation
uncertainty band. In the case of NNPDF2.3$_\text{coll}$, the larger
uncertainties arise from the lack of direct constraints on strangeness
in a collider-only fit.
\begin{table}[htbp]\renewcommand{\arraystretch}{1.2}
\begin{center}
 \topcaption{Predictions for $\SWc$ from \MCFM at NLO.
 Kinematic selection follows the experimental requirements: $\pt^\text{jet}>25\GeV$, $\abs{\eta^\text{jet}}<2.5$, and $\abs{\eta^\ell}<2.1$.
Partons are joined using an anti-\kt algorithm with a distance parameter of $1$.
Theoretical predictions are computed with \MCFM for two different thresholds in the lepton $\pt$: $\pt^{\ell}>25~(35)\GeV$
in the first (second) column of predictions.
For every PDF set, the central value of the prediction is given, together with the
relative uncertainty as prescribed from the PDF set.
The uncertainty associated with scale variations is $\pm 5\%.$
The last row in the table gives the experimental results presented in this document.}
\label{tab:MCFM_Sc}
\begin{tabular}{ccccc}
\hline\hline
          & \multicolumn{2}{c}{$\pt^\ell > 25\GeV$} & \multicolumn{2}{c}{$\pt^\ell > 35\GeV$}  \\ \cline{2-5}
PDF set   & $\SWc$ [pb] & $\Delta_{\mathrm{PDF}} [\%]$ & $\SWc$ [pb] & $\Delta_{\mathrm{PDF}} [\%]$ \\
\hline
MSTW08                          & $100.7$ & $^{+1.8}_{-2.2}$ & $78.7$ & $^{+1.8}_{-2.2}$ \\
CT10                            & $109.9$ & $^{+7.0}_{-5.8}$ & $87.3$ & $^{+7.1}_{-5.9}$ \\
NNPDF2.3                        & $\phantom{1}99.4$  & $\pm 4.2$ & $78.9$ & $\pm 4.2$ \\
NNPDF2.3$_\text{coll}$          & $129.9$ & $\pm 11.6$         & $102.7$ & $\pm 11.5$ \\ \hline
CMS                             & \multicolumn{2}{c}{$107.7 \pm 3.3~{\stat} \pm 6.9~{\syst}$} &
                                  \multicolumn{2}{c}{$84.1 \pm 2.0~{\stat} \pm 4.9~{\syst}$} \\
\hline\hline
\end{tabular}
\end{center}
\end{table}

\begin{figure}[htbp]
\begin{center}
    \includegraphics[width=0.45\textwidth]{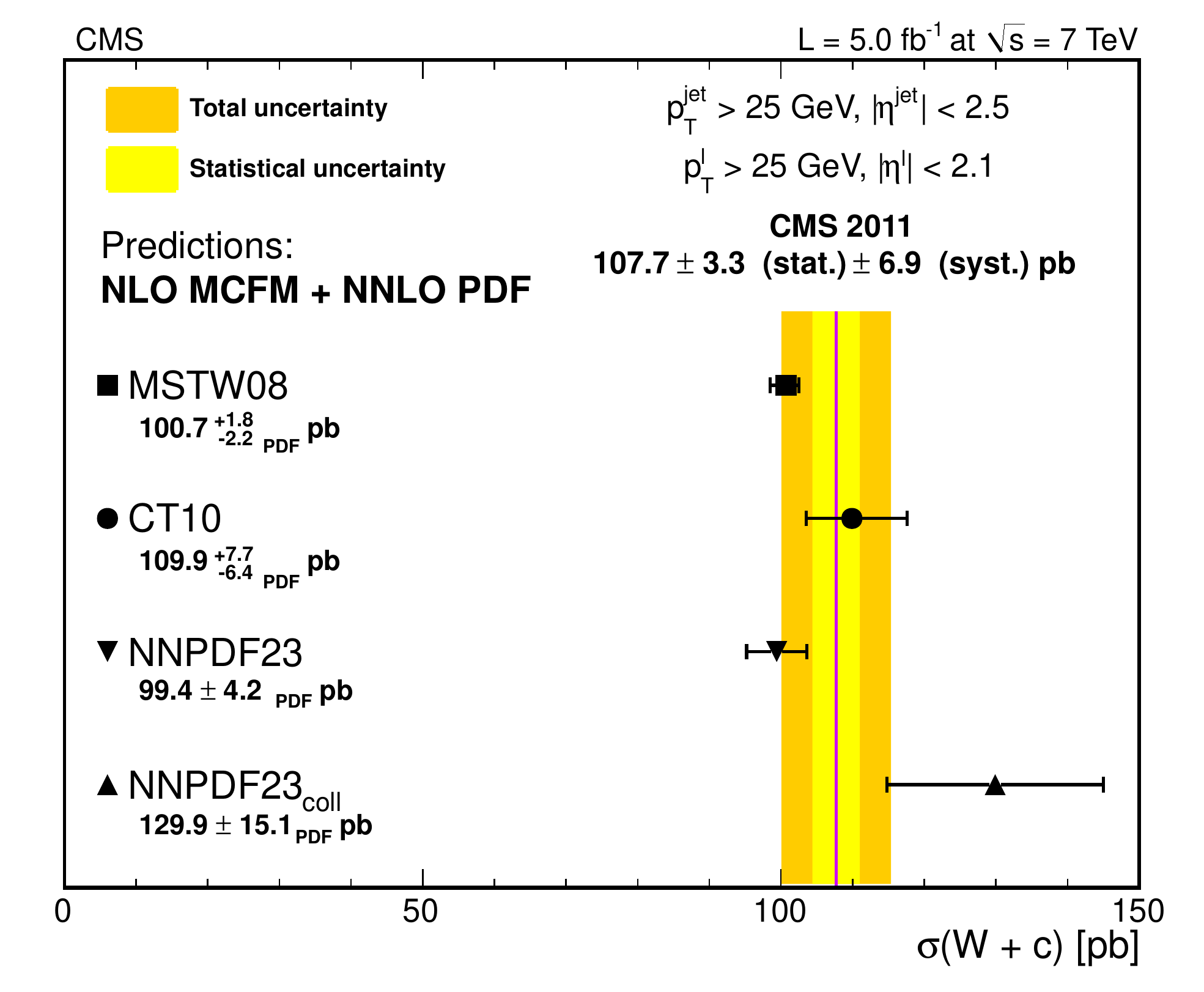} \\
    \includegraphics[width=0.45\textwidth]{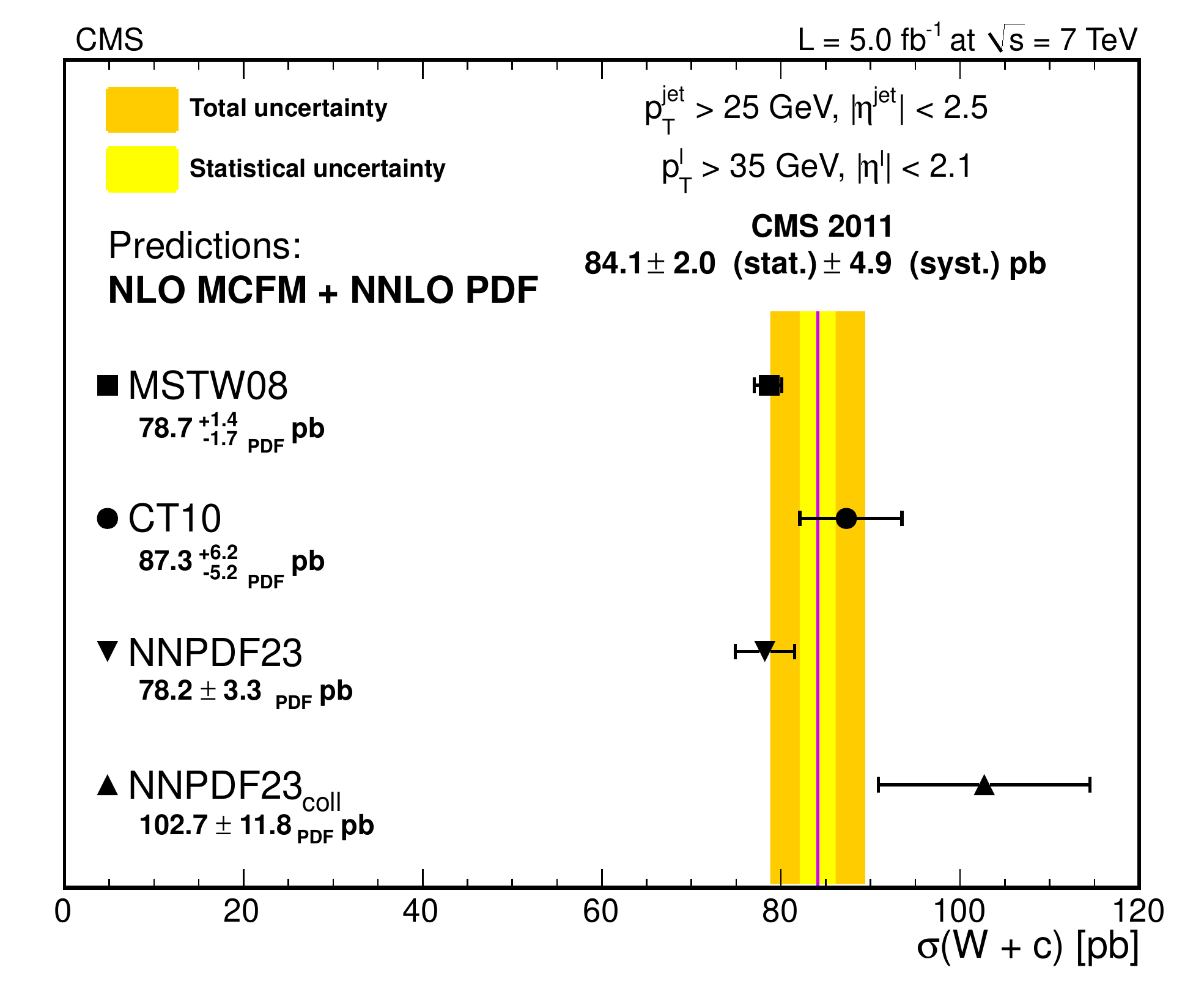}
    \caption{Comparison of the theoretical predictions for $\SWc$ computed with \MCFM
and several sets of PDFs with the average of the experimental measurements.
The top plot shows the predictions for a $\pt$ threshold of the lepton from the $\PW$-boson decay of $\pt^\ell>25\GeV$ and the bottom plot presents the
predictions for $\pt^\ell > 35\GeV$.
The uncertainty associated with scale variations is ${\pm}5\%.$
}
\label{fig:theo_Sc}
\end{center}
\end{figure}

These predictions are compared graphically to the experimental
measurement in Fig.~\ref{fig:theo_Sc}. Only PDF uncertainties are shown.
Scale uncertainties
in the total cross section are of the order of ${\pm}5\%$.
From Fig.~\ref{fig:theo_Sc} we see that measured $\Wc$ cross sections agree with the theoretical predictions using the
PDF sets introduced above within theoretical and experimental uncertainties.
The total cross sections for ABM11, JR09, and HERAPDF1.5
are respectively 98.9\unit{pb} (78.0\unit{pb}), 80.0\unit{pb} (63.4\unit{pb}) and 96.9\unit{pb} (76.7\unit{pb}) for a lepton $\pt$ threshold of $25~(35)\GeV$. As
discussed in~\cite{Ball:2012wy}, the strangeness in ABM11 and HERAPDF1.5 is close to that of MSTW and NNPDF, hence the
similarities in the predictions.

\subsection{Differential cross section~\label{sec:theo_diff}}

Predictions for the differential (both absolute and normalized) cross sections are obtained from analytical calculations from  \MCFM
using the same binning as in the data analysis:
{[0, 0.35],~[0.35, 0.7],~[0.7, 1.1],~[1.1, 1.6],~[1.6, 2.1]}.
Table~\ref{tab:MCFM_diff} presents the predictions for $\SWcdiffline$.
The differences among the central value of the predictions obtained with the various PDF sets are of the same order as
the associated uncertainties (at 68\% confidence level, CL).
As in the case of the inclusive cross section, the different size of the associated uncertainties arises from the different assumptions of
PDF groups about the strange quark and antiquark content
of the proton and from the different experimental inputs included~\cite{Charm_Eleni}.
As expected, PDF uncertainties increase at forward pseudorapidities, where the
range of Bjorken $x$  is outside that covered by available data
sensitive to strangeness.
Systematic uncertainties due to the scale variations are smaller than $1\%$ for all muon pseudorapidity bins.

\begin{table}[htbp]\renewcommand{\arraystretch}{1.2}
\begin{center}
 \topcaption{The $\SWcdiffline$ theoretical predictions calculated with \MCFM at NLO.
 Kinematic selection follows the experimental requirements: $\pt^\text{jet}>25\GeV$, $\abs{\eta^\text{jet}}<2.5$, and $\abs{\eta^\ell}<2.1$.
Partons are joined using an anti-\kt algorithm with a distance parameter of $1$.
Predictions for $\Wln$ when the transverse momentum of the lepton from the $\PW$ boson is larger than $25\GeV$ are given in the first block of the table.
The second block of predictions are for $\Wln$ production with $\pt^{\ell}>35\GeV$.
For every PDF set, the central value of the prediction is given, together with the
relative uncertainty as prescribed from the PDF set.
The uncertainty associated with scale variations is smaller than $1\%.$}
\label{tab:MCFM_diff}
\begin{tabular}{ccccccccc}
\hline\hline
 & \multicolumn{8}{c}{$\pt^\ell > 25\GeV$} \\ \hline
 & \multicolumn{2}{c}{MSTW08} & \multicolumn{2}{c}{CT10} &\multicolumn{2}{c}{NNPDF2.3} &\multicolumn{2}{c}{NNPDF2.3$_\text{coll}$} \\ \cline{2-9}
$[{\abs{\eta}}_\text{min},{\abs{\eta}}_\text{max}]$ & $\frac{1}{\sigma} \frac{\rd\sigma}{\rd\abs{\eta}}$ & $\Delta_{\mathrm{PDF}} [\%]$ & $\frac{1}{\sigma} \frac{\rd\sigma}{\rd\abs{\eta}}$ & $\Delta_{\mathrm{PDF}} [\%]$ &
                               $\frac{1}{\sigma} \frac{\rd\sigma}{\rd\abs{\eta}}$ & $\Delta_{\mathrm{PDF}} [\%]$ & $\frac{1}{\sigma} \frac{\rd\sigma}{\rd\abs{\eta}}$ & $\Delta_{\mathrm{PDF}} [\%]$ \\
\hline 
$[0, 0.35]$   & $0.596$ & $^{+0.5}_{-0.5}$ & $0.605$ & $^{+1.3}_{-2.3}$ & $0.612$ & $1.1$ & $0.569$ & $5.5$ \\
$[0.35, 0.7]$ & $0.566$ & $^{+0.4}_{-0.4}$ & $0.576$ & $^{+1.0}_{-1.8}$ & $0.590$ & $0.9$ & $0.556$ & $4.4$ \\
$[0.7, 1.1]$  & $0.518$ & $^{+0.2}_{-0.2}$ & $0.527$ & $^{+0.4}_{-0.7}$ & $0.521$ & $0.4$ & $0.513$ & $1.9$ \\
$[1.1, 1.6]$  & $0.446$ & $^{+0.3}_{-0.3}$ & $0.436$ & $^{+1.3}_{-0.8}$ & $0.429$ & $0.7$ & $0.448$ & $2.8$ \\
$[1.6, 2.1]$  & $0.327$ & $^{+0.9}_{-1.0}$ & $0.316$ & $^{+4.4}_{-2.4}$ & $0.314$ & $2.1$ & $0.354$ & $9.6$ \\
\hline\hline
& \multicolumn{8}{c}{$\pt^\ell > 35\GeV$} \\ \hline
& \multicolumn{2}{c}{MSTW08} & \multicolumn{2}{c}{CT10} &\multicolumn{2}{c}{NNPDF2.3} &\multicolumn{2}{c}{NNPDF2.3$_\text{coll}$} \\ \cline{2-9}
$[{\abs{\eta}}_\text{min},{\abs{\eta}}_\text{max}]$ & $\frac{1}{\sigma} \frac{\rd\sigma}{\rd\abs{\eta}}$ & $\Delta_{\mathrm{PDF}} [\%]$ & $\frac{1}{\sigma} \frac{\rd\sigma}{\rd\abs{\eta}}$ & $\Delta_{\mathrm{PDF}} [\%]$ &
                               $\frac{1}{\sigma} \frac{\rd\sigma}{\rd\abs{\eta}}$ & $\Delta_{\mathrm{PDF}} [\%]$ & $\frac{1}{\sigma} \frac{\rd\sigma}{\rd\abs{\eta}}$ & $\Delta_{\mathrm{PDF}} [\%]$ \\
\hline
$[0, 0.35]$   & $0.607$ & $^{+0.6}_{-0.5}$ & $0.615$ & $^{+1.4}_{-2.4}$ & $0.618$ & $1.2$ & $0.580$ & $5.0$  \\
$[0.35, 0.7]$ & $0.582$ & $^{+0.5}_{-0.4}$ & $0.588$ & $^{+1.0}_{-1.9}$ & $0.587$ & $0.9$ & $0.568$ & $3.8$  \\
$[0.7, 1.1]$  & $0.529$ & $^{+0.2}_{-0.2}$ & $0.532$ & $^{+0.4}_{-0.7}$ & $0.527$ & $0.4$ & $0.512$ & $2.5$  \\
$[1.1, 1.6]$  & $0.431$ & $^{+0.3}_{-0.3}$ & $0.428$ & $^{+1.5}_{-0.9}$ & $0.436$ & $0.8$ & $0.438$ & $1.4$  \\
$[1.6, 2.1]$  & $0.314$ & $^{+1.0}_{-1.2}$ & $0.304$ & $^{+4.9}_{-2.6}$ & $0.299$ & $2.3$ & $0.349$ & $11.4$  \\
\hline\hline
\end{tabular}
\end{center}
\end{table}

\begin{figure}[htbp]
\begin{center}
     \includegraphics[width=0.45\textwidth]{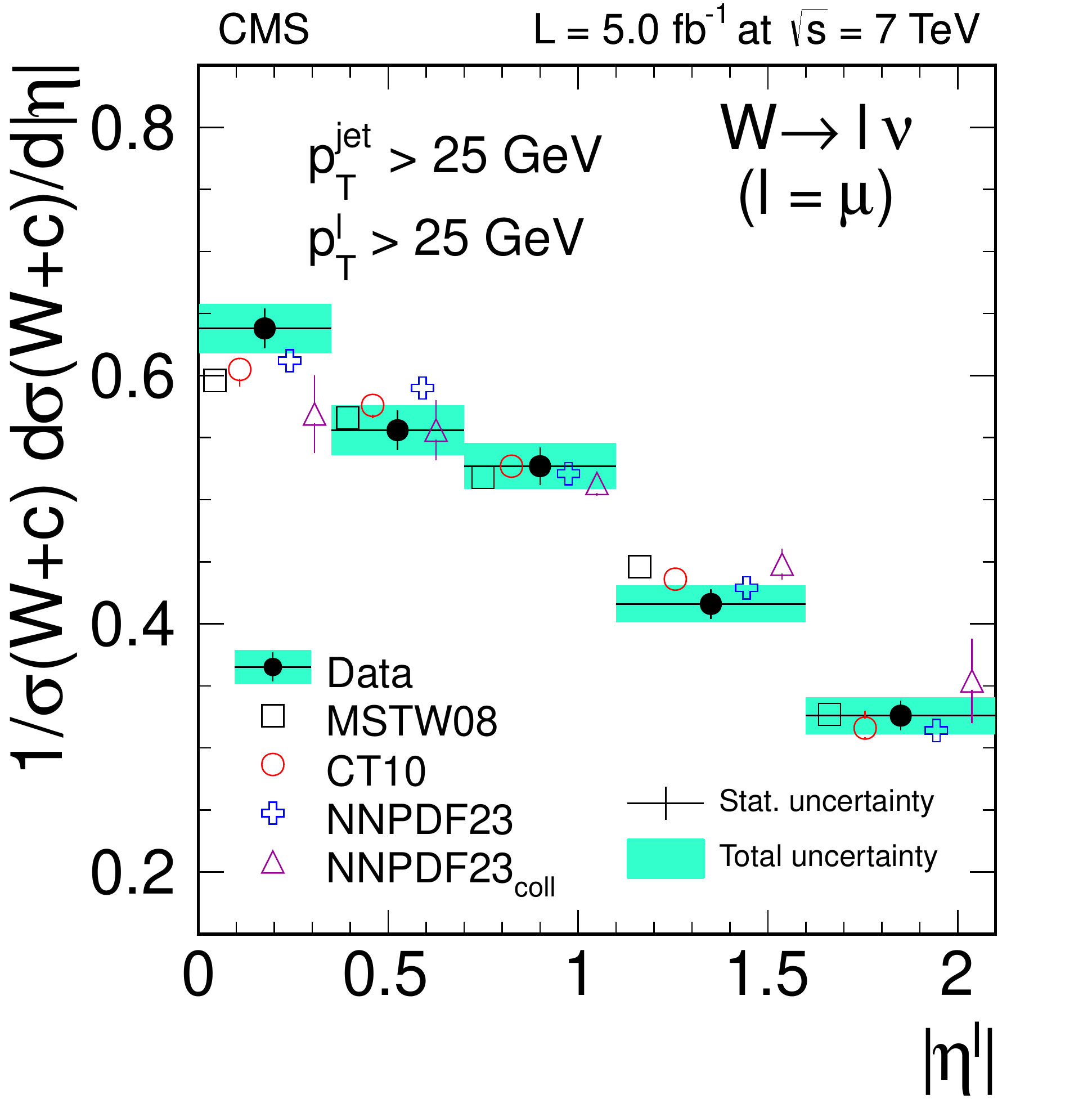}
     \includegraphics[width=0.45\textwidth]{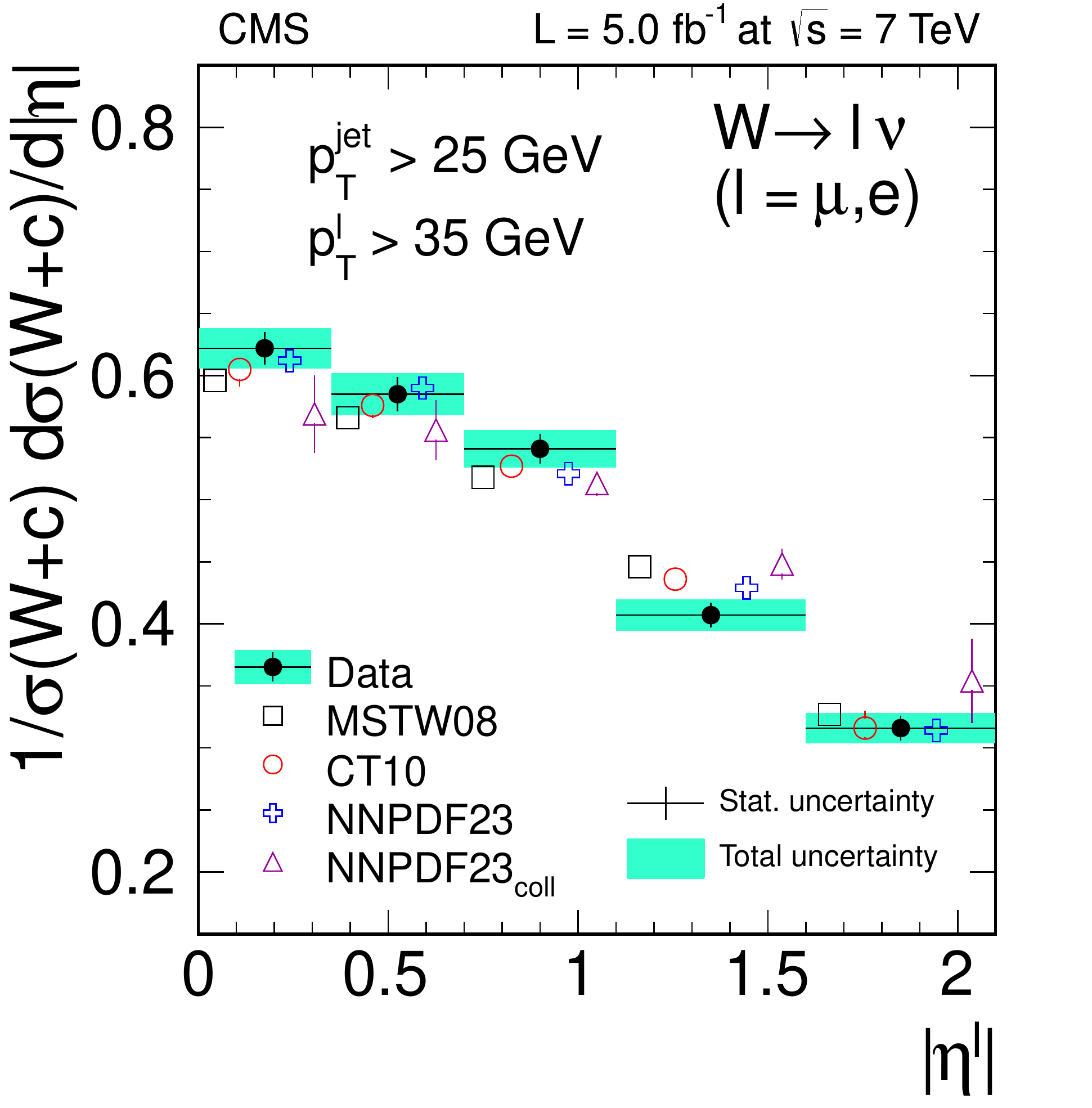}
    \caption{Normalized differential cross section, $\SWcdiffline$, as a function of the absolute value of the pseudorapidity of the lepton from the $\PW$ boson decay, compared
with the theoretical predictions.
Theoretical predictions at NLO are computed with \MCFM using four different PDF sets.
Kinematic selection follows the experimental requirements: $\pt^\text{jet}>25\GeV$, $\abs{\eta^\text{jet}}<2.5$, and $\abs{\eta^\ell}<2.1$.
The transverse momentum of the lepton is larger than $25\GeV$ in the left plot and larger than $35\GeV$ in the right plot.
The data points are the average of the results presented before with the three different samples:
inclusive three- and two-prong and semileptonic samples.
In the right plot the results obtained with the $\PWmn$ samples and $\Wen$ samples are combined.
Symbols showing the theoretical expectations are slightly displaced in the horizontal axis
for better visibility of the predictions.
The uncertainty associated with scale variations is smaller than $1\%.$
}
    \label{fig:Sc_w_th}
\end{center}
\end{figure}

\begin{figure}[htbp]
\begin{center}
      \includegraphics[width=0.45\textwidth]{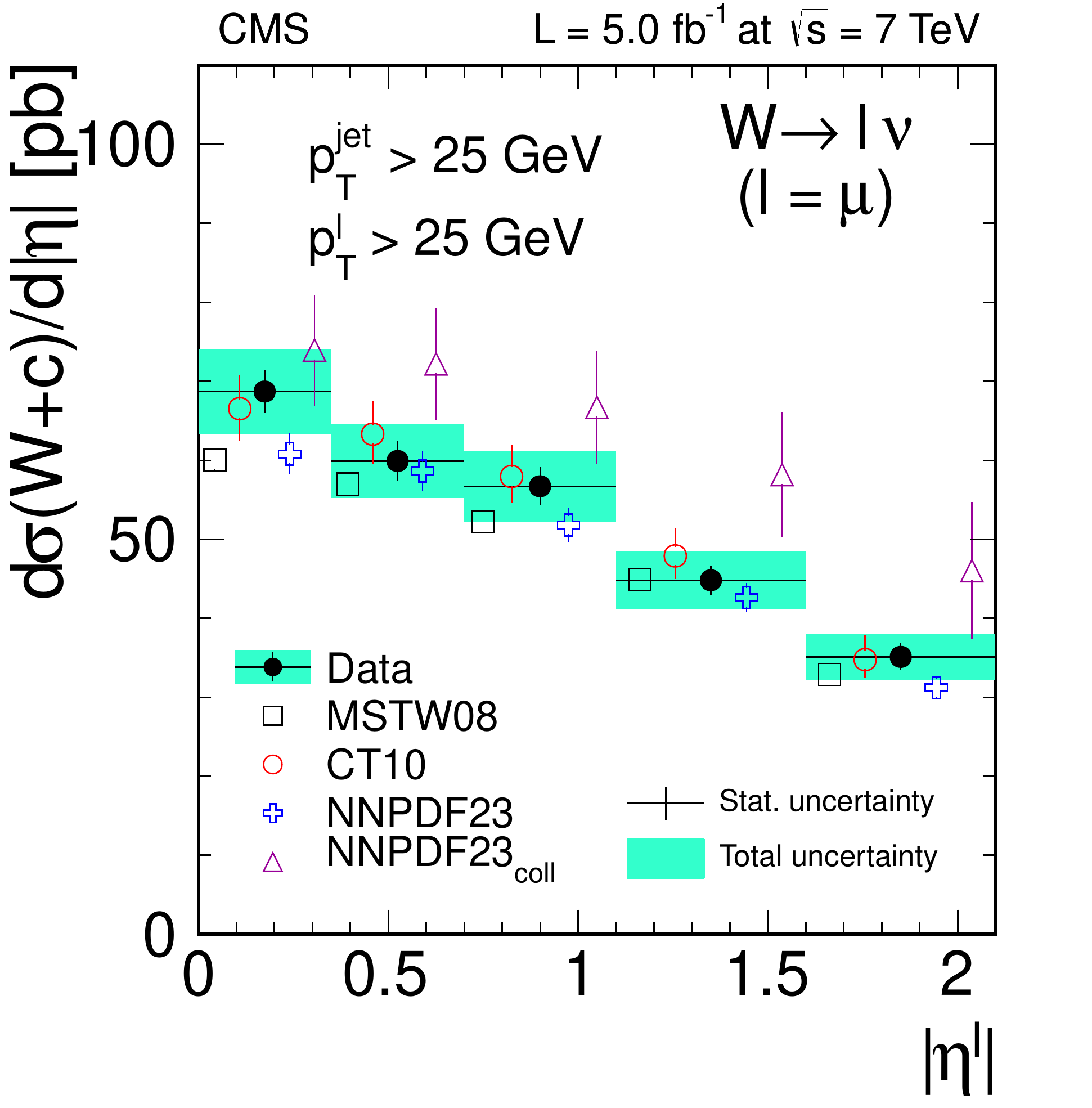}
      \includegraphics[width=0.45\textwidth]{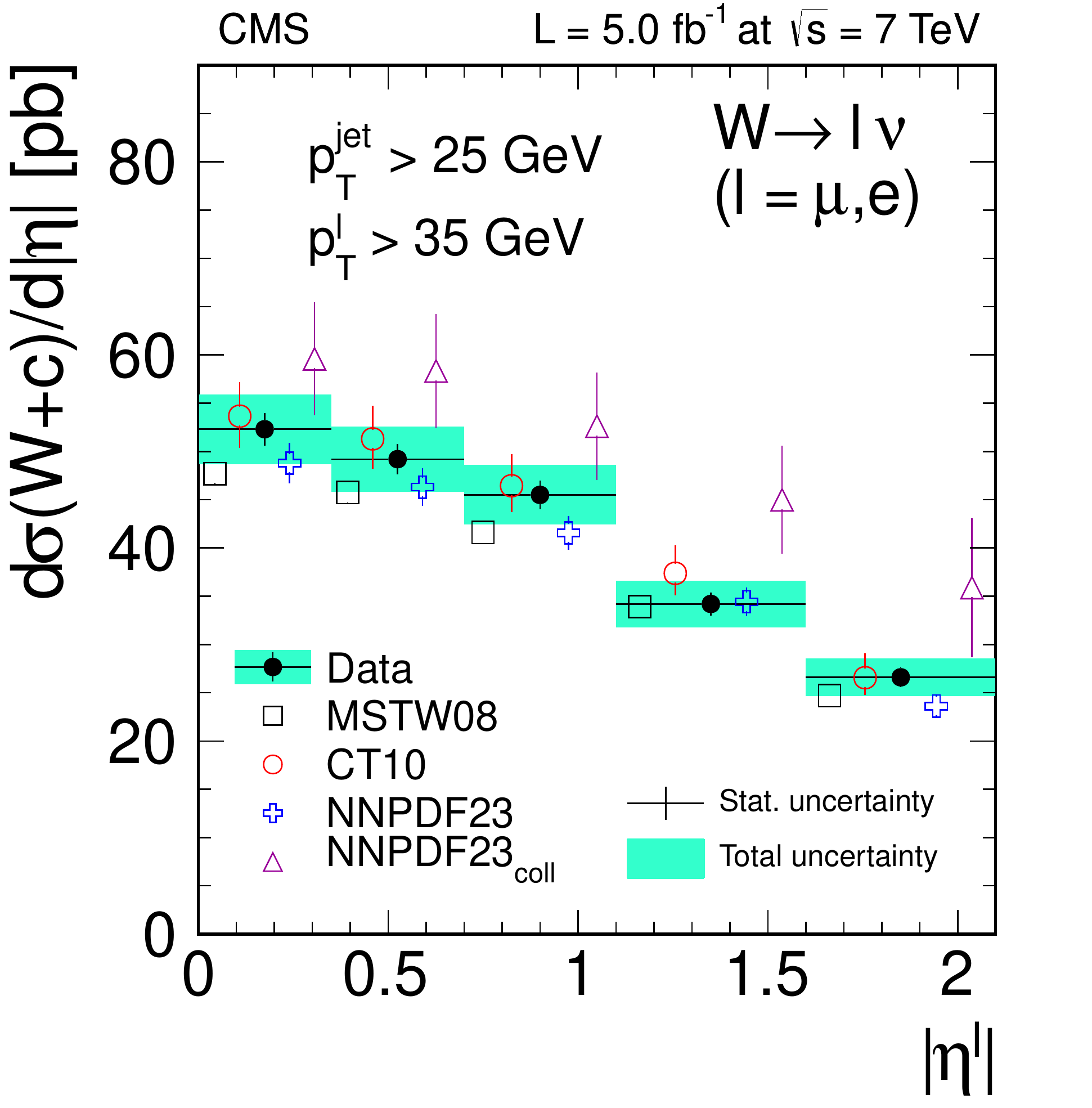}\\
    \caption{Differential cross section, $\SWcdifftline$, as a function of the absolute value of the pseudorapidity of the lepton from the $\PW$-boson decay, compared with
the theoretical predictions.
Theoretical predictions at NLO are computed with \MCFM and four different PDF sets.
Kinematic selection follows the experimental requirements: $\pt^\text{jet}>25\GeV$, $\abs{\eta^\text{jet}}<2.5$, and $\abs{\eta^\ell}<2.1$.
The transverse momentum of the lepton is larger than $25\GeV$ in the left plot and larger than $35\GeV$ in the right plot.
The data points are the average of the results from the
inclusive three- and two-prong and semileptonic samples.
In the right plot the results achieved with the $\PWmn$ samples and $\Wen$ samples are combined.
Symbols showing the theoretical expectations are slightly displaced in the horizontal axis
for better visibility of the predictions.
}
    \label{fig:Sctot_w_th}
\end{center}
\end{figure}

The theoretical predictions are compared with the average of the experimental measurements presented in Section~\ref{sec:differential_xec}.
Figure~\ref{fig:Sc_w_th} (Fig.~\ref{fig:Sctot_w_th}) compares the measurements and predictions for the normalized cross sections (absolute cross sections).
There is agreement between the measured distributions and the theoretical predictions.
We note that a comparison among the several predictions in Figs.~\ref{fig:Sc_w_th} and~\ref{fig:Sctot_w_th} may lead to different conclusions.
For instance, NNPDF2.3$_\text{coll}$ gives the smallest prediction in the first rapidity bin in Fig.~\ref{fig:Sc_w_th}, whereas it gives the highest value in
Fig.~\ref{fig:Sctot_w_th}. The normalized differential cross sections probe the shape of the strange-quark PDF whereas
the behaviour of the absolute differential cross sections is also driven by the overall magnitude of the strange-quark PDF.

\subsection{Charged cross section ratio~\label{sec:theo_ratio}}

Theoretical predictions for $\SWpc$ and $\SWmc$ production are computed independently under the same conditions explained before and for
the same lepton pseudorapidity intervals used in the analysis. Expectations for the cross section ratio $\SWpc/\SWmc$ are derived from them
and are presented in Table~\ref{tab:MCFM_Rcpm}.
The last row in each block of predictions gives the prediction of the charged cross section ratio for the full lepton pseudorapidity
interval, $\abs{\eta^\ell}<2.1$.
We note that this ratio is sensitive to the strangeness asymmetry in the proton,
but also to the down quark and antiquark asymmetry from the Cabibbo-suppressed process
$\Pg\cPqd\to \PW^- \cPqc$
($\Pg\cPaqd\to \PW^+\cPaqc$).
The $\cPqd$--$\cPaqd$ asymmetry is larger in absolute value than
the difference between strange quarks and antiquarks.

\begin{table}[htbp]\renewcommand{\arraystretch}{1.2}
\begin{center}
 \caption{
 Theoretical predictions for $\Rcpm(\eta^\ell) \equiv \SWpc(\abs{\eta^\ell})/\SWmc(\abs{\eta^\ell})$ calculated with \MCFM at NLO.
 Kinematic selection follows the experimental requirements: $\pt^\text{jet}>25\GeV$, $\abs{\eta^\text{jet}}<2.5$, and $\abs{\eta^\ell}<2.1$.
Partons are joined using an anti-\kt algorithm with a distance parameter of $1$.
Predictions for $\Wln$ when the transverse momentum of the lepton from the $\PW$ boson is larger than $25\GeV$ are given in the first block of the table.
The second block of predictions are for $\Wln$ production with $\pt^{\ell}>35\GeV$.
For each PDF set, the central value of the prediction is given, together with the
relative uncertainty as prescribed from the PDF set.
The uncertainty associated with scale variations are of the order of 1--2\%.}
\label{tab:MCFM_Rcpm}
\begin{tabular}{ccccccccc}
\hline\hline
& \multicolumn{8}{c}{$\pt^\ell > 25\GeV$} \\ \hline
& \multicolumn{2}{c}{MSTW08} & \multicolumn{2}{c}{CT10} &\multicolumn{2}{c}{NNPDF2.3} &\multicolumn{2}{c}{NNPDF2.3$_\text{coll}$} \\ \cline{2-9}
$[{\abs{\eta}}_\text{min},{\abs{\eta}}_\text{max}]$ & $\Rcpm$ & $\Delta_{\mathrm{PDF}} [\%]$ & $\Rcpm$ & $\Delta_{\mathrm{PDF}} [\%]$ &
                                 $\Rcpm$ & $\Delta_{\mathrm{PDF}} [\%]$ & $\Rcpm$ & $\Delta_{\mathrm{PDF}} [\%]$ \\
\hline
$[0, 0.35]$   & $0.944$ & $^{+1.2}_{-3.6}$ & $0.968$ & $^{+0.2}_{-0.2}$ & $0.993$ & $0.8$ & $0.959$ & $1.4$ \\
$[0.35, 0.7]$ & $0.941$ & $^{+1.3}_{-3.5}$ & $0.965$ & $^{+0.2}_{-0.2}$ & $0.920$ & $1.0$ & $0.975$ & $1.5$ \\
$[0.7, 1.1]$  & $0.918$ & $^{+1.7}_{-3.1}$ & $0.959$ & $^{+0.3}_{-0.3}$ & $0.949$ & $1.3$ & $0.948$ & $1.8$ \\
$[1.1, 1.6]$  & $0.871$ & $^{+2.4}_{-2.7}$ & $0.951$ & $^{+0.6}_{-0.5}$ & $0.893$ & $2.0$ & $0.913$ & $2.6$ \\
$[1.6, 2.1]$  & $0.854$ & $^{+3.1}_{-3.4}$ & $0.889$ & $^{+1.2}_{-0.9}$ & $0.842$ & $3.5$ & $0.893$ & $5.1$ \\ \hline
$[0, 2.1]$    & $0.906$ & $^{+1.9}_{-2.8}$ & $0.949$ & $^{+0.4}_{-0.4}$ & $0.922$ & $1.5$ & $0.937$ & $2.0$ \\
\hline\hline
& \multicolumn{8}{c}{$\pt^\ell > 35\GeV$} \\ \hline
& \multicolumn{2}{c}{MSTW08} & \multicolumn{2}{c}{CT10} &\multicolumn{2}{c}{NNPDF2.3} &\multicolumn{2}{c}{NNPDF2.3$_\text{coll}$} \\ \cline{2-9}
  $[|\eta_\text{min}|,|\eta_\text{max}|]$ & $\Rcpm$ & $\Delta_{\mathrm{PDF}} [\%]$ & $\Rcpm$ & $\Delta_{\mathrm{PDF}} [\%]$ &
                                  $\Rcpm$ & $\Delta_{\mathrm{PDF}} [\%]$ & $\Rcpm$ & $\Delta_{\mathrm{PDF}} [\%]$ \\
\hline
$[0, 0.35]$   & $0.949$ & $^{+1.2}_{-3.7}$ & $0.974$ & $^{+0.2}_{-0.2}$ & $0.972$ & $0.9$ & $1.009$ & $1.5$ \\
$[0.35, 0.7]$ & $0.932$ & $^{+1.4}_{-3.5}$ & $0.964$ & $^{+0.3}_{-0.3}$ & $0.957$ & $1.0$ & $0.984$ & $1.6$ \\
$[0.7, 1.1]$  & $0.902$ & $^{+1.8}_{-3.2}$ & $0.953$ & $^{+0.4}_{-0.3}$ & $0.953$ & $1.4$ & $0.927$ & $3.3$ \\
$[1.1, 1.6]$  & $0.882$ & $^{+2.5}_{-2.7}$ & $0.918$ & $^{+0.6}_{-0.5}$ & $0.909$ & $2.2$ & $0.886$ & $5.1$ \\
$[1.6, 2.1]$  & $0.845$ & $^{+3.4}_{-3.8}$ & $0.888$ & $^{+1.2}_{-1.0}$ & $0.831$ & $3.8$ & $0.877$ & $5.9$ \\ \hline
$[0, 2.1]$    & $0.904$ & $^{+2.0}_{-2.7}$ & $0.942$ & $^{+0.4}_{-0.4}$ & $0.923$ & $1.6$ & $0.936$ & $2.4$ \\
\hline \hline
\end{tabular}
\end{center}
\end{table}

Both the central values and the associated PDF uncertainties are quite different for the various sets of predictions.
These differences arise from the assumptions  underlying each global fit.
For instance, the CT10 set assumes equal content of strange quark and antiquark in the proton,
leading to a charged cross section ratio almost exclusively driven by the $\cPqd$--$\cPaqd$ asymmetry and with a very small PDF uncertainty in the
prediction.
On the other hand, both MSTW08 and NNPDF2.3 provide independent parametrizations of the strangeness asymmetry,
thus resulting in larger PDF uncertainties.
The MSTW08 and NNPDF2.3 predicted values for the $\SWpc/\SWmc$ ratio in the full pseudorapidity region are smaller than in the CT10 case.
As before, PDF uncertainties increase for large values of the lepton pseudorapidity.
Systematic uncertainties in the cross section ratio due to the scale variations are smaller than $1\%$ for the full lepton absolute pseudorapidity
range $[0., 2.1]$ and of the order of 1--2\% for the smaller pseudorapidity bins of the differential measurement.

Differences among the predictions are relatively large for some of the lepton pseudorapidity bins, ${\sim} 4$--$5\%$,
although this difference is covered by one standard deviation of the PDF uncertainties.
All PDF sets predict the decrease of the charged ratio with the absolute value of the lepton pseudorapidity as a consequence of the higher
\cPqd--\cPaqd\ asymmetry at large values of Bjorken $x$. The decrease with $\abs{\eta^\ell}$ is more pronounced in the case of NNPDF2.3.

\begin{figure}[htbp]
\begin{center}
    \includegraphics[height=0.5\textwidth,width=0.65\textwidth]{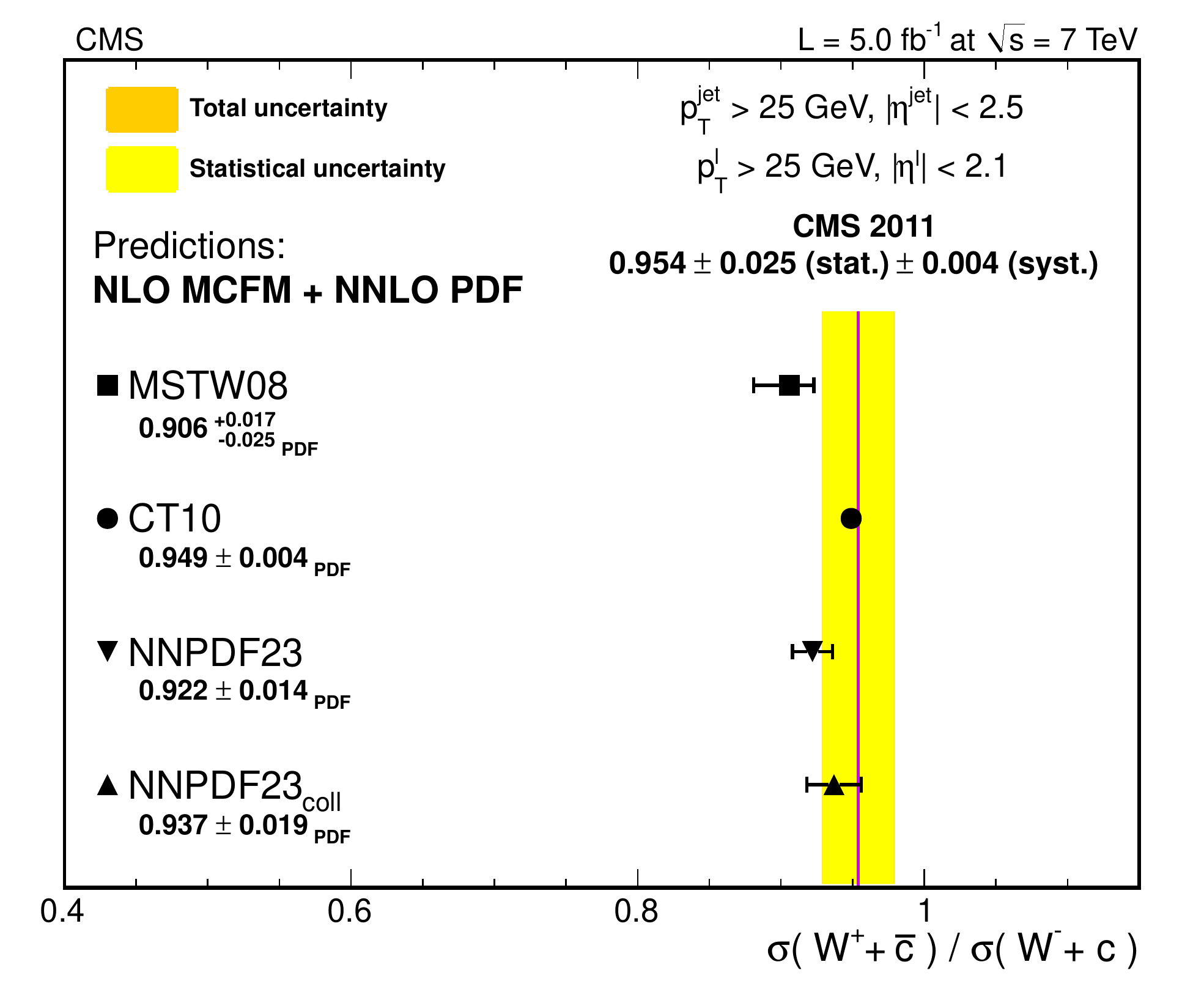} \\
    \includegraphics[height=0.5\textwidth,width=0.65\textwidth]{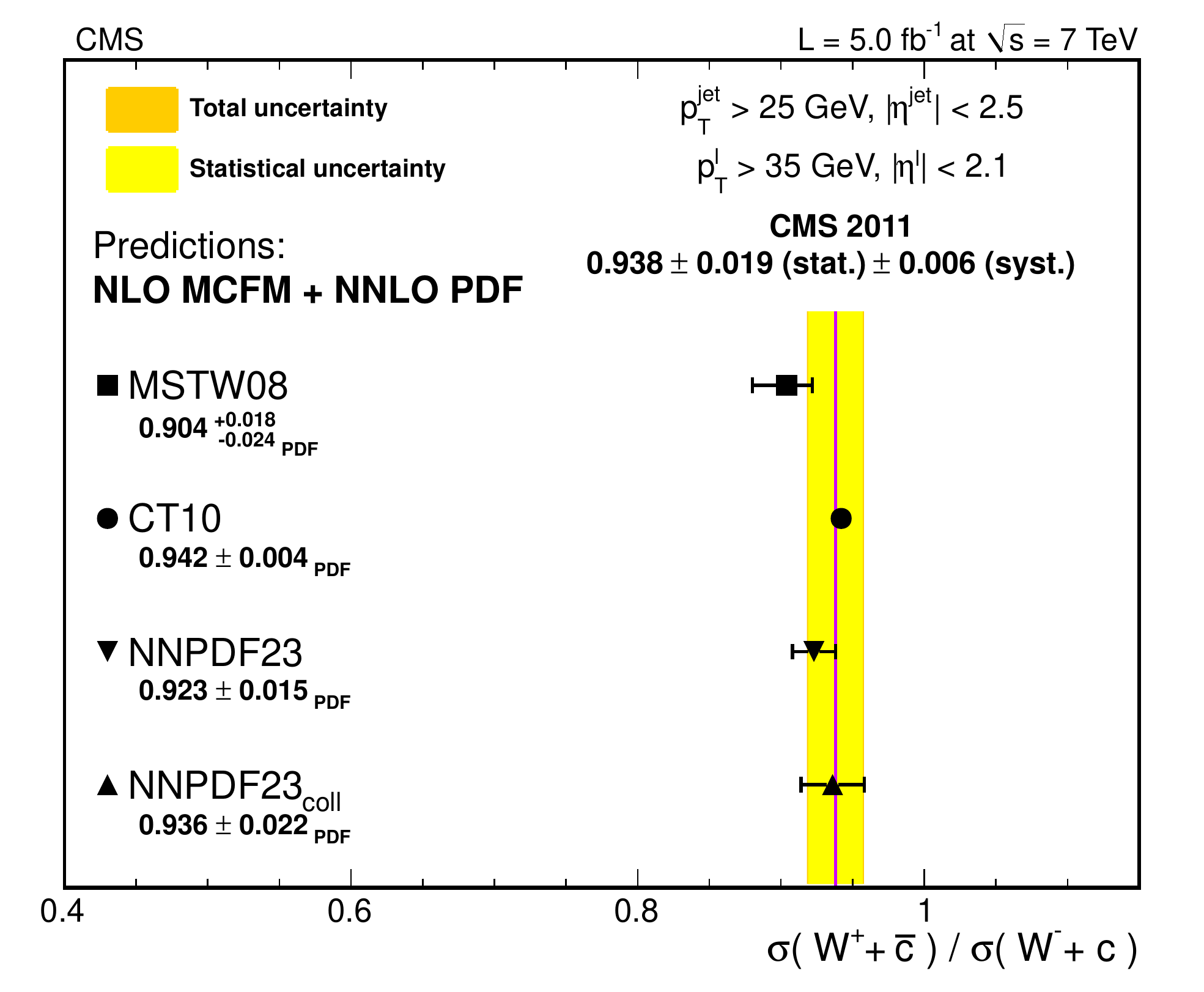}
    \caption{Comparison of the theoretical predictions for $\SWpc/\SWmc$ computed with \MCFM
and several PDF sets with the average of the experimental measurements.
The top plot compares the average of the measurements made in the muon channel for a $\pt$ threshold of the lepton from the $\PW$-boson decay of $\pt^\ell>25\GeV$. The bottom plot presents the average of the measurements in the muon and electron channel with the predictions for $\pt^\ell > 35\GeV$.
The uncertainty associated with scale variations is $\pm 1\%$.
}
\label{fig:Rpm_inc_w_th}
\end{center}
\end{figure}

\begin{figure}[htbp]
\begin{center}
     \includegraphics[width=0.45\textwidth]{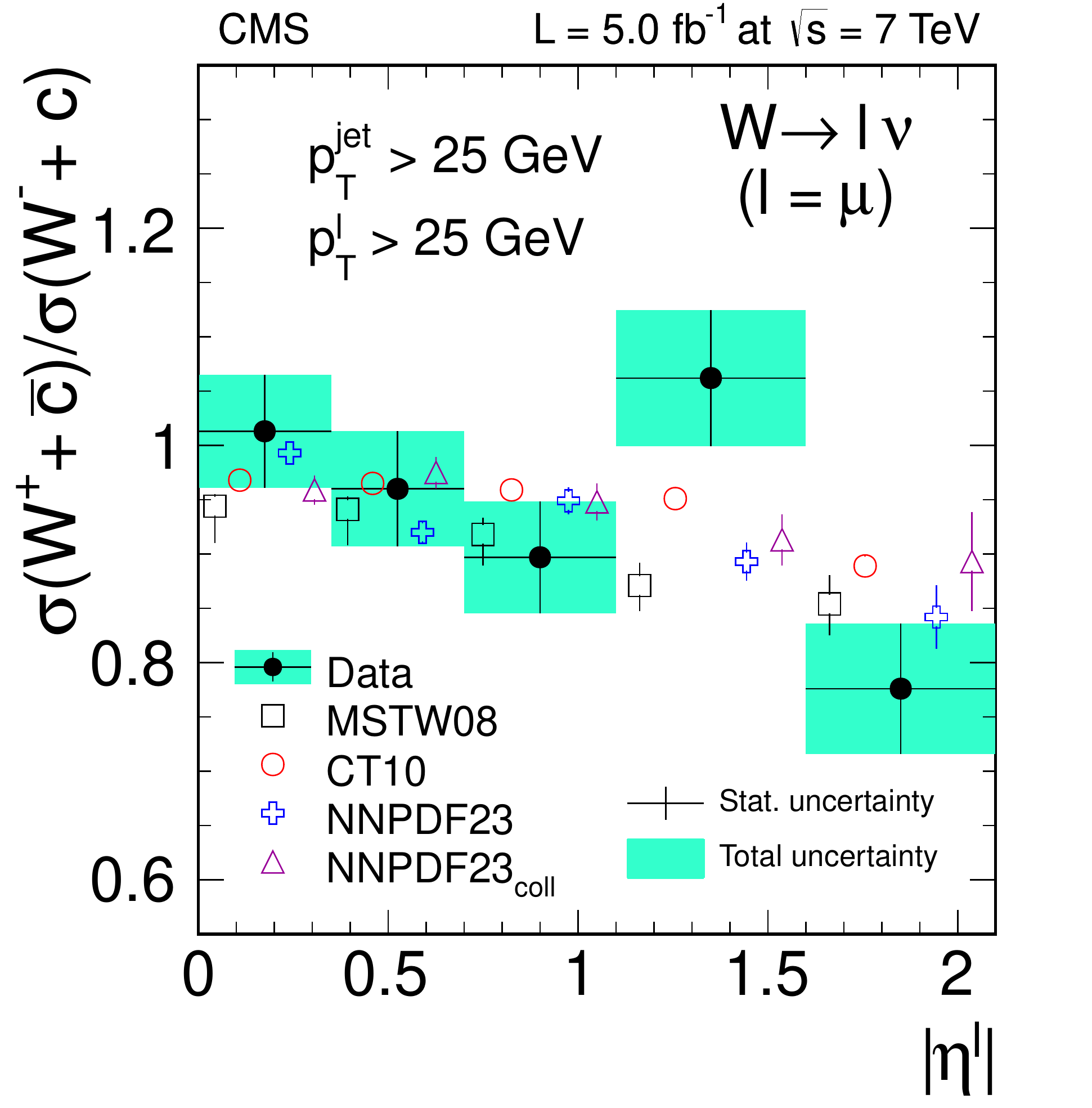}
     \includegraphics[width=0.45\textwidth]{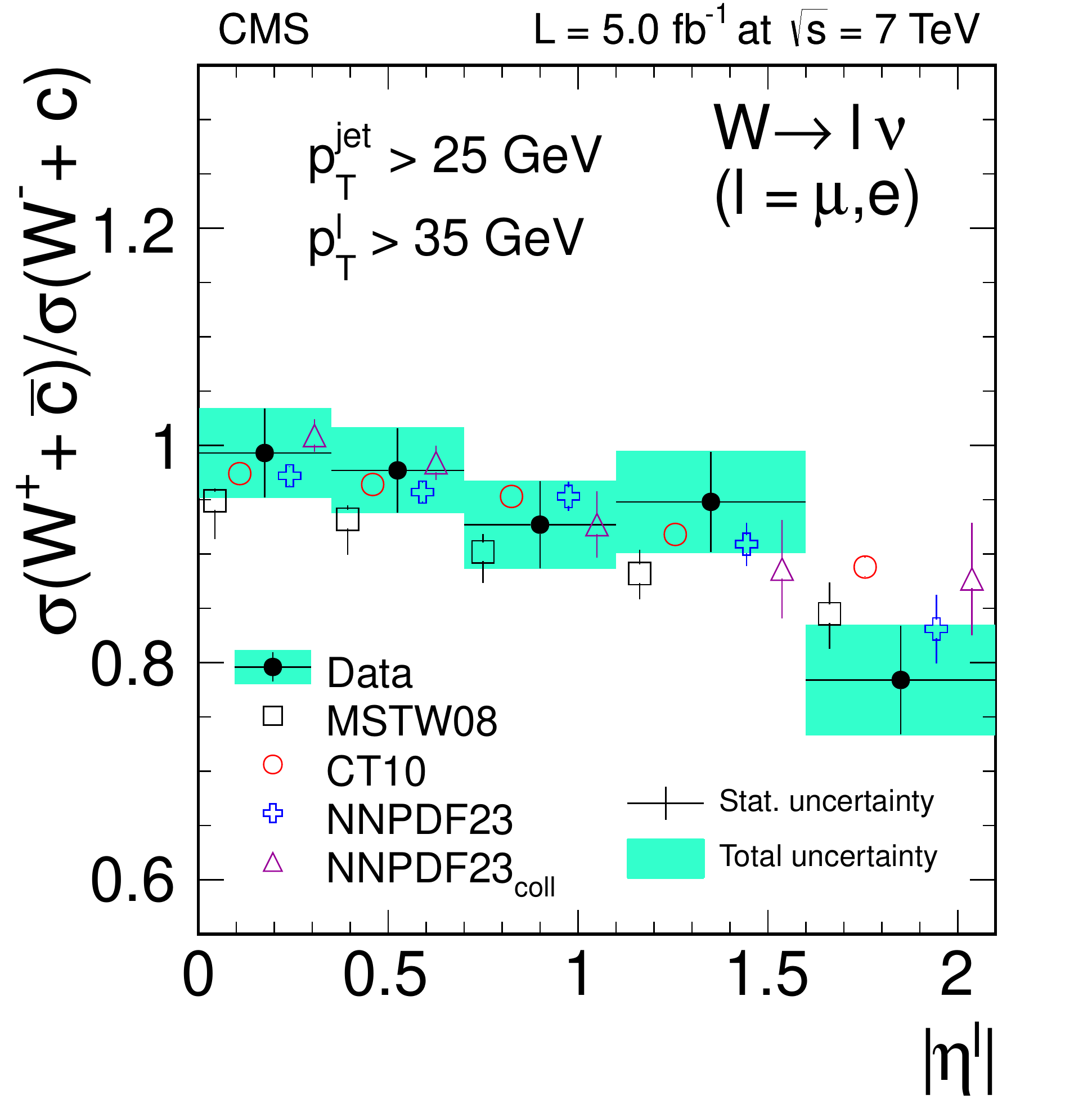}
    \caption{Cross section ratio, $\SWpc/\SWmc$, as a function of the absolute value of the pseudorapidity of the lepton from the $\PW$-boson decay.
Results for the $\pt^\ell > 25\GeV$ case are shown in the left plot (muon channel only). In the right plot, the transverse momentum of the lepton is larger than $35\GeV$.
The data points are the average of the results from the
inclusive three- and two-prong and semileptonic samples.
In the right plot the results obtained with the $\PWmn$ samples and $\Wen$ samples are combined.
Theoretical predictions at NLO computed with \MCFM and four different PDF sets are also shown.
The uncertainty associated with scale variations are of the order of 1--2\%.
Symbols showing the theoretical expectations are slightly displaced in the horizontal axis
for better visibility of the predictions.
}
    \label{fig:Rpm_w_th}
\end{center}
\end{figure}

Averaged cross section ratios obtained in Section~\ref{sec:ratio} are compared with theoretical predictions.
Figure~\ref{fig:Rpm_inc_w_th} shows the measurements and the predictions for the total cross section ratios and
Fig.~\ref{fig:Rpm_w_th} shows the cross section ratios as a function of the absolute value of the lepton pseudorapidity.

The theoretical predictions based on the CT10 PDF set agree with the measured cross section ratios. Predictions from NNPDF23 and NNPDF23$_\text{coll}$ are well within
the uncertainty of the measurements, whereas expectations using MSTW08 lie about 1.5 sigma below the measurements.
For the cross section ratio as a function of the absolute value of the lepton pseudorapidity, there is agreement between the measurements and the theoretical predictions, especially when
the transverse momentum of the lepton from the $\PW$-boson decay is larger than $35\GeV$.

\section{Summary and conclusions~\label{sec:summary}}

The associated production of a $\PW$ boson with a charm-quark jet in pp collisions at $\sqrt s = 7\TeV$ is experimentally established
for the first time, using a
data sample collected by the CMS experiment during the 2011 LHC run with an integrated luminosity of $5\fbinv$.
The signature of $\PW$-boson production together with a charm-quark jet is observed by identifying the leptonic decay of the $\PW$ boson
into a muon or an electron and a neutrino and the reconstruction of exclusive and inclusive final states from the decay of charm hadrons.
In total, distinct $\Wc$ signals are observed independently in six different final states.

The high performance of the CMS tracking detector and the algorithms devised for secondary-vertex reconstruction
allow the efficient selection of candidate samples with a displaced secondary vertex having three or two tracks corresponding
to the decay products of charm mesons. Clear signals of $\Dpm$ mesons are observed through the reconstruction of the decay mode
$\Dpm\to \PK^\mp \pi^\pm\pi^\pm$ in events with three-track secondary vertices and
from $\Dz$ production in the decay chain $\Dstar\to\Dz \pi^{\pm}$ with the subsequent decay $\Dz\to \PK^\mp\pi^\pm$ in events with two-track secondary vertices.
In addition, efficient muon identification among the particles constituting the jet leads to an independent $\Wc$ sample with an
identified muon from the semileptonic decay of the charm quark.

The analysis exploits the intrinsic charge correlation in $\Wc$ production between the charge of the $\PW$ boson and the charge of the c quark, which are always of opposite sign.
The $\PW$-boson decay into a well-identified charged lepton and the final-state mesons allow us to determine unequivocally the
signs of both the $\PW$ boson and the charm-quark jet candidates. Independent opposite-sign and same-sign samples of events are hence defined.
The background contributions from processes that are charge symmetric are subtracted in an essentially model-independent way through a same-sign sample
subtraction from the opposite-sign sample in the relevant variables used in the analysis.

The high purity of the resulting samples allows us to perform various measurements in an almost background-free environment. The sample of candidate
events from the semileptonic decay of charm mesons is affected by a larger background, mainly in the $\PWmn$ channel, but it provides
a larger statistical power so that the final precision attained in the measurements in the three charm meson final states is similar.
Furthermore, the large number of events in the inclusive three- and two-prong samples and in the semileptonic sample permit
us to perform differential measurements.

A detailed analysis of $\Wc$ production at $\sqrt s = 7\TeV$ is presented. The study is done
for the kinematic region $\pt^\text{jet}>25\GeV$, $\abs{\eta^\text{jet}}<2.5$, in the lepton pseudorapidity range $\abs{\eta^\ell}<2.1$,
and for two different thresholds for the transverse momentum of the lepton from the $\PW$-boson decay: $\pt^\ell > 25\GeV$ in the $\PW$-boson
muon decay channel only, and $\pt^\ell > 35\GeV$ in both the muon and the electron $\PW$-boson decay channels.
Results obtained in the three charm decay samples and in the two $\PW$-boson decay modes are fully consistent
and are thus combined to increase the final precision of the measurements.

The total $\Wc$ production cross sections are measured to be
\begin{align*}
\sigma(\Pp\Pp \rightarrow \mathrm{W+c + X}) \times \mathcal{B}(\PW \rightarrow \mu\nu) (\pt^\mu>25\GeV) & =  107.7 \pm 3.3\stat \pm 6.9\syst\unit{pb},\\
 & & \\
\sigma(\Pp\Pp \rightarrow \mathrm{W+c + X}) \times \mathcal{B}(\PW \rightarrow \ell\nu) (\pt^\ell>35\GeV) & =  84.1 \pm 2.0\stat \pm 4.9\syst\unit{pb}. \\
\end{align*}
Cross section ratios of the associated production of a positively charged $\PW$ boson with a $\bar\cPqc$ antiquark
and a negatively charged $\PW$ boson with a c quark are obtained:
\begin{align*}
\frac{\sigma(\Pp\Pp \rightarrow \mathrm{\PWp + \cPaqc +X})}{\sigma(\Pp\Pp \rightarrow \mathrm{\PWm + c + X})} (\pt^\mu>25\GeV) & =  0.954 \pm 0.025\stat \pm 0.004\syst, \\
 & & \\
\frac{\sigma(\Pp\Pp \rightarrow \mathrm{\PWp + \cPaqc +X})}{\sigma(\Pp\Pp \rightarrow \mathrm{\PWm + c + X})} (\pt^\ell>35\GeV) & =  0.938 \pm 0.019\stat \pm 0.006\syst.\\
\end{align*}
The measured cross section ratios are the first evidence for an asymmetry in the $\PWpc$ and $\PWmc$ production.
Total cross sections and cross section ratios are also measured as a function of the absolute value of the pseudorapidity
of the lepton from the $\PW$-boson decay, thus probing a wide range of Bjorken $x$ of the parton distribution of the proton.
These measurements provide the first direct constraint from LHC data on the strange quark and antiquark content of the proton and constitute
a valuable input for future global PDF analyses.

These measurements are compared with theoretical predictions calculated with \MCFM at next-to-leading order in perturbative QCD
using various sets of parton distribution functions. The PDF groups make different
assumptions in their global fits about the total strange-quark content of the proton and of the $\cPqs$--$\cPaqs$ asymmetry.
An overall agreement between the experimental results and the theoretical predictions is observed, which validates the
fitted strange quark and antiquark parton distribution functions at an energy significantly higher than those of previous
experiments.
In particular, the predicted total cross sections based on those PDF sets that include low-energy DIS data in their fits agree with the measurements.
Theoretical calculations also predict differential cross section shapes in agreement with the measured ones.
The observed $\PWmc$ yield is slightly larger than the $\PWpc$ yield, as expected from the dominance of the d quark over the $\cPqd$
antiquark in the proton.

\section*{Acknowledgements}

\hyphenation{Bundes-ministerium Forschungs-gemeinschaft Forschungs-zentren} We congratulate our colleagues in the CERN accelerator departments for the excellent performance of the LHC and thank the technical and administrative staffs at CERN and at other CMS institutes for their contributions to the success of the CMS effort. In addition, we gratefully acknowledge the computing centres and personnel of the Worldwide LHC Computing Grid for delivering so effectively the computing infrastructure essential to our analyses. Finally, we acknowledge the enduring support for the construction and operation of the LHC and the CMS detector provided by the following funding agencies: the Austrian Federal Ministry of Science and Research and the Austrian Science Fund; the Belgian Fonds de la Recherche Scientifique, and Fonds voor Wetenschappelijk Onderzoek; the Brazilian Funding Agencies (CNPq, CAPES, FAPERJ, and FAPESP); the Bulgarian Ministry of Education and Science; CERN; the Chinese Academy of Sciences, Ministry of Science and Technology, and National Natural Science Foundation of China; the Colombian Funding Agency (COLCIENCIAS); the Croatian Ministry of Science, Education and Sport; the Research Promotion Foundation, Cyprus; the Ministry of Education and Research, Recurrent financing contract SF0690030s09 and European Regional Development Fund, Estonia; the Academy of Finland, Finnish Ministry of Education and Culture, and Helsinki Institute of Physics; the Institut National de Physique Nucl\'eaire et de Physique des Particules~/~CNRS, and Commissariat \`a l'\'Energie Atomique et aux \'Energies Alternatives~/~CEA, France; the Bundesministerium f\"ur Bildung und Forschung, Deutsche Forschungsgemeinschaft, and Helmholtz-Gemeinschaft Deutscher Forschungszentren, Germany; the General Secretariat for Research and Technology, Greece; the National Scientific Research Foundation, and National Office for Research and Technology, Hungary; the Department of Atomic Energy and the Department of Science and Technology, India; the Institute for Studies in Theoretical Physics and Mathematics, Iran; the Science Foundation, Ireland; the Istituto Nazionale di Fisica Nucleare, Italy; the Korean Ministry of Education, Science and Technology and the World Class University program of NRF, Republic of Korea; the Lithuanian Academy of Sciences; the Mexican Funding Agencies (CINVESTAV, CONACYT, SEP, and UASLP-FAI); the Ministry of Business, Innovation and Employment, New Zealand; the Pakistan Atomic Energy Commission; the Ministry of Science and Higher Education and the National Science Centre, Poland; the Funda\c{c}\~ao para a Ci\^encia e a Tecnologia, Portugal; JINR, Dubna; the Ministry of Education and Science of the Russian Federation, the Federal Agency of Atomic Energy of the Russian Federation, Russian Academy of Sciences, and the Russian Foundation for Basic Research; the Ministry of Education, Science and Technological Development of Serbia; the Secretar\'{\i}a de Estado de Investigaci\'on, Desarrollo e Innovaci\'on and Programa Consolider-Ingenio 2010, Spain; the Swiss Funding Agencies (ETH Board, ETH Zurich, PSI, SNF, UniZH, Canton Zurich, and SER); the National Science Council, Taipei; the Thailand Center of Excellence in Physics, the Institute for the Promotion of Teaching Science and Technology of Thailand, Special Task Force for Activating Research and the National Science and Technology Development Agency of Thailand; the Scientific and Technical Research Council of Turkey, and Turkish Atomic Energy Authority; the Science and Technology Facilities Council, UK; the US Department of Energy, and the US National Science Foundation.

Individuals have received support from the Marie-Curie programme and the European Research Council and EPLANET (European Union); the Leventis Foundation; the A. P. Sloan Foundation; the Alexander von Humboldt Foundation; the Belgian Federal Science Policy Office; the Fonds pour la Formation \`a la Recherche dans l'Industrie et dans l'Agriculture (FRIA-Belgium); the Agentschap voor Innovatie door Wetenschap en Technologie (IWT-Belgium); the Ministry of Education, Youth and Sports (MEYS) of Czech Republic; the Council of Science and Industrial Research, India; the Compagnia di San Paolo (Torino); the HOMING PLUS programme of Foundation for Polish Science, cofinanced by EU, Regional Development Fund; and the Thalis and Aristeia programmes cofinanced by EU-ESF and the Greek NSRF.

\bibliography{auto_generated}   

\providecommand{\href}[2]{#2}\begingroup\raggedright\begin{thebibliography}{10}%
\makeatletter
\providecommand{\hrefCMSnoop }[0]{\@secondoftwo}%
\makeatother
\providecommand{\doi}{\texttt{doi:}\begingroup \urlstyle{tt}\Url}

\bibitem{Baur}
U.~Baur\hrefCMSnoop {} { {et~al.}, ``{The charm content of W + 1 jet events as
  a probe of the strange quark distribution function}'',} \textit{ Phys. Lett.
  B} \textbf{ 318} (1993) 544,
  \href{http://dx.doi.org/10.1016/0370-2693(93)91553-Y}{\doi{10.1016/0370-2693(93)91553-Y}},
  \href{http://www.arXiv.org/abs/hep-ph/9308370}{\texttt{
  arXiv:hep-ph/9308370}}.

\bibitem{Charm_Kusina}
A.~Kusina\hrefCMSnoop {} { {et~al.}, ``Strange quark parton distribution
  functions and implications for Drell-Yan boson production at the LHC'',}
  \textit{ Phys. Rev. D} \textbf{ 85} (2012) 094028,
  \href{http://dx.doi.org/10.1103/PhysRevD.85.094028}{\doi{10.1103/PhysRevD.85.094028}},
  \href{http://www.arXiv.org/abs/1203.1290}{\texttt{ arXiv:1203.1290}}.

\bibitem{Charm_Eleni}
\hrefCMSnoop {} {W.~J. Stirling and E.~Vryonidou, ``Charm Production in
  Association with an Electroweak Gauge Boson at the LHC'',} \textit{ Phys.
  Rev. Lett.} \textbf{ 109} (2012) 082002,
  \href{http://dx.doi.org/10.1103/PhysRevLett.109.082002}{\doi{10.1103/PhysRevLett.109.082002}},
  \href{http://www.arXiv.org/abs/1203.6781}{\texttt{ arXiv:1203.6781}}.

\bibitem{Ball:2012wy}
R.~D. Ball\hrefCMSnoop {} { {et~al.}, ``Parton distribution benchmarking with
  LHC Data'',} \textit{ JHEP} \textbf{ 04} (2013) 125,
  \href{http://dx.doi.org/10.1007/JHEP04(2013)125}{\doi{10.1007/JHEP04(2013)125}},
\href{http://www.arXiv.org/abs/1211.5142}{\texttt{ arXiv:1211.5142}}.

\bibitem{NNPDF:NuTeV}
\hrefCMSnoop {} {{ NNPDF} Collaboration, ``Precision determination of
  electroweak parameters and the strange content of the proton from neutrino
  deep-inelastic scattering'',} \textit{ Nucl. Phys. B} \textbf{ 823} (2009)
  195,
  \href{http://dx.doi.org/10.1016/j.nuclphysb.2009.08.003}{\doi{10.1016/j.nuclphysb.2009.08.003}},
  \href{http://www.arXiv.org/abs/0906.1958}{\texttt{ arXiv:0906.1958}}.

\bibitem{Rojo_Wmass}
\hrefCMSnoop {} {G.~Bozzi, J.~Rojo, and A.~Vicini, ``Impact of the parton
  distribution function uncertainties on the measurement of the W boson mass at
  the Tevatron and the LHC'',} \textit{ Phys. Rev. D} \textbf{ 83} (2011)
  113008,
  \href{http://dx.doi.org/10.1103/PhysRevD.83.113008}{\doi{10.1103/PhysRevD.83.113008}},
  \href{http://www.arXiv.org/abs/1104.2056}{\texttt{ arXiv:1104.2056}}.

\bibitem{Cabibbo}
\hrefCMSnoop {} {N.~Cabibbo, ``Unitary Symmetry and Leptonic Decays'',}
  \textit{ Phys. Rev. Lett.} \textbf{ 10} (1963) 531,
  \href{http://dx.doi.org/10.1103/PhysRevLett.10.531}{\doi{10.1103/PhysRevLett.10.531}}.

\bibitem{CDF-first}
\hrefCMSnoop {} {{ CDF} Collaboration, ``First Measurement of the W Boson
  Production in Association with a Single Charm Quark in $p\bar{p}$ Collisions
  at $\sqrt{s}= 1.96$ TeV'',} \textit{ Phys. Rev. Lett.} \textbf{ 100} (2008)
  091803,
  \href{http://dx.doi.org/10.1103/PhysRevLett.100.091803}{\doi{10.1103/PhysRevLett.100.091803}},
  \href{http://www.arXiv.org/abs/0711.2901}{\texttt{ arXiv:0711.2901}}.

\bibitem{CDF}
\hrefCMSnoop {} {{ CDF} Collaboration, ``Observation of the Production of a W
  Boson in Association with a Single Charm Quark'',} \textit{ Phys. Rev. Lett.}
  \textbf{ 110} (2013) 071801,
  \href{http://dx.doi.org/10.1103/PhysRevLett.110.071801}{\doi{10.1103/PhysRevLett.110.071801}},
  \href{http://www.arXiv.org/abs/1209.1921}{\texttt{ arXiv:1209.1921}}.

\bibitem{D0}
\hrefCMSnoop {} {{ D0} Collaboration, ``Measurement of the ratio of the $p
  \bar{p} \to W +$ c-jet cross section to the inclusive $p \bar{p} \to W +$
  jets cross section'',} \textit{ Phys. Lett. B} \textbf{ 666} (2008) 23,
  \href{http://dx.doi.org/10.1016/j.physletb.2008.06.067}{\doi{10.1016/j.physletb.2008.06.067}},
  \href{http://www.arXiv.org/abs/0803.2259}{\texttt{ arXiv:0803.2259}}.

\bibitem{JINST}
\hrefCMSnoop {} {{ CMS} Collaboration, ``The CMS experiment at the CERN LHC'',}
  \textit{ JINST} \textbf{ 3} (2008) S08004,
  \href{http://dx.doi.org/10.1088/1748-0221/3/08/S08004}{\doi{10.1088/1748-0221/3/08/S08004}}.

\bibitem{DQM}
\hrefCMSnoop {} {L.~Tuura, A.~Meyer, I.~Segoni, and G.~Della~Ricca, ``CMS data
  quality monitoring: Systems and experiences'',} \textit{ J. Phys.: Conf.
  Ser.} \textbf{ 219} (2010) 072020,
  \href{http://dx.doi.org/10.1088/1742-6596/219/7/072020}{\doi{10.1088/1742-6596/219/7/072020}}.

\bibitem{CMS-PAPER-MUO-10-004}
\hrefCMSnoop {} {{ CMS} Collaboration, ``Performance of CMS muon reconstruction
  in pp collision events at $\sqrt{s}$ = 7 TeV'',} \textit{ JINST} \textbf{ 7}
  (2012) P10002,
  \href{http://dx.doi.org/10.1088/1748-0221/7/10/P10002}{\doi{10.1088/1748-0221/7/10/P10002}},
  \href{http://www.arXiv.org/abs/1206.4071}{\texttt{ arXiv:1206.4071}}.

\bibitem{EGAMMA-PAS}
\href {http://cdsweb.cern.ch/record/1299116} {{ CMS} Collaboration, ``Electron
  reconstruction and identification at $\sqrt{s} = 7$ TeV'',} CMS Physics
  Analysis Summary CMS-PAS-EGM-10-004, 2010.

\bibitem{CMS-PAS-PFT-2010-003}
\href {http://cdsweb.cern.ch/record/1279347} {{ CMS} Collaboration,
  ``Particle-flow commissioning with muons and electrons from J/Psi, and W
  events at 7 TeV'',} CMS Physics Analysis Summary CMS-PAS-PFT-10-003, 2010.

\bibitem{antikt}
\hrefCMSnoop {} {M.~Cacciari, G.~P. Salam, and G.~Soyez, ``The anti-$k_t$ jet
  clustering algorithm'',} \textit{ JHEP} \textbf{ 04} (2008) 063,
  \href{http://dx.doi.org/10.1088/1126-6708/2008/04/063}{\doi{10.1088/1126-6708/2008/04/063}},
  \href{http://www.arXiv.org/abs/0802.1189}{\texttt{ arXiv:0802.1189}}.

\bibitem{PUsubtraction}
\hrefCMSnoop {} {M.~Cacciari and G.~P. Salam, ``Pileup subtraction using jet
  areas'',} \textit{ Phys. Lett. B} \textbf{ 659} (2008) 119,
  \href{http://dx.doi.org/10.1016/j.physletb.2007.09.077}{\doi{10.1016/j.physletb.2007.09.077}},
  \href{http://www.arXiv.org/abs/0707.1378}{\texttt{ arXiv:0707.1378}}.

\bibitem{PUsubtraction2}
\hrefCMSnoop {} {M.~Cacciari, G.~P. Salam, and G.~Soyez, ``The catchment area
  of jets'',} \textit{ JHEP} \textbf{ 04} (2008) 005,
  \href{http://dx.doi.org/10.1088/1126-6708/2008/04/005}{\doi{10.1088/1126-6708/2008/04/005}},
  \href{http://www.arXiv.org/abs/0802.1188}{\texttt{ arXiv:0802.1188}}.

\bibitem{CMS-PAPER-JME-10-011}
\hrefCMSnoop {} {{ CMS} Collaboration, ``Determination of jet energy
  calibration and transverse momentum resolution in CMS'',} \textit{ JINST}
  \textbf{ 6} (2011) P11002,
  \href{http://dx.doi.org/10.1088/1748-0221/6/11/P11002}{\doi{10.1088/1748-0221/6/11/P11002}},
  \href{http://www.arXiv.org/abs/1107.4277}{\texttt{ arXiv:1107.4277}}.

\bibitem{MADGRAPH5}
J.~Alwall\hrefCMSnoop {} { {et~al.}, ``MadGraph5: going beyond'',} \textit{
  JHEP} \textbf{ 06} (2011) 128,
  \href{http://dx.doi.org/10.1007/JHEP06(2011)128}{\doi{10.1007/JHEP06(2011)128}},
  \href{http://www.arXiv.org/abs/1106.0522}{\texttt{ arXiv:1106.0522}}.

\bibitem{Pythia}
\hrefCMSnoop {} {T.~Sj{\"o}strand, S.~Mrenna, and P.~Z. Skands, ``{PYTHIA} 6.4
  physics and manual'',} \textit{ JHEP} \textbf{ 05} (2006) 026,
  \href{http://dx.doi.org/10.1088/1126-6708/2006/05/026}{\doi{10.1088/1126-6708/2006/05/026}},
  \href{http://www.arXiv.org/abs/hep-ph/0603175}{\texttt{
  arXiv:hep-ph/0603175}}.

\bibitem{VJETS-paper}
\hrefCMSnoop {} {{ CMS} Collaboration, ``Jet production rates in association
  with {W} and {Z} bosons in pp collisions at $\sqrt{s} = 7$~{TeV}'',} \textit{
  JHEP} \textbf{ 01} (2012) 010,
  \href{http://dx.doi.org/10.1007/JHEP01(2012)010}{\doi{10.1007/JHEP01(2012)010}},
  \href{http://www.arXiv.org/abs/1110.3226}{\texttt{ arXiv:1110.3226}}.

\bibitem{POWHEG}
\hrefCMSnoop {} {S.~Alioli, P.~Nason, C.~Oleari, and E.~Re, ``NLO vector-boson
  production matched with shower in POWHEG'',} \textit{ JHEP} \textbf{ 07}
  (2008) 060,
  \href{http://dx.doi.org/10.1088/1126-6708/2008/07/060}{\doi{10.1088/1126-6708/2008/07/060}},
  \href{http://www.arXiv.org/abs/0805.4802}{\texttt{ arXiv:0805.4802}}.

\bibitem{CT10}
H.-L. Lai\hrefCMSnoop {} { {et~al.}, ``New parton distributions for collider
  physics'',} \textit{ Phys. Rev. D} \textbf{ 82} (2010) 074024,
  \href{http://dx.doi.org/10.1103/PhysRevD.82.074024}{\doi{10.1103/PhysRevD.82.074024}},
  \href{http://www.arXiv.org/abs/1007.2241}{\texttt{ arXiv:1007.2241}}.

\bibitem{CTEQ6L1}
J.~Pumplin\hrefCMSnoop {} { {et~al.}, ``New generation of parton distributions
  with uncertainties from global QCD analysis'',} \textit{ JHEP} \textbf{ 07}
  (2002) 012,
  \href{http://dx.doi.org/10.1088/1126-6708/2002/07/012}{\doi{10.1088/1126-6708/2002/07/012}},
  \href{http://www.arXiv.org/abs/hep-ph/0201195}{\texttt{
  arXiv:hep-ph/0201195}}.

\bibitem{Z2Tune}
\hrefCMSnoop {} {{ CMS} Collaboration, ``Measurement of the underlying event
  activity at the {LHC} with {$\sqrt{s} = 7\TeV$} and comparison with
  {$\sqrt{s} = 0.9\TeV$}'',} \textit{ JHEP} \textbf{ 09} (2011) 109,
  \href{http://dx.doi.org/10.1007/JHEP09(2011)109}{\doi{10.1007/JHEP09(2011)109}},
  \href{http://www.arXiv.org/abs/1107.0330}{\texttt{ arXiv:1107.0330}}.

\bibitem{fewz}
\hrefCMSnoop {} {K.~Melnikov and F.~Petriello, ``Electroweak gauge boson
  production at hadron colliders through $\mathcal{O}({\alpha_s}^{2})$'',}
  \textit{ Phys. Rev. D} \textbf{ 74} (2006) 114017,
  \href{http://dx.doi.org/10.1103/PhysRevD.74.114017}{\doi{10.1103/PhysRevD.74.114017}},
  \href{http://www.arXiv.org/abs/hep-ph/0609070}{\texttt{
  arXiv:hep-ph/0609070}}.

\bibitem{MSTW08}
\hrefCMSnoop {} {A.~D. Martin, W.~J. Stirling, R.~S. Thorne, and G.~Watt,
  ``Parton distributions for the LHC'',} \textit{ Eur. Phys. J. C} \textbf{ 63}
  (2009) 189,
  \href{http://dx.doi.org/10.1140/epjc/s10052-009-1072-5}{\doi{10.1140/epjc/s10052-009-1072-5}},
  \href{http://www.arXiv.org/abs/0901.0002}{\texttt{ arXiv:0901.0002}}.

\bibitem{Czakon:2013goa}
\hrefCMSnoop {} {M.~Czakon, P.~Fiedler, and A.~Mitov, ``{The total top quark
  pair production cross-section at hadron colliders through ${\cal
  O}(\alpha_{\rm S}^4)$}'',} \textit{ Phys. Rev. Lett.} \textbf{ 110} (2013)
  252004,
  \href{http://dx.doi.org/10.1103/PhysRevLett.110.252004}{\doi{10.1103/PhysRevLett.110.252004}},
  \href{http://www.arXiv.org/abs/1303.6254}{\texttt{ arXiv:1303.6254}}.

\bibitem{MCFM}
\hrefCMSnoop {} {J.~M. Campbell and R.~Ellis, ``MCFM for the Tevatron and the
  LHC'',} \textit{ Nucl. Phys. Proc. Suppl.} \textbf{ 205} (2010) 10,
  \href{http://dx.doi.org/10.1016/j.nuclphysbps.2010.08.011}{\doi{10.1016/j.nuclphysbps.2010.08.011}},
  \href{http://www.arXiv.org/abs/1007.3492}{\texttt{ arXiv:1007.3492}}.

\bibitem{GEANT4}
\hrefCMSnoop {} {{ GEANT4} Collaboration, ``GEANT4: A simulation toolkit'',}
  \textit{ Nucl. Instrum. and Methods A} \textbf{ 506} (2003) 250,
  \href{http://dx.doi.org/10.1016/S0168-9002(03)01368-8}{\doi{10.1016/S0168-9002(03)01368-8}}.

\bibitem{WZ-paper}
\hrefCMSnoop {} {{ CMS} Collaboration, ``Measurement of the inclusive {W} and
  {Z} production cross sections in pp collisions at {$\sqrt{s}=7\TeV$} with the
  CMS experiment'',} \textit{ JHEP} \textbf{ 10} (2011) 132,
  \href{http://dx.doi.org/10.1007/JHEP10(2011)132}{\doi{10.1007/JHEP10(2011)132}},
  \href{http://www.arXiv.org/abs/1107.4789}{\texttt{ arXiv:1107.4789}}.

\bibitem{Vertex}
\hrefCMSnoop {} {W.~Waltenberger, R.~Fr{\"u}hwirth, and P.~Vanlaer, ``Adaptive
  vertex fitting'',} \textit{ J. Phys. G} \textbf{ 34} (2007) N343,
  \href{http://dx.doi.org/10.1088/0954-3899/34/12/N01}{\doi{10.1088/0954-3899/34/12/N01}}.

\bibitem{CMS-PAPER-BTV-12-001}
\hrefCMSnoop {} {{ CMS} Collaboration, ``Identification of b-quark jets with
  the {CMS} experiment'',} \textit{ JINST} \textbf{ 8} (2013) P04013,
  \href{http://dx.doi.org/10.1088/1748-0221/8/04/P04013}{\doi{10.1088/1748-0221/8/04/P04013}},
  \href{http://www.arXiv.org/abs/1211.4462}{\texttt{ arXiv:1211.4462}}.

\bibitem{PDG}
\hrefCMSnoop {} {{Particle Data Group, J. Beringer} {et~al.}, ``Review of
  Particle Physics'',} \textit{ Phys. Rev. D} \textbf{ 86} (2012) 010001,
  \href{http://dx.doi.org/10.1103/PhysRevD.86.010001}{\doi{10.1103/PhysRevD.86.010001}}.

\bibitem{charm_OPAL_Dpm}
\hrefCMSnoop {} {{ OPAL} Collaboration, ``{A study of charm hadron production
  in $\cPZ \to {\rm c\bar c}$ and $\cPZ \to {\rm b\bar b}$ decays at LEP}'',}
  \textit{ Z. Phys. C} \textbf{ 72} (1996) 1,
  \href{http://dx.doi.org/10.1007/s002880050218}{\doi{10.1007/s002880050218}}.

\bibitem{charm_ALEPH}
\hrefCMSnoop {} {{ ALEPH} Collaboration, ``Study of charm production in Z
  decays'',} \textit{ Eur. Phys. J. C} \textbf{ 16} (2000) 597,
  \href{http://dx.doi.org/10.1007/s100520000421}{\doi{10.1007/s100520000421}},
  \href{http://www.arXiv.org/abs/hep-ex/9909032}{\texttt{
  arXiv:hep-ex/9909032}}.

\bibitem{charm_DELPHI_Dpm}
\hrefCMSnoop {} {{ DELPHI} Collaboration, ``Measurements of the Z partial decay
  width into ${\rm c\bar c}$ and multiplicity of charm quarks per b decay'',}
  \textit{ Eur. Phys. J. C} \textbf{ 12} (2000) 225,
  \href{http://dx.doi.org/10.1007/s1005299000228}{\doi{10.1007/s1005299000228}}.

\bibitem{charm_OPAL_Dstar}
\hrefCMSnoop {} {{ OPAL} Collaboration, ``{Measurement of ${\rm f}({\rm c}\to
  {\rm D}^{*} + X)$, ${\rm f}({\rm b}\to {\rm D}^{*}+ X)$ and $\Gamma({\rm
  c\bar c})/\Gamma({\rm had})$ using ${\rm D}^{*\pm}$ mesons}'',} \textit{ Eur.
  Phys. J. C} \textbf{ 1} (1998) 439,
  \href{http://dx.doi.org/10.1007/s100520050095}{\doi{10.1007/s100520050095}},
  \href{http://www.arXiv.org/abs/hep-ex/9708021}{\texttt{
  arXiv:hep-ex/9708021}}.

\bibitem{charm_DELPHI_Dstar}
\hrefCMSnoop {} {{ DELPHI} Collaboration, ``{Determination of $P({\rm c}\to
  {\rm D}^{*+})$ and $BR({\rm c}\to\ell^+)$ at LEP 1}'',} \textit{ Eur. Phys.
  J. C} \textbf{ 12} (2000) 209,
  \href{http://dx.doi.org/10.1007/s100529900227}{\doi{10.1007/s100529900227}}.

\bibitem{charm-fractions}
\hrefCMSnoop {} {L.~Gladilin, ``Charm Hadron Production Fractions'',} (1999).
  \href{http://www.arXiv.org/abs/hep-ex/9912064}{\texttt{
  arXiv:hep-ex/9912064}}.

\bibitem{CMS-PAS-TRK-10-002}
\href {http://cds.cern.ch/record/1279139} {{ CMS} Collaboration, ``Measurement
  of Tracking Efficiency'',} CMS Physics Analysis Summary CMS-PAS-TRK-10-002,
  2010.

\bibitem{CMS-PAS-SMP-12-008}
\href {http://cds.cern.ch/record/1434360} {{ CMS} Collaboration, ``Absolute
  Calibration of the Luminosity Measurement at CMS: Winter 2012 Update'',} CMS
  Physics Analysis Summary CMS-PAS-SMP-12-028, 2012.

\bibitem{mu_scale}
A.~Bodek\hrefCMSnoop {} { {et~al.}, ``{Extracting muon momentum scale
  corrections for hadron collider experiments}'',} \textit{ Eur. Phys. J. C}
  \textbf{ 72} (2012) 2194,
  \href{http://dx.doi.org/10.1140/epjc/s10052-012-2194-8}{\doi{10.1140/epjc/s10052-012-2194-8}},
  \href{http://www.arXiv.org/abs/1208.3710}{\texttt{ arXiv:1208.3710}}.

\bibitem{CMS-PAPER-EGM-11-001}
\hrefCMSnoop {} {{CMS Collaboration}, ``{Energy calibration and resolution of
  the CMS electromagnetic calorimeter in pp collisions at $\sqrt{s}$ = 7
  TeV}'',} (2013). \href{http://www.arXiv.org/abs/1306.2016}{\texttt{
  arXiv:1306.2016}}. Submitted to JINST.

\bibitem{Ball:2011uy}
\hrefCMSnoop {} {{ NNPDF} Collaboration, ``{Unbiased global determination of
  parton distributions and their uncertainties at NNLO and at LO}'',} \textit{
  Nucl. Phys. B} \textbf{ 855} (2012) 153,
  \href{http://dx.doi.org/10.1016/j.nuclphysb.2011.09.024}{\doi{10.1016/j.nuclphysb.2011.09.024}},
  \href{http://www.arXiv.org/abs/1107.2652}{\texttt{ arXiv:1107.2652}}.

\bibitem{CMS-PAPER-SMP-12-001}
\hrefCMSnoop {} {{ CMS} Collaboration, ``Measurement of the electron charge
  asymmetry in inclusive {W} production in pp collisions at {$\sqrt{s} =
  7\TeV$}'',} \textit{ Phys. Rev. Lett.} \textbf{ 109} (2012) 111806,
  \href{http://dx.doi.org/10.1103/PhysRevLett.109.111806}{\doi{10.1103/PhysRevLett.109.111806}},
  \href{http://www.arXiv.org/abs/1206.2598}{\texttt{ arXiv:1206.2598}}.

\bibitem{CMS-PAPER-MUO-10-001}
\hrefCMSnoop {} {{ CMS} Collaboration, ``Measurement of the charge ratio of
  atmospheric muons with the {CMS} detector'',} \textit{ Phys. Lett. B}
  \textbf{ 692} (2010) 83,
  \href{http://dx.doi.org/10.1016/j.physletb.2010.07.033}{\doi{10.1016/j.physletb.2010.07.033}},
  \href{http://www.arXiv.org/abs/1005.5332}{\texttt{ arXiv:1005.5332}}.

\bibitem{MCFM_WplusC}
\hrefCMSnoop {} {J.~M. Campbell and F.~Tramontano, ``Next-to-leading order
  corrections to Wt production and decay'',} \textit{ Nucl. Phys. B} \textbf{
  726} (2005) 109,
  \href{http://dx.doi.org/10.1016/j.nuclphysb.2005.08.015}{\doi{10.1016/j.nuclphysb.2005.08.015}},
  \href{http://www.arXiv.org/abs/hep-ph/0506289}{\texttt{
  arXiv:hep-ph/0506289}}.

\bibitem{NNPDF23}
\hrefCMSnoop {} {{ NNPDF} Collaboration, ``Parton distributions with LHC
  data'',} \textit{ Nucl. Phys. B} \textbf{ 867} (2013) 244,
  \href{http://dx.doi.org/10.1016/j.nuclphysb.2012.10.003}{\doi{10.1016/j.nuclphysb.2012.10.003}},
  \href{http://www.arXiv.org/abs/1207.1303}{\texttt{ arXiv:1207.1303}}.

\bibitem{ABM}
\hrefCMSnoop {} {S.~Alekhin, J.~Blumlein, and S.~Moch, ``{Parton distribution
  functions and benchmark cross sections at next-to-next-to-leading order}'',}
  \textit{ Phys. Rev. D} \textbf{ 86} (2012) 054009,
  \href{http://dx.doi.org/10.1103/PhysRevD.86.054009}{\doi{10.1103/PhysRevD.86.054009}},
  \href{http://www.arXiv.org/abs/1202.2281}{\texttt{ arXiv:1202.2281}}.

\bibitem{JR}
\hrefCMSnoop {} {P.~Jimenez-Delgado and E.~Reya, ``{Dynamical
  next-to-next-to-leading order parton distributions}'',} \textit{ Phys. Rev.
  D} \textbf{ 79} (2009) 074023,
  \href{http://dx.doi.org/10.1103/PhysRevD.79.074023}{\doi{10.1103/PhysRevD.79.074023}},
  \href{http://www.arXiv.org/abs/0810.4274}{\texttt{ arXiv:0810.4274}}.

\bibitem{CooperSarkar}
\href {http://pos.sissa.it/archive/conferences/134/320/EPS-HEP2011_320.pdf}
  {A.~M. Cooper-Sarkar, ``{PDF Fits at HERA}'',} in \textit{ XXIst
  International Europhysics Conference on High Energy Physics, EPS-HEP2011},
  p.~320.
\newblock 2011.
\newblock \href{http://www.arXiv.org/abs/1112.2107}{\texttt{ arXiv:1112.2107}}.

\bibitem{Radescu}
\href {http://pos.sissa.it/archive/conferences/120/168/ICHEP%202010_168.pdf}
  {V.~Radescu, ``{Combination and QCD Analysis of the HERA Inclusive Cross
  Sections}'',} in \textit{ 35th International Conference of High Energy
  Physics, ICHEP 2010}, p.~168.
\newblock 2010.

\end{thebibliography}\endgroup

\clearpage
\appendix
\section{Normalized differential cross section and cross section ratios as a function of the
lepton pseudorapidity~\label{app:tables_diff}}

\begin{table}[htbp]
\begin{center}
    \caption{Estimated number of $\OSSS$ events in the inclusive three-prong sample (defined in Section~\ref{sec:inclusive}).
The estimated numbers of remaining background events after SS subtraction is given in the third column.
The normalized differential cross section as a function of the absolute value of the  lepton pseudorapidity is shown in the last column.
The first two blocks of the table present the results from the $\PWmn$ sample, with $\pt^\mu > 25\GeV$  and
$\pt^\mu > 35\GeV$. The results from the $\Wen$ sample, with $\pt^\Pe > 35\GeV$ are given in the lowest block of the table.
The first error in the normalized differential cross section is due to the statistical size of the data sample and the second one is
the systematic uncertainty from to the sources discussed in Section~\ref{sec:syst_diff}.
    }
    \label{tab:Dpm_differential}
\begin{tabular}{cccc}
\hline\hline
\multicolumn{4}{c}{$\PWmn$, $\pt^\mu > 25\GeV$} \\ \hline
$[{\abs{\eta}}_\text{min},{\abs{\eta}}_\text{max}]$ & $N_\text{sel}$ & $N_\text{bkg}$ & $\SWcdiff$ \\
\hline
$[0, 0.35]$   &  $1697 \pm 83$ &  $86  \pm 49$ & $0.64 \pm 0.03 \pm 0.02$  \\
$[0.35, 0.7]$ &  $1596 \pm 86$ &  $63  \pm 46$ & $0.57 \pm 0.03 \pm 0.02$  \\
$[0.7, 1.1]$  &  $1558 \pm 83$ &  $113 \pm 52$ & $0.52 \pm 0.03 \pm 0.02$  \\
$[1.1, 1.6]$  &  $1495 \pm 85$ &  $142 \pm 56$ & $0.40 \pm 0.02 \pm 0.02$  \\
$[1.6, 2.1]$  &  $1133 \pm 72$ &  $72  \pm 43$ & $0.34 \pm 0.02 \pm 0.01$  \\
\hline \hline
\multicolumn{4}{c}{$\PWmn$, $\pt^\mu > 35\GeV$} \\ \hline
$[{\abs{\eta}}_\text{min},{\abs{\eta}}_\text{max}]$ & $N_\text{sel}$ & $N_\text{bkg}$ & $\SWcdiff$ \\
\hline
$[0, 0.35]$   &  $1390 \pm 75$ & $ 37 \pm 37$ & $0.65 \pm 0.03 \pm 0.02$  \\
$[0.35, 0.7]$ &  $1323 \pm 76$ & $ 40 \pm 37$ & $0.58 \pm 0.03 \pm 0.02$  \\
$[0.7, 1.1]$  &  $1252 \pm 74$ & $ 87 \pm 45$ & $0.51 \pm 0.03 \pm 0.02$  \\
$[1.1, 1.6]$  &  $1224 \pm 75$ & $ 90 \pm 45$ & $0.40 \pm 0.02 \pm 0.02$  \\
$[1.6, 2.1]$  &  $899  \pm 63$ & $ 16 \pm 30$ & $0.34 \pm 0.02 \pm 0.02$  \\
\hline\hline
\multicolumn{4}{c}{$\Wen$, $\pt^\Pe > 35\GeV$} \\ \hline
$[{\abs{\eta}}_\text{min},{\abs{\eta}}_\text{max}]$ & $N_\text{sel}$ & $N_\text{bkg}$ & $\SWcdiff$ \\
\hline

$[0, 0.35]$   &  $950 \pm 65$ & $219\pm44$ &  $0.56 \pm 0.05 \pm 0.03$  \\
$[0.35, 0.7]$ &  $955 \pm 64$ & $182\pm44$ &  $0.60 \pm 0.05 \pm 0.03$  \\
$[0.7, 1.1]$  &  $940 \pm 64$ & $178\pm44$ &  $0.51 \pm 0.04 \pm 0.03$  \\
$[1.1, 1.6]$  &  $741 \pm 55$ & $97\pm38$ &   $0.50 \pm 0.04 \pm 0.03$  \\
$[1.6, 2.1]$  &  $437 \pm 50$ & $100\pm33$ &  $0.28 \pm 0.04 \pm 0.03$ \\
\hline\hline
\end{tabular}
\end{center}
\end{table}

\begin{table}[htbp]
\begin{center}
    \caption{Estimated number of $\OSSS$ events in the inclusive two-prong sample (defined in Section~\ref{sec:inclusive}).
The estimated numbers of remaining background events after SS subtraction is given in the third column.
The normalized differential cross section as a function of the absolute value of the lepton pseudorapidity is shown in the last column.
The first two blocks of the table present the results from the $\PWmn$ sample, with $\pt^\mu > 25\GeV$  and
$\pt^\mu > 35\GeV$. The results from the $\Wen$ sample, with $\pt^\Pe > 35\GeV$ are given in the lowest block of the table.
The first error in the normalized differential cross section is due to the statistical size of the data sample and the second one is
the systematic uncertainty from to the sources discussed in Section~\ref{sec:syst_diff}.
}
    \label{tab:Dstar_differential}
\begin{tabular}{c c c c}
\hline\hline
\multicolumn{4}{c}{$\PWmn$, $\pt^\mu > 25\GeV$} \\ \hline
$[{\abs{\eta}}_\text{min},{\abs{\eta}}_\text{max}]$ & $N_\text{sel}$ & $N_\text{bkg}$ & $\SWcdiff$ \\
\hline
$[0, 0.35]$   & $1815 \pm 96$  & $ 210 \pm 65$ & $0.68 \pm 0.04 \pm 0.03$  \\
$[0.35, 0.7]$ & $1609 \pm 98$  & $ 303 \pm 67$ & $0.52 \pm 0.04 \pm 0.03$  \\
$[0.7, 1.1]$  & $1657 \pm 98$  & $ 325 \pm 67$ & $0.51 \pm 0.03 \pm 0.02$  \\
$[1.1, 1.6]$  & $1675 \pm 103$ & $ 265 \pm 71$ & $0.44 \pm 0.03 \pm 0.02$  \\
$[1.6, 2.1]$  & $1097 \pm 91$  & $ 159 \pm 63$ & $0.32 \pm 0.03 \pm 0.02$  \\
\hline\hline
\multicolumn{4}{c}{$\PWmn$, $\pt^\mu > 35\GeV$} \\ \hline
$[{\abs{\eta}}_\text{min},{\abs{\eta}}_\text{max}]$ & $N_\text{sel}$ & $N_\text{bkg}$ & $\SWcdiff$ \\
\hline
$[0, 0.35]$   & $1517 \pm 86$ & $ 170 \pm 56$ & $0.66 \pm 0.04 \pm 0.03$  \\
$[0.35, 0.7]$ & $1364 \pm 87$ & $ 200 \pm 58$ & $0.54 \pm 0.04 \pm 0.03$  \\
$[0.7, 1.1]$  & $1407 \pm 86$ & $ 256 \pm 58$ & $0.51 \pm 0.03 \pm 0.02$  \\
$[1.1, 1.6]$  & $1381 \pm 90$ & $ 218 \pm 61$ & $0.42 \pm 0.03 \pm 0.02$  \\
$[1.6, 2.1]$  & $919  \pm 79$ & $  94 \pm 50$ & $0.33 \pm 0.03 \pm 0.02$  \\
\hline\hline
\multicolumn{4}{c}{$\Wen$, $\pt^\Pe > 35\GeV$} \\ \hline
$[{\abs{\eta}}_\text{min},{\abs{\eta}}_\text{max}]$ & $N_\text{sel}$ & $N_\text{bkg}$ & $\SWcdiff$ \\
\hline

$[0, 0.35]$   & $931\pm61 $ &  $153 \pm 42$ & $ 0.59 \pm 0.04 \pm 0.03 $  \\
$[0.35, 0.7]$ & $944\pm62 $ &  $200 \pm 42$ & $ 0.58 \pm 0.04 \pm 0.03 $  \\
$[0.7, 1.1]$  & $1031\pm63 $ & $128 \pm 43$ & $ 0.59 \pm 0.04 \pm 0.03 $  \\
$[1.1, 1.6]$  & $655\pm55 $ &  $155 \pm 38$ & $ 0.39 \pm 0.04 \pm 0.03 $  \\
$[1.6, 2.1]$  & $476\pm52 $ &  $83 \pm 35$ & $  0.32 \pm 0.04 \pm 0.03 $  \\

\hline\hline
\end{tabular}
\end{center}
\end{table}

\begin{table}[htbp]
\begin{center}
    \caption{Estimated number of $\OSSS$ events in the semileptonic sample (defined in Section~\ref{sec:dileptons}).
The estimated numbers of remaining background events after SS subtraction is given in the third column.
The normalized differential cross section as a function of the absolute value of the lepton pseudorapidity is shown in the last column.
The first two blocks of the table present the results from the $\PWmn$ sample, with $\pt^\mu > 25\GeV$  and
$\pt^\mu > 35\GeV$. The results from the $\Wen$ sample, with $\pt^\Pe > 35\GeV$ are given in the lowest block of the table.
The first error in the normalized differential cross section is due to the statistical size of the data sample and the second one is
the systematic uncertainty from to the sources discussed in Section~\ref{sec:syst_diff}.
    }
    \label{tab:dilepton_differential}
\begin{tabular}{c c c  c }
\hline\hline
\multicolumn{4}{c}{$\PWmn$, $\pt^\mu > 25\GeV$} \\ \hline
$[{\abs{\eta}}_\text{min},{\abs{\eta}}_\text{max}]$ & $N_\text{sel}$ & $N_\text{bkg}$ & $\SWcdiff$ \\
\hline
$[0, 0.35]$   & $3059 \pm 88$ & $ 941 \pm 66$ & $0.62 \pm 0.02 \pm 0.02$  \\
$[0.35, 0.7]$ & $3068 \pm 89$ & $1008 \pm 69$ & $0.57 \pm 0.02 \pm 0.02$  \\
$[0.7, 1.1]$  & $2976 \pm 89$ & $ 902 \pm 68$ & $0.54 \pm 0.02 \pm 0.02$  \\
$[1.1, 1.6]$  & $3004 \pm 93$ & $1040 \pm 72$ & $0.42 \pm 0.02 \pm 0.01$  \\
$[1.6, 2.1]$  & $2071 \pm 79$ & $ 687 \pm 63$ & $0.32 \pm 0.02 \pm 0.01$  \\
\hline\hline
\multicolumn{4}{c}{$\PWmn$, $\pt^\mu > 35\GeV$} \\ \hline
$[{\abs{\eta}}_\text{min},{\abs{\eta}}_\text{max}]$ & $N_\text{sel}$ & $N_\text{bkg}$ & $\SWcdiff$ \\
\hline
$[0, 0.35]$   & $2435 \pm 77$ & $ 751 \pm 59$ & $0.62 \pm 0.03 \pm 0.02$  \\
$[0.35, 0.7]$ & $2483 \pm 79$ & $ 823 \pm 61$ & $0.57 \pm 0.02 \pm 0.02$  \\
$[0.7, 1.1]$  & $2425 \pm 79$ & $ 713 \pm 59$ & $0.56 \pm 0.02 \pm 0.02$  \\
$[1.1, 1.6]$  & $2444 \pm 81$ & $ 891 \pm 62$ & $0.41 \pm 0.02 \pm 0.01$  \\
$[1.6, 2.1]$  & $1673 \pm 68$ & $ 578 \pm 54$ & $0.31 \pm 0.02 \pm 0.01$  \\
\hline\hline
\multicolumn{4}{c}{$\Wen$, $\pt^\Pe > 35\GeV$} \\ \hline
$[{\abs{\eta}}_\text{min},{\abs{\eta}}_\text{max}]$ & $N_\text{sel}$ & $N_\text{bkg}$ & $\SWcdiff$ \\
\hline

$[0, 0.35]$   & $1607 \pm 64 $ &  $213 \pm 43$ & $ 0.62\pm0.03 \pm0.02$  \\
$[0.35, 0.7]$ & $1574 \pm 64 $ &  $163 \pm 43$ & $ 0.62\pm0.03 \pm0.02$  \\
$[0.7, 1.1]$  & $1633 \pm 66 $ &  $208 \pm 46$ & $ 0.55\pm0.02 \pm0.02$  \\
$[1.1, 1.6]$  & $1078 \pm 58 $ &  $198 \pm 39$ & $ 0.38\pm0.02 \pm0.01$  \\
$[1.6, 2.1]$  & $815  \pm 54 $ &  $103 \pm 35$ & $ 0.31\pm0.02 \pm0.01$  \\

\hline\hline
\end{tabular}
\end{center}
\end{table}

\begin{table}[htbp]
\begin{center}
    \caption{Cross section ratios $\SWpc/\SWmc$ as a function of the absolute value of the lepton pseudorapidity from the $\PW$-boson decay
for the three samples: inclusive three-prong and two-prong and semileptonic.
The first two blocks of the table present the results from the $\PWmn$ sample, with $\pt^\mu > 25\GeV$  and
$\pt^\mu > 35\GeV$. The results from the $\Wen$ sample, with $\pt^\Pe > 35\GeV$ are given in the lowest block of the table.
    The last row of each block gives the cross section ratio for the full lepton pseudorapidity range [0., 2.1].
    The first error is due to the statistical size of the data sample and the second one is the systematic uncertainty due to the sources discussed in Section~\ref{sec:ratio}.
    }
    \label{tab:Rpm}
\begin{tabular}{c c c c }
\hline\hline
\multicolumn{4}{ c }{$\PWmn$, $\pt^\mu > 25\GeV$} \\ \hline
$[{\abs{\eta}}_\text{min},{\abs{\eta}}_\text{max}]$ & Three-prong sample & Two-prong sample & Semileptonic sample \\
\hline

$[0, 0.35]$   &  $0.877 \pm 0.087 \pm 0.004$ & $1.213 \pm 0.129 \pm 0.005$ & $1.047 \pm 0.076 \pm 0.010$  \\
$[0.35, 0.7]$ &  $0.973 \pm 0.104 \pm 0.005$ & $0.882 \pm 0.109 \pm 0.005$ & $0.990 \pm 0.075 \pm 0.009$  \\
$[0.7, 1.1]$  &  $0.837 \pm 0.091 \pm 0.006$ & $1.023 \pm 0.121 \pm 0.007$ & $0.890 \pm 0.071 \pm 0.015$  \\
$[1.1, 1.6]$  &  $0.999 \pm 0.114 \pm 0.007$ & $1.043 \pm 0.127 \pm 0.007$ & $1.114 \pm 0.089 \pm 0.030$  \\
$[1.6, 2.1]$  &  $0.898 \pm 0.115 \pm 0.010$ & $0.784 \pm 0.134 \pm 0.012$ & $0.709 \pm 0.078 \pm 0.028$  \\ \hline
$[0, 2.1]$    &  $0.915 \pm 0.045 \pm 0.003$ & $0.999 \pm 0.055 \pm 0.004$ & $0.959 \pm 0.035 \pm 0.009$  \\
\hline\hline
\multicolumn{4}{ c }{$\PWmn$, $\pt^\mu > 35\GeV$} \\ \hline
$[{\abs{\eta}}_\text{min},{\abs{\eta}}_\text{max}]$ & Three-prong sample & Two-prong sample & Semileptonic sample \\
\hline
$[0, 0.35]$   &  $0.844 \pm 0.092 \pm 0.005$ & $1.202 \pm 0.137 \pm 0.009$ & $0.991 \pm 0.080 \pm 0.009$  \\
$[0.35, 0.7]$ &  $0.912 \pm 0.106 \pm 0.006$ & $0.988 \pm 0.126 \pm 0.007$ & $1.044 \pm 0.085 \pm 0.011$  \\
$[0.7, 1.1]$  &  $0.801 \pm 0.096 \pm 0.006$ & $1.039 \pm 0.127 \pm 0.008$ & $0.933 \pm 0.080 \pm 0.016$  \\
$[1.1, 1.6]$  &  $0.946 \pm 0.117 \pm 0.007$ & $1.028 \pm 0.133 \pm 0.008$ & $1.030 \pm 0.088 \pm 0.028$  \\
$[1.6, 2.1]$  &  $0.873 \pm 0.124 \pm 0.010$ & $0.791 \pm 0.140 \pm 0.013$ & $0.779 \pm 0.089 \pm 0.031$  \\ \hline
$[0, 2.1]$    &  $0.873 \pm 0.047 \pm 0.003$ & $1.021 \pm 0.059 \pm 0.004$ & $0.965 \pm 0.038 \pm 0.009$  \\
\hline\hline
\multicolumn{4}{ c }{$\Wen$, $\pt^\Pe > 35\GeV$} \\ \hline
$[{\abs{\eta}}_\text{min},{\abs{\eta}}_\text{max}]$ & Three-prong sample & Two-prong sample & Semileptonic sample \\
\hline
$[0, 0.35]$   & $1.097 \pm 0.148 \pm 0.016$ & $0.924 \pm 0.123 \pm 0.012$ & $1.042 \pm 0.083 \pm 0.014$  \\
$[0.35, 0.7]$ & $0.990 \pm 0.133 \pm 0.014$ & $1.070 \pm 0.141 \pm 0.015$ & $0.832 \pm 0.068 \pm 0.011$  \\
$[0.7, 1.1]$  & $0.996 \pm 0.136 \pm 0.014$ & $1.054 \pm 0.130 \pm 0.014$ & $0.899 \pm 0.074 \pm 0.013$  \\
$[1.1, 1.6]$  & $0.920 \pm 0.137 \pm 0.013$ & $0.871 \pm 0.148 \pm 0.012$ & $0.865 \pm 0.095 \pm 0.014$  \\
$[1.6, 2.1]$  & $0.619 \pm 0.154 \pm 0.009$ & $0.581 \pm 0.142 \pm 0.008$ & $0.964 \pm 0.127 \pm 0.016$  \\ \hline
$[0, 2.1]$    & $0.953 \pm 0.063 \pm 0.013$ & $0.929 \pm 0.061 \pm 0.012$ & $0.917 \pm 0.038 \pm 0.012$  \\

\hline\hline
\end{tabular}
\end{center}
\end{table}

\cleardoublepage \section{The CMS Collaboration \label{app:collab}}\begin{sloppypar}\hyphenpenalty=5000\widowpenalty=500\clubpenalty=5000\textbf{Yerevan Physics Institute,  Yerevan,  Armenia}\\*[0pt]
S.~Chatrchyan, V.~Khachatryan, A.M.~Sirunyan, A.~Tumasyan
\vskip\cmsinstskip
\textbf{Institut f\"{u}r Hochenergiephysik der OeAW,  Wien,  Austria}\\*[0pt]
W.~Adam, T.~Bergauer, M.~Dragicevic, J.~Er\"{o}, C.~Fabjan\cmsAuthorMark{1}, M.~Friedl, R.~Fr\"{u}hwirth\cmsAuthorMark{1}, V.M.~Ghete, N.~H\"{o}rmann, J.~Hrubec, M.~Jeitler\cmsAuthorMark{1}, W.~Kiesenhofer, V.~Kn\"{u}nz, M.~Krammer\cmsAuthorMark{1}, I.~Kr\"{a}tschmer, D.~Liko, I.~Mikulec, D.~Rabady\cmsAuthorMark{2}, B.~Rahbaran, C.~Rohringer, H.~Rohringer, R.~Sch\"{o}fbeck, J.~Strauss, A.~Taurok, W.~Treberer-Treberspurg, W.~Waltenberger, C.-E.~Wulz\cmsAuthorMark{1}
% ...
\vskip\cmsinstskip
% ...
1:~~Also at Vienna University of Technology, Vienna, Austria\\
2:~~Also at CERN, European Organization for Nuclear Research, Geneva, Switzerland\\
% ...

\end{sloppypar}
\end{document}